\documentclass[11pt]{article}

\usepackage{geometry}                
\usepackage{xcolor}
\usepackage{graphicx}
\usepackage{amssymb}

\usepackage{url}                
\usepackage{hyperref}       
\usepackage{booktabs}      
\usepackage{amsfonts}      
\usepackage{nicefrac}        
\usepackage{microtype}     

\usepackage [round,authoryear]{natbib}
\usepackage{mathtools}

\newcommand{\figurecontent}[1]{#1}    

 \newcommand{\comment}[1]{}   

\newcommand{\mktall}[1]{}
\renewcommand{\mktall}[1]{\mbox{\rule[-1ex]{0ex}{2ex}{#1}}}

\newcommand{\mypar}[1]{ \qquad  \\ \noindent{\bf #1}}

\def\det{\mathop{\rm det}\nolimits}
\def\tr{\mathop{\rm tr}\nolimits}

\def\argmax{\mathop{\rm argmax}\nolimits}

\newcommand{\Eqn}[1]{Eq.~(\ref{#1})}
\newcommand{\Fig}[1]{Fig.~(\ref{#1})}

\renewcommand{\Hat}[1]{\hat{\bf #1}}
\renewcommand{\Vec}[1]{{\bf #1}}

\newcommand{\R}[1]{\mathbb{R}^{#1}}

\newcommand{\Sphere}[1]{\mathbf{S}^{#1}}

\newcommand{\SO}[1]{\mathbf{SO}({#1})}

\newcommand{\RP}[1]{\mathbf{RP}^{#1}}

\newcommand{\SLTC}[1]{\mathbf{SL2}({\mathbb{C}})}

\newlength{\ourfigwidthw}
\newlength{\ourfigwidth}
\newlength{\ourfigwidthh}
\newlength{\ourfigwidtht}
\newcommand{\showfile}[1]{}
  \renewcommand{\showfile}[1]{{\tt #1}\\[.2in]}

\newcommand{\ednote}[1]{}
  \renewcommand{\ednote}[1]{{\bf\large{\color{red} Note to Editor:}\ \/}%
       {\color{red} {#1}}}
     \renewcommand{\ednote}[1]{}

\newcommand{\earl}[1]{}
  \renewcommand{\earl}[1]{{\bf\large{\color{red} Note to Editor:}\ \/}%
       {\color{red} {#1}}}

\newcommand{\advanced}[1]{}
\renewcommand{\advanced}[1]{\begin{quote}
  \rule{0.75\ourfigwidth}{.5pt}\\ \nopagebreak
   $\dagger$ {#1}\\ \nopagebreak
   \rule{0.75\ourfigwidth}{.5pt}
   \end{quote}. }

\newcommand{\opt}{\mbox{\small opt}}
 \newcommand{\ID}{\mbox{\tiny ID}}
   \renewcommand{\t}{}
 \renewcommand{\t}{\mbox{\footnotesize t}}   
 
 \newcommand{\Rot}{R}

\def\ShowFiles{1}

\def\ShowFiles{0}

\title{  
   Exploring the Adjugate Matrix Approach\\  to Quaternion Pose Extraction}

  \author{ Andrew J. Hanson \\ Indiana University \\ Bloomington, IN\\ 
                 \texttt{hansona@indiana.edu} \\
    \and  Sonya M. Hanson\\  The Flatiron Institute\\ New York, NY \\
          \texttt{shanson@flatironinstitute.org}}
 \date{\today}

\begin{document}

\maketitle

\begin{abstract} 

\qquad 

Quaternions are important for a wide variety of rotation-related problems in computer
graphics, machine vision, and robotics.
We study the nontrivial geometry of the relationship between quaternions and rotation matrices
by exploiting the adjugate matrix of the characteristic equation of a related eigenvalue problem
to obtain the manifold of the space of a quaternion eigenvector.  We argue that quaternions  parameterized by their corresponding rotation matrices cannot be expressed, for example, 
in machine learning tasks, as single-valued functions:  the quaternion solution must instead
be treated as a manifold, with different algebraic solutions for each of several single-valued sectors
represented by the adjugate matrix.    We conclude with novel constructions exploiting
the quaternion adjugate variables to revisit  several classic  pose estimation applications:
2D point-cloud matching, 2D point-cloud-to-projection matching, 
3D point-cloud matching,   3D orthographic point-cloud-to-projection matching, 
and 3D perspective point-cloud-to-projection matching.   We find an exact solution
to the 3D orthographic least squares pose extraction problem, and apply it successfully
also to the perspective pose extraction problem with results that improve on existing methods.


\qquad 

   \end{abstract}

\section{Introduction}
\label{intro.sec}

\begin{figure}[h!]
\vspace{-0.0in}
\figurecontent{
\centering
   \includegraphics[width=5.5in]{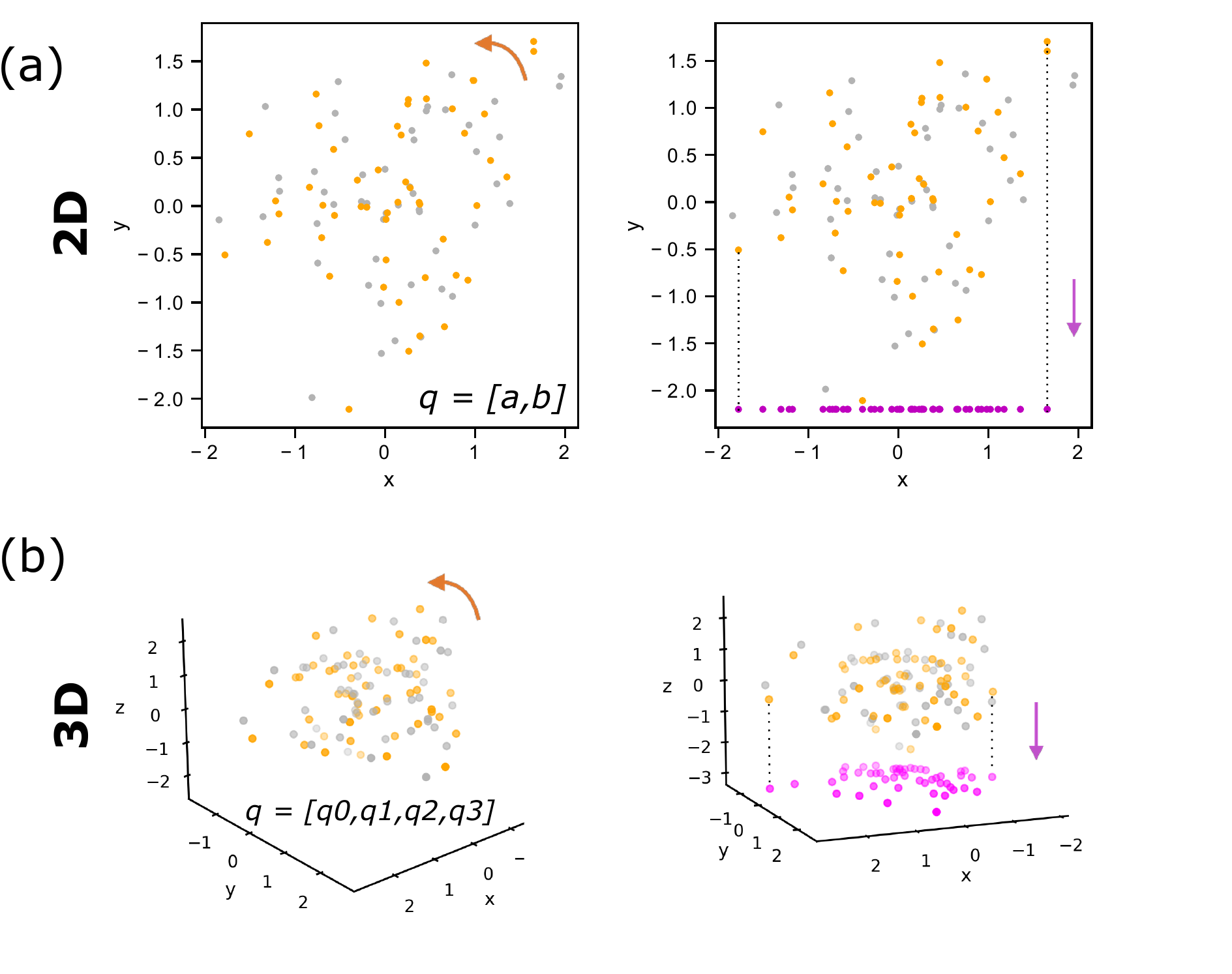}  } 
    \caption[]{\ifnum\ShowFiles=1 {\bf   PointClouds-dir/imgspdf/Figure1-v1.pdf. \\ }\fi  
    \ifnum\ShowFiles=1 {\it  old:  PointClouds-dir/imgspdf/new-figures-mar18-fig1.pdf. \\ }\fi \ 
 \footnotesize {\bf The fundamental point-cloud matching problems in 2D and 3D.} 
 Grey points indicate a reference point-cloud, orange points indicate a rotated point-cloud, and magenta points are projections. In this paper rotations are represented by quaternion rotation parameters: $q=[a,b]$ for 2D rotations and $q = [q_0,q_1,q_2,q_3]$ for 3D rotations.
 (a) Left: Alignment between a 2D reference cloud and a 2D test cloud differing by a rotation.
   Right: Alignment between a 2D reference cloud and a rotated cloud producing a 1D projected image. 
    (b) Left: Alignment between a 3D reference cloud and a 3D test cloud differing by a rotation.
   Right: Alignment between a 3D reference cloud and a rotated cloud producing a 2D projected image. 
} 
 \label {2D3DRMSDPose.fig}  
 \end{figure}

We address the task of understanding whether there are obstacles to using quaternions
to represent orientation space, typical examples being the determination of 
the optimal rotation to align two matched 2D or 3D point clouds, or find the pose
of the 2D or 3D point cloud that produced a given projected image; see \Fig{2D3DRMSDPose.fig}.  Our approach is to carefully study
how to compute the optimal quaternion corresponding to a given
measurement of a (typically inexact) rotation matrix.


The optimal-quaternion  extraction problem is universal, and occurs in many frameworks, reflecting the appealing fact that unit quaternions form a smooth manifold that parameterizes rotations free of Euler angle issues such as gimbal lock (see, e.g.  \citet{HansonQuatBook:2006,Hanson:ib5072}). 
Our investigation is motivated particularly by what has been referred to as the ``quaternion discontinuity problem''
in the context of machine learning in rotation space by, e.g., 
\citet{Saxena-Rot-Opt2009,zhou2019continuity,Peretroukhin2020,zhao2020quaternion,xiang2020revisiting}.
Understanding such potential issues is important, as the determination of orientation and
pose is widespread in machine learning applications, including self-driving vehicles,
drone navigation, and general problems of understanding 3D space, evaluating 3D models,
 and the extraction of 3D information from 2D data.

 Our first contribution is to show explicitly how  the topological properties of a quaternion, understood
 as a multi-sector manifold, resolve the questions regarding discontinuities posed in certain
 sectors of the machine learning literature.    Traditional
computational algorithms (see, e.g., \cite{Shepperd1978,SarabandiARK2018,HansonQuatBook:2006})  for 
extracting a corresponding quaternion from a 3D rotation matrix have always included
specific methods to account for possible singularities and discontinuities in the mapping,
but have not been clearly incorporated into some of the recent literature in which such
issues have been encountered.  We show how to exploit
 a classic linear algebra construction known formally 
as the \emph{adjugate matrix};
remarkably, the adjugate embodies an alternative set of quaternion-related variables that has
surprising use cases, greatly clarifies how the traditional quaternion extraction 
algorithms avoid singularities, and enables the exact solution of certain
challenging pose-estimation problems.
The adjugate suggests a new appreciation of the  variational method
for  quaternion extraction  introduced by \citet{BarItzhack2000},
and enables unique ways of applying least squares
methods to solve 3D rotation discovery problems.   In particular, we can use the
adjugate variables to find a closed-form formula solving a least squares optimization
formula for pose estimation.
We are thus able to explain the origin and resolution of the discontinuity problem,
and to further exploit our technology to provide novel insights into pose estimation.

\qquad 
 
 {\bf Outline:} 
 In the following, we illustrate our arguments beginning with a simplified and intuitive 2D rotation
 framework that exhibits essentially all the relevant properties.  We explore three ways of looking
 at the 2D problem in preparation for a parallel treatment of the application-relevant 
 3D quaternion-rotation case:
 we begin by considering 2D rotations as special cases of 3D rotations
 to produce a pair of formulas for the $2\times 2$ orthonormal
 2D rotation matrix.  We gain new insights by solving
 these for the 2D quaternion variables directly. We then explore
 the 2D version of a variational method due to 
 \citet{BarItzhack2000}, minimizing a difference measure between the  
 two relevant matrices.  Next, we replace the ideal, error-free 2D rotation matrix by a noisy version for
 which each matrix element is still close to a rotation matrix element, but is treated algebraically
 as distinct.   The variational methods expose in further detail how to 
 understand the multivalued nature of the 2D 
 quaternion extraction  problem.  We  repeat a similar analysis to derive the corresponding
 3D results, determining how to extract complete 
 quaternion information without  singularities or discontinuities from 3D rotation matrices, 
 for both ideal and  noisy data elements. Finally, we study some applications of the
 adjugate variables as parameters replacing the quaternions themselves in rotation
 optimization applications, and show in particular how the problem of estimating the pose
 in 3D space of a 2D point image relative to its corresponding 3D point cloud can be computed
 in closed form using only rational polynomials combined with a Bar-Itzhack optimization.

\qquad


\section{Fundamental Background} 
\label{fundamentals.sec}

   The arguments we present in this paper rely on a short list of key background concepts.  Our description of the problem at hand relies on the relationship between two ways of representing a rotation matrix.  Furthermore, error-robust quaternion extraction can be done in two basic ways. These four concepts are described below:\\[0.05in]
   
\noindent{\bf Representing rotations:}
\begin{itemize}
 \item   {\bf Quaternions parameterize a rotation in terms of a point on a topological three-sphere.}
    Any 4D vector $q$ with unit length\footnote{Henceforth, we will always be
     assuming unit-length quaternions.}, $q\cdot q = 1$, is a quaternion
      point on the unit three-sphere $\Sphere{3}$, and corresponds
       exactly to a \emph{proper} 3D  rotation matrix $R=R(q)$ through the equation
       \begin{equation}
    R(q) =  \left[ \!\!
 \begin{array}{ccc}
 {q_0}^2+{q_1}^2-{q_2}^2 - {q_3}^2 & 2 q_1 q_2  -2 q_0 q_3 
& 2 q_1 q_3 +2 q_0 q_2   \\
 2 q_1 q_2 + 2 q_0 q_3  &  {q_0}^2-{q_1}^2 + {q_2}^2 - {q_3}^2   
      &  2 q_2 q_3 - 2 q_0 q_1  \\
  2 q_1 q_3 - 2 q_0 q_2  &  2 q_2 q_3 + 2 q_0 q_1  & {q_0}^2 - {q_1}^2 - {q_2}^2 + {q_3}^2
\end{array} \!\!  \right]   .
\label{Rofqq.eq}
\end{equation}
  This fundamental equation is quadratic in $q$,  so $R(q) = R(-q)$, and thus
  every possible  rotation is represented \emph{twice} in the manifold $\Sphere{3}$.
   Alternatively,  one can say that every possible rotation appears \emph{once} in a 
   hyperhemisphere of $\Sphere{3}$,   the solid three-ball $\mathbf{B}^{3}$ 
   that can be drawn in ordinary 3D space.  In mathematical terms, \Eqn{Rofqq.eq}
   can also be described in terms of the group $\SO{3}$, whose
  topological  manifold is $\RP{3}$, the real projective 3-space, but
   we will not need to consider that feature in our treatment.

    \item  {\bf Quaternions encompass the axis-angle  rotation parameterization.}
      The axis-angle  representation  $R=R(\theta,\Hat{n})$ parameterizes any 3D rotation
       in terms of the unit eigenvector $\Hat{n}$ of $R$, the direction fixed by the rotation, and
       the angle $\theta$ of that rotation.  With $c =\cos \theta$ and $s =\sin \theta$,
       any rotation matrix can be written explicitly using axis-angle parameters as
 \begin{equation}
      R(\theta,\Hat{n}) =   \left[  
           \begin{array}{ccc}
          c  +(1- c  )\, {\hat{n}_1}^{\ 2} & (1- c  )\, \hat{n}_1 \hat{n}_2 - s \,  \hat{n}_3 &
                  (1- c  ) \, \hat{n}_1 \hat{n}_3+ s \, \hat{n}_2 \\
          (1- c  )\, \hat{n}_1 \hat{n}_2+ s  \, \hat{n}_3 &  c  +  (1- c  )\,  {\hat{n}_2}^{\ 2} &
                  (1- c  ) \, \hat{n}_2 \hat{n}_3 - s  \, \hat{n}_1 \\
         (1- c  ) \, \hat{n}_1 \hat{n}_3 - s   \,\hat{n}_2 &  (1- c  )\, \hat{n}_2 \hat{n}_3   + s \, \hat{n}_1&
                  c  + (1- c  ) \, {\hat{n}_3}^{\ 2} \\
       \end{array}    \right]    \  . 
\label{R.axisangle.eq}
\end{equation}
        Choosing the quaternion parameterization  
        \begin{equation}
        q(\theta,\Hat{n}) = ( \cos (\theta/2),\, \sin(\theta/2) \, \Hat{n} )
        \label{qthetanhat}
        \end{equation}
 and substituting it into  \Eqn{Rofqq.eq}  gives exactly   \Eqn{R.axisangle.eq}, double
          covered with $0 \leq \theta < 4 \pi$.
\end{itemize}

 \qquad 
 
 \pagebreak 
\noindent{\bf Error-robust quaternion extraction:}  
\begin{itemize}     
\item  {\bf Classical error-robust quaternion extraction is hard.}
    The rotation matrices described in \Eqn{Rofqq.eq}  and  \Eqn{R.axisangle.eq} are exact. 
    However, extracting the optimal quaternion representation of an inexact measured rotation 
    matrix $R(m)$, with measured numerical matrix elements $m_{ij}$,  is not an exact procedure.
     The task can be defined as the problem of recovering the best axis-angle parameters
          describing $R(m)$.  This is well-known to be a subtle multi-step process
          \citep[see, e.g.,][]{Shepperd1978,HansonQuatBook:2006,SarabandiARK2018}.   
                   In order to account for all possible anomalies, however rare, the classical
         procedure must check for zeros, conducting several separate checks for small numbers (see 
         Appendix  \ref{RotToQuatCode.app} for detailed pseudocode and examples).
           This algorithm reliably generates the axis-angle parameters  $\cos \theta$, $\sin \theta$,
           and $\Hat{n}$ needed to define a provably optimal $q(\theta, \Hat{n})$
           from the numerical data in $R(m)$. 
            Note that  this procedure assumes $R(m)$ is perfectly described 
            by \Eqn{R.axisangle.eq}, and thus does not  gracefully handle finding the 
            quaternion that is the best
            approximation for an imperfectly measured $R(m)$; this can be achieved using  
            the variational method of 
             \citet{BarItzhack2000}, which  will play a significant role in our narrative.

 \item {\bf The adjugate matrix approach to eigenvectors is important.}
       A standard method for finding the optimal quaternion corresponding to a rotation aligning two 3D point clouds
       uses the quaternion eigenvector corresponding to the maximal eigenvalue of the
        $4 \times 4$ \emph{profile matrix}   $M$ \citep[see, e.g.,][]{Horn1987,Hanson:ib5072}.
       In our treatment we will take advantage of a construction called the `adjugate'. 
       (Further details may be found in Appendix \ref{adjugatematrix.app}.)
       The adjugate of any square matrix $S$ is built from the matrix's transposed cofactors, 
       which facilitate the construction of the inverse through the following identity:
                  \begin{equation}
                        S \cdot \mbox{Adjugate}(S) = \det S\; I_{4} \ .     
                       \label{adjdef.eq}
           \end{equation}
       To see how this is exploited, we consider the characteristic matrix of $M$
          \begin{equation}
                          \chi = \left[ M - \lambda\, I_{4}  \right] \ ,
                    \label{char.eq}
          \end{equation}
    whose characteristic equation $\det \chi=0$, quartic in $\lambda$, 
      determines the eigenvalues of $M$.  
      We then insert the maximal eigenvalue $\lambda_{\opt}$
       into the matrix,  setting $\chi = \chi(\lambda_{\opt} )$, and multiply that
        characteristic matrix by its adjugate as follows:
          \begin{equation}
                        \chi \cdot \mbox{Adjugate}(\chi) = \det \chi\; I_{4} =0 \ .
                    \label{chiadj.eq}
          \end{equation}  
     This allows us, via \Eqn{char.eq}, to write the solved eigensystem of $M$ as
               \begin{equation}
                   M \cdot \mbox{Adjugate}(\chi) - \lambda_{\opt} \mbox{Adjugate}(\chi) =0 \ .
                  \label{adjfinal.eq}
          \end{equation}  
    We see that each of the adjugate's four column vectors is an 
    eigenvector of the single eigenvalue $\lambda_{\opt}$;  thus the adjugate provides
   \emph{four}  parallel solutions to the same eigensystem, and hence embodies
   four apparently equivalent  unnormalized  optimal quaternions $q_{\opt}$.
\end{itemize}

Starting from this list of observations,
 we will use the relationships between quaternions and rotation matrices
to show there are no singularity-free single functions relating a quaternion to a measured rotation
matrix, but that an adjugate matrix, listing four alternatives (technically eight, taking into account the
quaternion sign ambiguity) describing the entire quaternion manifold $\Sphere{3}$, always contains 
at least one normalizable column that produces a valid quaternion. 

 \qquad

\section{Two-Dimensional Rotations and the Quaternion Map }
\label{2DRotations.sec}

While quaternions are the most robust possible representation of rotations,
computing the relationship between a rotation matrix and a quaternion 
exposes unexpected singularities. 
The singularities we wish to investigate occur already for 2D rotations and
their quaternion counterparts, so we begin our journey in the simpler 2D space.
The context in the back of our minds is the exploration of  data sets of (reference, sample) pairs of
matched ND point clouds $(r,s)$ related by a single rotation matrix $R$,
\[  s = R \cdot r + \mbox{\it $<\!$ noise $\!>$} \ , \]
or
\[  s = R \cdot (r + \mbox{\it $<\!$ noise $\!>$} )\ , \]
depending on how one views the application context.
We will resolve apparent discrepancies in the use of quaternions 
to solve such problems first in 2D by introducing the adjugate variables.

\qquad

\subsection{Direct Solution of the Two-Dimensional Problem}
\label{2DdirectSoln.sec}

We begin with the simplified context of 2D rotations, obtained by
setting $q_{1} = q_{2} =0$  and $\hat{n}_{1} = \hat{n}_{2} =0$ in 
 Eqs.~(\ref{Rofqq.eq}) and  (\ref{R.axisangle.eq}) to restrict rotations
 to the $(x,y)$ plane, that is, fixing the $\Hat{z}$ axis.  For convenience,
we define  $c= \cos \theta$, $s = \sin \theta$,  $a = q_{0} =  \cos(\theta/2)$, and   $b =q_{3} = \sin(\theta/2)$, so $c^2 + s^2 =a ^2+b^2 =1$.
We observe that,  taking the range $0^{\circ} \le \theta < 720^{\circ}$,  $a$ and $b$  
cover the $(a,b)$ circle  only once, while the $(c,s)$ pair covers its circle  twice
over this range.
With these parameterizations, we can now construct ideal algebraic forms
of either a quaternion-parameterized 2D rotation matrix or  
a standard 2D rotation matrix as follows:    
\begin{align} \label{2DabRotN1.eq}
  \Rot(a,b) = &  \left[ \begin{array}{cc}  a^2 - b^2 & - 2 a b \\
                       2 a b &   a^2 - b^2 \end{array} \right] \\[0.1in]
\Rot(c,s) =& \left[ \begin{array}{cc} c & - s \\ s & c  \end{array} \right] 
 \label{2DabRotN2.eq}  \ .    
        \end{align}
  We easily verify that $ \det \Rot(a,b) = \det \Rot(c,s) = 1$, and also that $\Rot\cdot \Rot^{\t} = I_{2}$, where
     $I_{2}$ is the 2D identity matrix.   The most important property of  $\Rot(c,s)$ is that  its matrix elements
    correspond to a   \emph{numerically measurable} rotation matrix, as do noisy versions of $\Rot(c,s)$,
    which we will distinguish by the notation $\Rot(m)$ for separate treatment,
    while we will   see that $\Rot(a,b)$ has some intriguing ambiguities.

Now suppose we erase the formulas for $\Rot$ in terms of $\theta$, and think only
of the algebraic expressions in Eqs.~(\ref{2DabRotN1.eq},\,\ref{2DabRotN2.eq}), 
assuming that we have some
sound way of measuring this 2D rotation to determine the numerical values
of $(c,s)$.  Then we can find expressions for the now-abstract
variables $(a,b)$ in several ways. We begin by noting that both 
constraints $\Rot(a,b) = \Rot(c,s)$ and $\Rot(a,b)\cdot \Rot(c,s)^{\t}=I_{2}$
produce the same equations,   \begin{equation}
\  \begin{array}{rcl}a^2 - b^2    &=& c\\  2 a b & =& s \ , \end{array} 
  \label{R2solveabcs.eq}
  \end{equation} 
  which are simply the trigonometric half-angle formulas.
  We now take an important step: assuming the constraint $a^2 + b^2 = 1$,
  we can eliminate either $a^2$ or $b^2$, and complete our solution in terms of
  the measured rotation transformation parameters $(c,s)$ in two very distinct ways:%
  \begin{align}
\left. \begin{array}{rcl}
\mbox{\rm (1) Eliminate\ $b^2$:}\\[0.05in]
a^2 - b^2   = 2 a^2 -1 &=& c\\  2 a b & =& s \end{array} \right\}
  \ \  \mbox{\rm Solve for $(a^2,\, a b)$ } \ \ \rightarrow \ \ 
  a^2   =  \frac{ 1+c }{2}, \ \ a b  = \frac{s}{2} \ ,
  \label{R2solveasq.eq} \\
  \ \  \mbox{\rm\emph{Normalize} or solve for $(a,b)$ } \ \ \rightarrow \ \ 
  a   = \pm\frac{\sqrt{1+c}}{\sqrt{2}}, \ \ b  = \pm \frac{s}{\sqrt{2} \sqrt{1+c}} \ ,
  \label{R2solvea.eq}
  \end{align} 
  \begin{align}
 \left. \begin{array}{rcl} 
 \mbox{\rm (2) Eliminate\ $a^2$:}\\[0.05in]
a^2 - b^2   = 1-2 b^2   &= & c   \\   2 a b  &= & s \end{array} \right\}
 \ \  \mbox{\rm Solve for $(a b, \,b^2)$ }  \ \ \rightarrow \ \ 
   a b =     \frac{s}{2} ,  \ \   b^2 =    \frac{ 1-c}{2}  \ ,
    \label{R2solvebsq.eq} \\
 \ \  \mbox{\rm \emph{Normalize} or  solve for $(a,b)$ }  \ \ \rightarrow \ \ 
   a  =  \pm  \frac{s}{\sqrt{2} \sqrt{1-c}},   \ \   b =   \pm  \frac{\sqrt{1-c}}{\sqrt{2}}   
    \label{R2solveb.eq} \ .
  \end{align} 

The second set of solutions, Eqs.~(\ref{R2solvea.eq}) and (\ref{R2solveb.eq}), is seen to
be the same as the result  of normalizing Eqs.~(\ref{R2solveasq.eq}) and (\ref{R2solvebsq.eq}),
and these normalized forms are algebraically identical 
if we multiply them by the ratios  $\sqrt{1-c}/\sqrt{1-c}$ \, or \, $\sqrt{1+c}/\sqrt{1+c}$, 
 respectively.   \emph{This clearly gives situations that require multiplying by $0/0$:}
  the first normalized solution is impossible for $a\sim 0$, or $c \sim -1$, a perfectly
  legal rotation, and the second solution is impossible for $b \sim 0$, or
  $c \sim +1$, also perfectly legal!  In addition, \emph{both signs} in
  Eqs.~(\ref{R2solvea.eq}) and (\ref{R2solveb.eq}) are valid,
  as we have the same rotation $\Rot(a,b)$ if  $(a,b) \rightarrow (-a,-b)$.
    The problem, actually an important \emph{feature},
   is that one normalized solution, \Eqn{R2solvea.eq}, fails in one experimentally
  measurable domain, and the other, \Eqn{R2solveb.eq},  fails in a \emph{different} experimentally 
  measurable domain.  \emph{Both must be considered together}, along with their
  opposite signs, in order to completely cover the full multivalued 
  $720^{\circ}$  range of $\theta$ parameterizing $(a,b)$.  
  Those familiar with the long-standing quaternion extraction method of \citet{Shepperd1978}
   may recognize some basic features appearing in a novel context here, and in the
   full quaternion treatment later on:  there is in effect a condition on \emph{which} rotation matrix
   elements can be trusted to produce a regular quaternion.
   
\begin{figure}[h!]
\vspace{-0.2in}
\figurecontent{
\centering
  \includegraphics[width=6.1in]{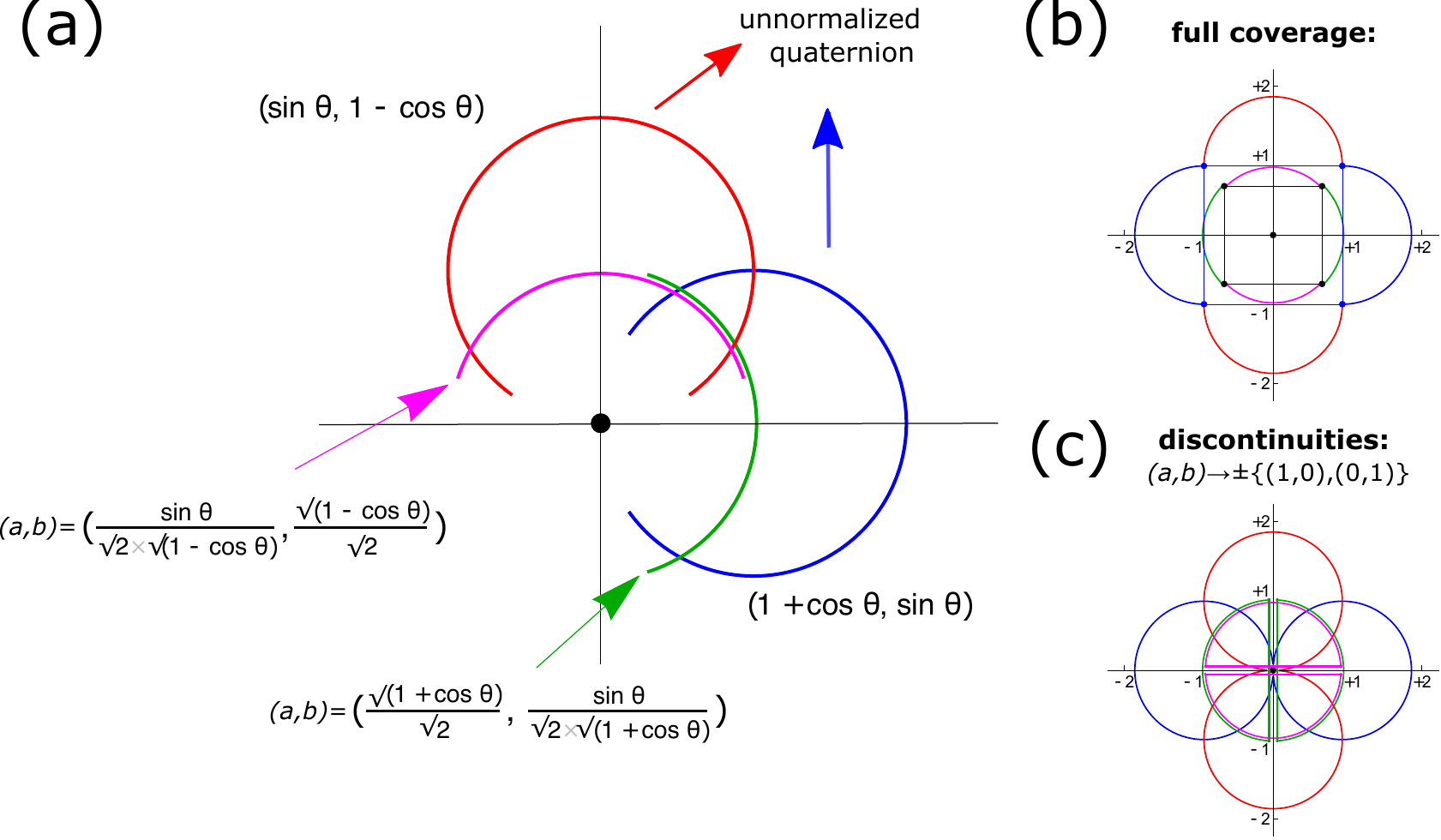} 
  }
    \caption[]{\ifnum\ShowFiles=1  {\bf Figure2-v1.pdf } \fi  
      \ifnum\ShowFiles=1   {\it   PointClouds-dir/imgspdf/new-figures-mar18-fig2.pdf. }\fi  
   \ifnum\ShowFiles=1 {\it originals: adj2UR-labeled.eps. }\fi  
   \ifnum\ShowFiles=1 {\it originals: Adj2Dab-picAL.eps,  Adj2Dab-picBx.eps,  Adj2Dab-picF.eps. }\fi  
   \footnotesize  {\bf Quaternion maps have discontinuities in 2D rotation space. } (a) The overlapping $(a,b)$ regions in the positive quadrants. 
  The normalized $x$ axis region is in green, derived from the unnormalized region in blue,
centered at $(1,0)$); this sector is regular at $0^{\circ}$, and singular at $\pm 180^{\circ}$ 
(remember that for $(a,b)$, the range of $\theta$ is $720^{\circ}$).
The  normalized  $y$ axis region is in magenta,  with corresponding singularites at $0^{\circ}$ and $360^{\circ}$.  derived from the unnormalized
region in red, centered at $(0,1)$), which
is regular at $180^{\circ}$ but singular at $0^{\circ}$ and $360^{\circ}$.  They overlap
in the neighborhood of $90^{\circ}$, so, together, they are regular over an entire range in the parameters
of $(a,b)$ that covers $360^{\circ}$.
  (b) Each of the four unnormalized maps that cover the full quaternion space has a singularity in
 the normalization. 
   The blue circles  are the paths of $\pm(1+ c,\, s)$, mapping to the green half-circles in $(a,b)$, failing at $c=-1,\,a=0$.
 The red circles are the paths of $\pm(s,\,1- c )$, mapping to the magenta half-circles in $(a,b)$,
 which fail at $c=+1,\, b=0$.  The curves along the positive axes, extending from $-45^{\circ}$ to $135^{\circ}$ actually  cover the whole rotation space;   all four  together  span
 the entire quaternion space.  (c)  The discontinuities appear as singular
 limits when the outer circles close in on the origin, and the divide-by-zeros at those limits
 break the normalized circles half-way through, indicated by the four ``half-pie'' shapes,
 two in each color. }
    \label {2Dquaternionsingularities.fig}  
 \end{figure}

   The left side of \Fig{2Dquaternionsingularities.fig}  shows how Eqs.~(\ref{R2solveasq.eq}) and (\ref{R2solvebsq.eq})
 describe unit circles passing through the origin,  with \emph{distinct centers} at $(1,0)$, $(0,1)$,
 while their normalizations, Eqs.~(\ref{R2solvea.eq}) and (\ref{R2solveb.eq}) for $a$ and $b$,
 are unit  \emph{half circles} centered at $(0,0)$, covering the positive $x$ axis and the positive $y$ axis,
 respectively.  The two domains of the $(a,b)$ solutions
 overlapping in  the first quadrant work  \emph{together} to cover each other's singular normalization
locations.  This shows us unequivocally how the variables $(a,b)$ and their multiple
solutions describe a \emph{manifold}, a topological space that cannot be described
by a single function, but requires overlapping descriptions. Incorporating both signs 
of the circles  of Eqs.~(\ref{R2solveasq.eq}) and (\ref{R2solvebsq.eq}), passing through the origin 
but with \emph{distinct centers} at $(1,0)$, $(0,1)$, $(-1,0)$, and $(0,-1)$,  gives 
the completed picture illustrated in the right side of  \Fig{2Dquaternionsingularities.fig}.  The non-singular almost-half-circles 
(one sees a ``half-pie'' shape) resulting from  normalization cover the
entire range of four progressively overlapping domains that \emph{together}
describe the possible values of $(a,b)$ over the whole range $0^{\circ} \le \theta < 720^{\circ}$
 with four local non-singular options.

 Observe that, given the variables $(a,b)$,  the four
singularities in their solutions in terms of $(c,s)$ occur when one variable or
the other vanishes, $(a,b) \to (0,\pm 1)$ and $(a,b) \to (\pm1, 0)$.  We shall see
in the full quaternion case that similar singularities occur in  14 submanifolds,
the loci where any legal combination of quaternion elements vanishes,  thus
requiring similar multiple overlapping representations.

 
 
  \comment{ 
\begin{figure}[h!]
\vspace{0.0in}
\figurecontent{
\centering
 \includegraphics[width=2. in ]{2DarcsA.eps} \hspace{.2in}
    \includegraphics[width=2. in ]{2DarcsB.eps}  \\
    \hspace{1.5in}  (a) \hfill (b) \hspace*{2in} \\
   \includegraphics[width=2. in ]{2DarcsC.eps} \hspace{.2in}
    \includegraphics[width=2. in ]{2DarcsD.eps}  \\
   \hspace{1.5in} (c) \hfill (d) \hspace*{2in} \\
  \vspace{0in} }
\caption[]{\ifnum\ShowFiles=1 {\bf 2DarcsA.eps, 2DarcsB.eps, 2DarcsC.eps,
2DarcsC.eps. }\fi
 \footnotesize  The ranges of the 2D pair of adjugate functions, which are circles
  passing through zero,  compared with their normalizations to the corresponding
  quaternions, which fail at $180^{\circ}$ for the blue circle, and at $0^{\circ}$ for the red
  arc.  Together, however, the two can cover the whole needed quaternion range.}
\label{2Darcs.fig}
\end{figure}   
  }  

\qquad 
    
    \comment{ Consider better intro:

[see comment] Next, we reexamine the question of solving the equations $\Rot(c,s) = \Rot(a,b)$ for $(a,b)$
using a variational method that will lead us directly to the  approach of 
 \citet{BarItzhack2000}  for full 3D quaternion rotations.
  
    } 
    
\subsection{Variational Approach: the  Bar-Itzhack Method in 2D}

  We now investigate the fact that the results of Section  \ref{2DdirectSoln.sec}  can
    be seen in an alternative light by using a variational
    approach \citep{BarItzhack2000} rather than an algebraic approach. 
      Seeing the quaternion appear as the solution to an
    eigensystem gives us  a novel way, exploiting the adjugate formula introduced
    in Section  \ref{fundamentals.sec},  to see how singularities can be systematically
    resolved in this problem.

  We begin by replacing the direct algebraic solution of the equations $\Rot(c,s) = \Rot(a,b)$ 
  with what is essentially a least-squares formulation,
minimizing the difference between the two matrices expressed using the Fr\"{o}benius norm,
\begin{equation} \label{BIFrobNorm2.eq}
\begin{split}
\mbox{ L}_{\mbox{\footnotesize \bf Fr\"{o}benius}} =&
 \tr\left( \left(\Rot(a,b) - \Rot(c,s)\right) \cdot \left(\Rot(a,b) - \Rot(c,s)\right)^{\t}\right) \\
  = & \tr\left( I_{2} + I_{2} - 2 \Rot(a,b)\cdot \Rot(c,s)^{\t} \right)  \ .
  \end{split}  
 \end{equation}
 At this point we can discard the constants and rephrase the problem of minimizing
 the least-squares version of the Fr\"{o}benius norm by a maximization of
 the cross-term, which we choose to write as
 \begin{equation} \label{BIFrobDelta2.eq}
\Delta_\mathbf{F} =   \tr \frac{1}{2}\Rot(a,b) \cdot \Rot(c,s)^{\t} \ .
\end{equation}
The task is now to maximize $\Delta_\mathbf{F}$ by finding $(a,b)_{\opt}$  such that
\begin{alignat}{3} \label{argmaxFrob2D.eq}
(a,b) _{\opt} & = \   \raisebox{-.9em}{$\stackrel{\textstyle\argmax}{\textstyle(a,b)} $}
     \left(\tr \frac{1}{2}\Rot(a,b) \cdot \Rot(c,s)^{\t}    \right)\\[0.0in] \label{qMcsMabq2.eq}
     & \hspace{-.45in} \mbox{\it -- now regroup quadratic terms in $(a,b)$ into  left and right vectors:}
         \nonumber \\[0.0in]
   & = \;   \raisebox{-.9em}{$\stackrel{\textstyle\argmax}{\textstyle(a,b)} $}
      \left( [a,b] \cdot K(c,s) \cdot [a,b]^{\t} \right)
\ \equiv \   \raisebox{-.9em}{$\stackrel{\textstyle\argmax}{\textstyle(a,b)} $}
        \left( [a,b] \cdot  \left[ \begin{array}{cc} c &  s \\ s & -c  \end{array} \right] \cdot  [a,b]^{\t}  \right)  .
 \end{alignat} 
 We refer to the matrix $K(c,s)$, which has  eigenvalues $ \lambda = \pm1$,
  as the \emph{profile matrix} of the  Bar-Itzhack optimization problem.   
    From classical linear algebra \citep{Golub-vanLoan-MatrixComp}, we know that
  the task of  maximizing  $\Delta_\mathbf{F}$ is equivalent to identifying 
    the maximal eigenvalue  of the symmetric real matrix  $K(c,s)$ and its eigenvector.  The 
    necessary eigenvalue is just $\lambda = +1$,  and   the corresponding eigenvector  
   is just  $(a,b)_{\opt}= \left(\; \cos(\theta/2),\sin(\theta/2)\;\right)$;
    this choice  for $(a,b)$ minimizes the Fr\"{o}benius norm
   \Eqn{BIFrobNorm2.eq}, and in fact sets it to zero in this simplified example.   But this misses
 the  crucial information noted in Eqs.~(\ref{R2solvea.eq}) and (\ref{R2solveb.eq}).  A more formal
and generalizable way to find the  eigenvector, which can have any non-zero scale without changing the eigenvalue,  is to form the characteristic equation's matrix, as noted in Section \ref{fundamentals.sec},
 by subtracting the eigenvalue $\lambda = +1$ from the diagonal,
\begin{equation}
\chi(c,s) \, = \, K(c,s) - (+1) I_{2} = \left[ \begin{array}{cc} c - 1&  s \\ s & -c -1 \end{array} \right] \ ,
\end{equation}
and extract the eigenvector from $\chi(c,s)$.
We now encounter the main point of this approach: the multiple forms of the solutions
for $(a,b)$ that we found by direct calculation earlier now appear \emph{automatically} in the variational
version.  The crucial fact is that, given that the determinant of $\chi$ vanishes, $\det \chi \equiv 0$,
\emph{both adjugate columns} of the matrix $\chi$ are \emph{unnormalized} eigenvectors of the 
given maximal eigenvalue.

  For our case, we see that   the
  two copies of the (unnormalized) maximal eigenvector are the two columns
  of the adjugate of $\chi$:
   \begin{equation} \label{2Dadjugate.eq}
    \mbox{Adjugate} (\chi) =  \left\{ \begin{array}{ccc}
  \left[  \begin{array}{c} 
      -1- c\\  - s \end{array} \right] & , &
    \left[ \begin{array}{c} -  s \\ -1+c \end{array}\right] \end{array} \right\} \ .
     \end{equation}
       Since the eigenvectors are insensitive to overall
  sign and scale, we are free to multiply  by $(-1/2)$ to get a more
  convenient form of the adjugate eigenvectors, which is
 \begin{align} \label{2DadjugateN.eq}
    \mbox{AdjEigVectors} (\chi) =& \left\{ \begin{array}{ccc}
  \frac{1}{2} \left[  \begin{array}{c} 
      1+ c\\  s \end{array} \right]  & ,  &
   \frac{1}{2}  \left[ \begin{array}{c} s \\ 1-c \end{array}\right] \end{array} \right\}  
   \end{align}
   We finally expose the \emph{Adjugate Matrix} for our eigenvector representation
   by using the angle $\theta$ to rewrite \Eqn{2DadjugateN.eq} in terms
   of our 2D quaternion $(a,b)$,
   \begin{align}      \label{2DadjugateNab.eq}
    \mbox{Adjugate Matrix}  =&   
    \left[  \begin{array}{cc}   a^2  &  a b \\
      a b   & b^2 \end{array}\right]    \ .
     \end{align}
    The problem here is by now familiar: neither eigenvector is a complete
    solution, as when we normalize, we find
    \begin{equation} \label{2DNormAdjugate.eq}
    \mbox{Normalized AdjEigVectors} (\chi) =  \left\{ \begin{array}{ccc}
  \left[    \begin{array}{c} 
     \frac{\textstyle\sqrt{1+c}}{\textstyle\sqrt{2}}\\[0.15in] 
         \frac{\textstyle s}{\mktall{\rule{0in}{1em} }  \textstyle \sqrt{2} \sqrt{1+c}} \end{array}   \right] & , &
    \left[ \begin{array}{c} \frac{\textstyle s}{\mktall{\rule{0in}{1em} }\textstyle \sqrt{2}\sqrt{1-c}} \\[0.2in] 
     \frac{\textstyle\sqrt{1-c}}{\textstyle\sqrt{2}} \end{array}\right] 
           \end{array} \right\} \ . 
     \end{equation} 
    The first column is singular at $c=-1 $, the second column at $c=+1$, both
    completely legal points, but neither normalized adjugate column is a valid
    quaternion-like 2-vector for  the \emph{entire}  range of the data $(c,s)$.

    \qquad
    
    From \Eqn{2DadjugateNab.eq}, we can see clearly that 
    both columns normalize to the eigenvector $(a,b)$ since $a^2 + b^2 = 1$.
    But that eigenvector is multiplied by $a$ in the first case, so no normalization
    is possible as $a\to 0$, and in the second column no normalization is
    possible as $b \to 0$. \emph{Both pre-normalization columns} of the adjugate matrix must
    be included to cover the entire space of rotations;  technically, due to the $(a,b) \to (-a,-b)$ equivalence,
    the full topological space of $(a,b)$  of course actually has four natural components.
    (See Figure   \ref{2Dquaternionsingularities.fig}.)
    To reiterate, the reason for
    this is that only $a^2$, $b^2$, and $ab$, the quadratic forms, can actually
    be expressed in terms of the measurable rotation matrix data $(c,s)$.
    We must have access to the \emph{entire} adjugate matrix because the entire
    circular quaternion path of $(a,b)$ describes a multivalued \emph{manifold}, 
     not a single-valued \emph{function}.
    With the conventional configuration of neural nets that only approximate functions,
    one can never determine a global solution for $(a,b)$ directly, but must
    target the multiple-valued  \emph{adjugate matrix} of locally
    normalizable solutions, with the choice of adjugate column occurring
    as a final data-driven post-processing step.  It is worthwhile noting
    that this is essentially the same type of process involved in the classic
    multiple-choice  Shepperd algorithm \citep{Shepperd1978} for extracting a quaternion
    from a 3D rotation matrix, but with the clearer
    purely linear algebraic adjugate-matrix context that arises naturally in the 
    Bar-Itzhack variational approach.

       \qquad

\subsection{Bar-Itzhack Errorful Measurement Strategy in 2D}

We have thus far assumed that measurements of a rotation matrix resulted in a perfect
orthonormal matrix.  That strategy allowed us to clearly expose the requirement to
use a multivalued formula to find the 2D quaternion analog $(a,b)$ from  the parameters $(c,s)$ 
of an ideal orthonormal measured rotation matrix.  The same basic approach is valid also for
inaccurate measurements that report rotation matrix elements that are not orthonormal.
 The basic ideas appear in the original work of \citep{Shepperd1978} and of \citet{BarItzhack2000}, 
 while further details of the 2D case appear in \citet{Haralick-pose-1989} and in
  the Supplementary  
 Material of \citet{Hanson:ib5072}.  
 
       
To see how to work with inaccurate  rotation matrix measurements, we introduce the ``measured
matrix data''  $R(m)$,
\begin{equation}
  R(m) = \left[
\begin{array}{cc}
m_{11} & m_{12}   \\
m_{21} & m_{22}  \\
\end{array}    \label{rotmat2Mij.eq}
\right] \ .
\end{equation} 
Comparing this to our ideal quadratic quaternion-like target for the same matrix, \Eqn{2DabRotN2.eq},
we immediately encounter the problem that we have two variables in $R(a,b)$ and four variables 
in $R(m)$.   We cannot solve directly for the 2D $(a,b)$ quadratic forms, and, unsurprisingly,
this problem still arises in 3D, as we will see below.

  \qquad 
  
 
We now argue that for noisy data, we gain clarity by applying the
 Bar-Itzhack optimization approach,  and that its validity is
much easier to understand and justify.   We begin with \Eqn{rotmat2Mij.eq} and
insert it into the Fr\"{o}benius norm for the distance between $R(m)$ and
\Eqn{2DabRotN2.eq} for $R(a,b)$, yielding
\begin{align} 
  S_{\mbox{\bf \footnotesize F} } =& \tr\left( R(a,b) - R(m)) \right)\cdot \left( R(a,b) - R(m))\right)^{\t} \nonumber\\
 = & \  (a^2 - b^2 - m_{11})^2 + (2 a b +m_{12})^2 + (2 a b - m_{21})^2 + (a^2 - b^2 - m_{22})^2 
      \nonumber \\
  = &  \ 2 + \sum_{i,j} \left( m_{ij}\right)^2 - 2(a^2 - b^2)\left(m_{11} + m_{22} \right)
       + 4 a b \left( m_{12} - m_{21} \right)  \ .
\label{FrobOfBI2D.eq}
\end{align}
  We strip the constants and change the sign to turn the problem of minimizing
   $S_{\mbox{\bf \footnotesize F} }$
  to the equivalent problem of maximizing the cross term, which, as before,
  we can now write immediately as this  matrix   product defining the profile
  matrix $K(m)$:
   \begin{align}  
  \Delta_{\mbox{\bf \footnotesize F} } =&  \   (a^2 - b^2)\left(m_{11} + m_{22} \right)
       + 2 a b \left( m_{21} - m_{12} \right) \nonumber \\
        = &\  [a \ b] \cdot  \left[\begin{array}{cc} \left(m_{11} + m_{22} \right)
 &   \left( m_{21} - m_{12} \right) \\
          \left(m_{21} - m_{12} \right) & - \left(m_{11} + m_{22} \right) \end{array} \right] 
        \cdot [a \  b ]^{\t}   \nonumber \
       \  \equiv  \   [a \ b] \cdot  K(m_{ij})   \cdot  [a \  b ]^{\t}   \  .
\end{align}
The solution to our optimization problem
\begin{align}
[a\ b] _{\opt} & =\raisebox{-.9em}{$\stackrel{\textstyle{\argmax}}{\textstyle(a,b)}$} \;
 [a \ b] \cdot   K(m)   \cdot   [a \  b ]^{\t} ] 
  \label{argmaxK2D.eq} 
  \end{align}
is then reduced to finding the eigenvector $[a,b]_{\opt}$ corresponding to the
maximal eigenvalue of $  K(m) $.  Since $\tr\left[ K(m)\right]=0 $, the
eigenvalues are an opposite-sign pair, with the maximal eigenvalue being the positive choice
\begin{equation}
\lambda_{\mbox{\rm max}} =  \sqrt{\left(m_{11} + m_{22} \right)^2 + \left(m_{12} - m_{21} \right)^2 }
\label{maxEigK2D.eq}
\end{equation}
Note that the appearance of only the antisymmetric part of the off-diagonal terms
in $R(m)$ is an inevitable consequence of the optimization of the Fr\"{o}benius norm
 \Eqn{FrobOfBI2D.eq}.

 The final step  is to cover the entire manifold of solutions for the multivalued
 $(a,b)$ using the adjugate matrix of the maximal eigensystem,
 \begin{align}
 \MoveEqLeft[35]{  \mbox{\rm Adj}(\,\left[ K(m) - \lambda_{\mbox{\rm max}} I_{2} \right] \,) =  } \nonumber\\
    =  \left\{ \left[ \begin{array}{c} - \left(m_{11} + m_{22} \right) - \lambda_{\mbox{\rm max}} \\
   m_{12} - m_{21} \end{array} \right]  \  , \ 
    \left[ \begin{array}{c}  m_{12} - m_{21}\\
    \left(m_{11} + m_{22} \right) - \lambda_{\mbox{\rm max}}
     \end{array} \right] \right\}
     \label{Adjmat2Da.eq}
\end{align}
We see immediately the automatic appearance of a pair of solutions for $[a,b]_{\opt}$ that have
singularities in different places when normalized, thus covering, with their negative counterparts,
the entire manifold of $(a,b)$.   Since the eigenvectors represented by the columns of the adjugate
matrix are insensitive to rescaling, we can transform \Eqn{Adjmat2Da.eq} to a more readable form
by changing the sign and denoting the trace and antisymmetric off-diagonal terms as follows:
\begin{alignat}{6} \label{dtLambda.eq}
d =& \frac{1}{2}\left( m_{11} + m_{22} \right) & \hspace{.5in} & t =& \frac{1}{2}\left( m_{21} - m_{12} \right)
 & \hspace{.5in}  &\lambda =& \sqrt{d^2+t^2}   \ .
\end{alignat}
We note that if our matrix were to be a pure rotation, we would have $d= \cos \theta$
and $t = \sin\theta$, so $\lambda = 1$.
We find the unnormalized eigenvector pairs
\begin{alignat}{4}
\mbox{\rm Adjugate\  eigenvectors} &= & \left\{ \pm \,\left[ \begin{array}{c}   
    \sqrt{t^2 + d^2 } + d\\ t
     \end{array} \right]  \right.
 &, \ \left.
 \pm\,  \left[ \begin{array}{c} t \\[0.05in]
    \sqrt{t^2 + d^2 }- d
     \end{array} \right]   \right\} \\[.1in]
 &=& \left\{ \pm \,\left[ \begin{array}{c}   
    \lambda + d\\ t
     \end{array} \right]  \right.
 &, \   \left.
 \pm\,  \left[ \begin{array}{c}  t \\
    \lambda-  d
     \end{array} \right]   \right\}    
     \ .
     \label{Adjmat2Db.eq}
\end{alignat}
 Equation (\ref{Adjmat2Db.eq}) is the exact analog for errorful rotations of the adjugate matrix 
 \Eqn{2DadjugateNab.eq}.  
 Normalizing \Eqn{Adjmat2Db.eq} to obtain quaternions, we see as before that 
 the two versions are equivalent except at singular points, but that one is
 always  a computable  normalized eigenvector:
 \begin{alignat}{4}
 \mbox{\rm normalized  Adj   eigenvectors} &= & 
  \left\{\pm \, \left[
\begin{array}{c}
 \frac{\mktall{\rule{0in}{1em} }\sqrt{\textstyle \lambda+d}}{\mktall{\rule{0in}{1.2em} }\textstyle \sqrt{2} \sqrt{\lambda}} \\[.25in]
   \frac{\textstyle t}{\mktall{\rule{0in}{1.2em} }\textstyle \sqrt{2} \sqrt{\lambda ( \lambda+d)}}
       \end{array} \right]  \right.
 &, \ & \left.
 \pm\,  \left[ \begin{array}{c}
 \frac{\textstyle t}{\mktall{\rule{0in}{1.2em} }\textstyle \sqrt{2} \sqrt{\lambda \left( \lambda - d \right)}} \\[.25in]
     \frac{\mktall{\rule{0in}{1em} }\textstyle \sqrt{\lambda -d }}{\mktall{\rule{0in}{1.2em} }\textstyle \sqrt{2} \sqrt{\lambda}} \\
\end{array}\right]
\right\} \ .
  \label{NormAdjmat2Db.eq}
  \end{alignat}
    \qquad


Observe that the numerical eigenvalue $\lambda$ appears throughout
in just such a way that if we have $\lambda = 1$,
 we have precisely the previous solution \Eqn{2DNormAdjugate.eq}
 for the normalized set of maximal eigenvectors.   
 It is worth noting that while \Eqn{dtLambda.eq} was mentioned in the
 supplement to \citep{Hanson:ib5072}, the importance of the adjugate
 and the singularities in the rotation-to-quaternion mappings 
 of Eqs.~(\ref{Adjmat2Db.eq})  and  (\ref{NormAdjmat2Db.eq})  were yet
 to be recognized.
 
 %

\qquad
%
\qquad

 \subsection{Summary}
 
 In this section, we have worked through the simple case of two dimensional rotations
 by an angle $\theta$, in parallel with how that corresponds to the 2D simplification of
 the quaternion-parameterized rotation matrix in \Eqn{Rofqq.eq} to a 2D version
 written in terms of the reduced quaternion  $(a,b)$.   This has led us to the understanding
 that in order to solve for $(a,b)$ in terms of the elements of a 2D measured rotation matrix,
 we must have two separate sectors, one regular at $(1,0)$ and singular at $(0,1)$, and the
 other reversed. To account for the full quaternion space where both $(a,b)$ and $(-a,-b)$ 
 correspond to the same 2D rotation $R(a,b)$, we in fact need four sectors covering
 the full circular manifold of the quaternion space.  We have also seen that the techniques
 that reveal the manifold properties of the quaternion space 
 also give us a way to find the \emph{optimal} exact rotation corresponding
 to a noisy set of rotation matrix data.  In the next Section, we perform a parallel analysis
 for 3D rotations and full quaternions, deriving and studying the results previewed in
 Section \ref{fundamentals.sec}.   We will again find that the clearest understanding of this problem is
 based on the adjugate matrix arising in the Bar-Itzhack optimization algorithm, and that
 noisy input data in particular are most clearly treated in this way.
 
 \qquad
 
 
 \section{ Three-Dimensional Rotations and the  Quaternion Map}
\label{3DRotations.sec}
 
 We now turn to the realistic case of interest, how to correctly determine a quaternion
 corresponding to a measured 3D rotation matrix, and presenting a complete derivation
 and explanation of the results summarized in Section  \ref{fundamentals.sec}.   As in
 Section \ref{2DRotations.sec}, we will begin with a direct derivation using only the
 symbolic forms of the quaternion-rotation problem, followed by the Bar-Itzhack
 variational version of the same problem, which will exhibit some new features.
 Again, the multivalued target required for, e.g., a rigorous formulation of a machine
 learning process, will be expressed in terms of an adjugate matrix.
 We  conclude with the corresponding variational treatment of noisy inexact rotation measurements
 producing the optimal true rotation matrix deduced from the noisy measured approximate
 rotation matrix data.
 
 \subsection{Direct Solution of the Three-Dimensional Problem}
 \label{3DdirectSoln.sec}
 
A \emph{proper}  orthonormal 3D rotation matrix   can be written as a quadratic form
 $R(q)$ in the quaternion elements $q = (q_{0},q_{1},q_{2},q_{3})$, with $q \cdot q = 1$,
 and identified with the axis-angle form $R(\theta,\Hat{n})$ as follows
\begin{align}
  R(q) =& R(\theta,\Hat{n})  
  \end{align}
  \begin{align}
\MoveEqLeft [6]   \left[  \begin{array}{ccc}
 {q_0}^2+{q_1}^2-{q_2}^2 - {q_3}^2 & 2 q_1 q_2  -2 q_0 q_3 
& 2 q_1 q_3 +2 q_0 q_2   \\
 2 q_1 q_2 + 2 q_0 q_3  &  {q_0}^2-{q_1}^2 + {q_2}^2 - {q_3}^2   
      &  2 q_2 q_3 - 2 q_0 q_1  \\
 2 q_1 q_3 - 2 q_0 q_2  &  2 q_2 q_3 + 2 q_0 q_1 
 & {q_0}^2 - {q_1}^2 - {q_2}^2 + {q_3}^2 
\end{array}
\right]  \nonumber \\[0.05in]
&=   \left[  
\begin{array}{ccc}
  c  +(1- c  )\, {\hat{n}_1}^{\ 2} & (1- c  )\, \hat{n}_1 \hat{n}_2 - s \,  \hat{n}_3 &
     (1- c  ) \, \hat{n}_1 \hat{n}_3+ s \, \hat{n}_2 \\
 (1- c  )\, \hat{n}_1 \hat{n}_2+ s  \, \hat{n}_3 &  c  +  (1- c  )\,  {\hat{n}_2}^{\ 2} &
      (1- c  ) \, \hat{n}_2 \hat{n}_3 - s  \, \hat{n}_1 \\
 (1- c  ) \, \hat{n}_1 \hat{n}_3 - s   \,\hat{n}_2 &  (1- c  )\, \hat{n}_2 \hat{n}_3   + s \, \hat{n}_1&
       c  + (1- c  ) \, {\hat{n}_3}^{\ 2} \\
\end{array}
    \right]     ,
\label{qrot.axisangle.eq}
\end{align}
where $\Hat{n}\cdot \Hat{n}=1$  and we abbreviate $c = \cos \theta$, $s = \sin \theta$.  
   As noted earlier, $\Hat{n}$ is the fixed axis about which we rotate by the angle $\theta$,
and parameterizing the quaternion as
\begin{equation}
q(\theta,\Hat{n}) = \left(\cos(\theta/2),\,\hat{n}_{1} \sin(\theta/2),\,
    \hat{n}_{2} \sin(\theta/2),\, \hat{n}_{3} \sin(\theta/2) \right) 
    \label{qAAform.eq}
    \end{equation}
    exactly reproduces  $R(\theta,\Hat{n})$.
We can easily rearrange \Eqn{qrot.axisangle.eq} (or just apply $q(\theta,\Hat{n})$)
to produce an explicit solution for the 10 quadratic forms in $q$,  written in terms
 a $4\times 4$ symmetric matrix that will turn out to be important in our narrative:
\begin{align}
  \MoveEqLeft[20] \left[
\begin{array}{cccc}
 {q_0}^2 & q_0 q_1 & q_0 q_2 & q_0 q_3 \\
 q_0 q_1 & {q_1}^2 & q_1 q_2 & q_1 q_3 \\
 q_0 q_2 & q_1 q_2 & {q_2}^2 & q_2 q_3 \\
 q_0 q_3 & q_1 q_3 & q_2 q_3 & {q_3}^2 \\
\end{array}
\right]    \!\! 
=  \frac{1}{2}
\left[
 \begin{array}{cccc}
 1+c & s \, n_{1}    & s \, n_{2}    & s \, n_{3}   \\
 s \, n_{1}    & (1-c)\, { n_{1}} ^2 & (1-c) \, n_{1}   n_{2}  & (1-c) \, n_{1}   n_{3}  \\
 s \, n_{2}   & (1-c)  \,n_{1}   n_{2}  & (1-c) \, {n_{2}} ^2 & (1-c) \, n_{2}   n_{3}  \\
 s \,  n_{3}    & (1-c) \, n_{1}   n_{3}  & (1-c) \, n_{2}   n_{3}  & (1-c)\, {n_{3}} ^2 \\
\end{array}
\right]  .
\label{4DQQAA.eq}
\end{align}
 Equation (\ref{4DQQAA.eq}) is the full quaternion analog of the unnormalized 2D
Equations~(\ref{R2solveasq.eq}) and (\ref{R2solvebsq.eq}), where we
  note that $(c,s)= (\cos\theta, \sin\theta)$ are in terms of the full
  angle, not the half-angle quaternion form.  In Eqs.~(\ref{R2solvea.eq}) and (\ref{R2solveb.eq}),
  we wrote out the explicit normalized quaternions and noted the two distinct
  singularities at $c=+1,\,c = -1$.  Here we choose to analyze the
  normalization factors separately to clarify the analysis.  Obviously
  each row of \Eqn{4DQQAA.eq} is normalized to a \emph{symbolically
  correct} quaternion $q = (q_{0},q_{1},q_{2},q_{3})$  by normalizing, in row-wise
  sequence, by  \\[-0.35in]  
  
  \begin{Large} 
  \begin{equation}    
  \begin{array}{cccccc}
  \mbox{\normalsize row 0:} & \frac{1}{q_{0}} & = & \frac{1}{\cos(\theta/2)} & = & \sqrt{\frac{2}{1+c}}\ \\
  \mbox{\normalsize row 1:} & \frac{1}{q_{1}} & = & \frac{1}{n_{1}\sin(\theta/2)} & = & \frac{1}{n_{1}} \sqrt{\frac{2}{1-c}}  \ \\
  \mbox{\normalsize row 2:} & \frac{1}{q_{2}} & = & \frac{1}{n_{2}\sin(\theta/2)} & = &  \frac{1}{n_{2}} \sqrt{\frac{2}{1-c}}\  \\
  \mbox{\normalsize row 3:} & \frac{1}{q_{3}} & = & \frac{1}{n_{3}\sin(\theta/2)} & = &  \frac{1}{n_{3}} \sqrt{\frac{2}{1-c}}   \ . \\
  \end{array}    
    \label{QQAAnorms.eq} 
     \end{equation}
  \end{Large}

 \noindent
We observe  that there are singularities in the normalization factors at new locations \emph{in addition}
to the $c= \pm 1$ singularities appearing in the 2D case.  This is easy to understand: our expression for
$q(\theta,\Hat{n})$ is arbitrary, and any permutation of the parameter elements is equally valid, so the
singularities \emph{should} be spread among the components without singling out any given one.
We can conclude that the 4D analog of the 2D pair of  $(a,b)\to \pm \{(1,0),(0,1) \}$  singularities in the
normalization is in fact this set of 14 distinct cases:
\begin{equation}   
\begin{array}{rcl}
\mbox{one\ zero} &\to & \{ (0, x,y,z),\, (x,0,y,z),\, (x,y,0,z),\,(x,y,z,0) \} \\
\mbox{two\ zeroes} &\to & \{ (0, 0,x,y),\, (0,x, 0,y ),\,  (0,x,y,0 ),\, (x,0,0,y),\,(x,0,y,0) ,\,(x,y,0,0)\} \\
\mbox{three\ zeroes} &\to &\pm  \{ (1,0,0,0),\, (0,1,0,0), \, (0,0,1,0),\,( 0,0,0,1) \} 
\end{array} \ ,   
\end{equation}
where $x^2+y^2 +z^2 =1$ in the first line and $x^2+y^2=1$ in the second line to preserve $q\cdot q =1 $.
Since the first two cases are spheres $\Sphere{2}$ and $\Sphere{1}$,
 the sign option $\pm$ in the third case is included; in fact, 
a more general way to think of the points $\pm1$ is as  just another sphere, 
the zero-sphere $\Sphere{0}$
solving $x^2 = 1$.
The key of course is that, since $q\cdot q = 1$, there always has to be at least one element
with $|q_{i}| \ge 1/2$, and thus we can always find a row that is normalizable.  These fourteen unnormalizable regions of the quaternion solution for a given rotation are illustrated
 in Figure  \ref{ZeroAdjSingAB.fig}.   In Appendix \ref{AdjZeroMatrices.app}, \Eqn{tableofSing.eq},
 we list the
explicit  adjugate matrices observed when each of these anomalies occurs,
along with visualizations  in Figures (\ref{obliqueS2Axes.fig}) and (\ref{4D8arcs.fig}).

A standard choice for selecting a well-defined solution from the $4\times 4$ matrix of alternate quaternion
solutions, in parallel to the algorithm of \citet{Shepperd1978}, is to note that the diagonal of the left-hand
side of \Eqn{4DQQAA.eq} is simply $\{ {q_0}^2,{q_1}^2,{q_2}^2,{q_3}^2\}$, so if we identify
the ordinal location $k$ of the maximal diagonal ${q_{k}}^2$, that is the row we normalize:
\begin{equation}
\mbox{\bf Normalizable solution:\  \ } q_{\opt}  = \frac{\pm 1}{|q_{k}|} ( {q_0}q_{k} ,{q_1}q_{k} ,{q_2}q_{k},{q_3}q_{k}) \ .
\label{optimalNormq.eq}
\end{equation}
The sign is of course arbitrary, though a standard choice is to make $q_{0}> 0$ when possible.
The significance of this for our problem is that, because the quaternion sphere $\Sphere{3}$ is a topological
manifold that cannot be described by a single function produced by a neural net, any algorithm that
needs to find \emph{a universally applicable quaternion} must produce a \emph{list of four candidates corresponding to the columns of \Eqn{4DQQAA.eq}},
remembering that there are 14 ways to fail normalizing any single one, and choose a normalizable
candidate from that list to produce a usable quaterniion via \Eqn{optimalNormq.eq}.

\qquad

\qquad

\subsection{Variational Approach: Bar-Itzhack in 3D}
\label{3DVariationalSoln.sec}

 The variational method we  presented for finding $(a,b)_{\opt}$ from experimental
 2D rotation matrix data extends straightforwardly to the 3D case to determine $q_{\opt}$.   The basic
 ideas appear in \citet{BarItzhack2000}, with some refinements in Appendix C and
 the Supplementary Material
 of \cite{Hanson:ib5072}; see also \cite{SarabandiR2Q2018,SarabandiToR2020}.
 
 The full 3D problem consists of finding the
 4D quaternion  $q_{\opt} = (q_0, \,q_1,\, q_2,\, q_3)$  such that $R(q_{\opt})$ 
 best describes a measured 3D numerical rotation  matrix $R$. 
 We begin as before with the ideal symbolic case, which is now  $R=R(\theta,\Hat{n})$.
 The task is to exploit quaternion features to minimize the Fr\"{o}benius norm of  the
  difference between the two matrices, where our initial optimization measure is
 \begin{equation} \label{BIFrobNorm.eq}
\begin{split}
{\mathbf L}_{\textstyle\mbox{Fr\"{o}benius}} =&
 \tr\left( \left(R(q) - \Rot(\theta,\Hat{n})\right) \cdot \left(R(q) - \Rot(\theta,\Hat{n})\right)^{\t}\right) \\
  = & \tr\left( I_{3} + I_{3} - 2 R(q)\cdot \Rot(\theta,\Hat{n})^{\t} \right)  \ .
  \end{split}  
 \end{equation}
 At this point we can discard the constants and rephrase the problem of minimizing
 the least-squares version of the Fr\"{o}benius norm in terms of maximizing 
 the cross-term, which we choose to write as
 \begin{equation} \label{BIFrobDelta.eq}
\Delta_\mathbf{F} =  \tr  R(q) \cdot R(\theta,\Hat{n})^{\t}  =   q \cdot  K(\theta,\Hat{n}) \cdot q\ .
\end{equation}
where expanding $R(q)$ using the form of \Eqn{qrot.axisangle.eq} allows us to rewrite the
trace in \Eqn{BIFrobDelta.eq} as matrix product using the \emph{profile matrix}
\begin{equation} \label{BIProfileA.eq}   
  K_{0}(\theta,\Hat{n})  
 =     \left[ 
\begin{array}{cccc}
  2  c+1 &  2 s \, \hat{n}_{1}  &  2  s  \, \hat{n}_{2}   &   2 s  \, \hat{n}_{3}   \\
 2   s  \, \hat{n}_{1}   &  2 (1-c) \,  \hat{n}_{1}^2 - 1  &  2(1-c) \,  \hat{n}_{1} \hat{n}_{2}  &  2(1-c) \,  \hat{n}_{1} \hat{n}_{3} \\
  2 s \,  \hat{n}_{2}   &  2 ( 1-c) \,  \hat{n}_{1} \hat{n}_{2} & 2  (1-c) \,  \hat{n}_{2}^2 - 1    & 2( 1-c) \,  \hat{n}_{2} \hat{n}_{3} \\
   2  s  \, \hat{n}_{3}   & 2  (1-c) \,  \hat{n}_{1} \hat{n}_{3} & 2  (1-c) \,  \hat{n}_{2} \hat{n}_{3} & 2  (1-c) \,  \hat{n}_{3}^2 - 1 \\
\end{array}
\right]
  \ .
  \end{equation}
The task is now to maximize $\Delta_\mathbf{F}$ by finding $q_{\opt}$  such that
\begin{equation}  \label{argmaxFrob3D.eq}
q _{\opt}   =\ \raisebox{-.9em}{$\stackrel{\textstyle\argmax}{\textstyle q} $}
     \left(\tr R(q) \cdot R(\theta,\Hat{n})^{\t}    \right)
      \ = \  \raisebox{-.75em}{$\stackrel{\textstyle\argmax}{\textstyle q} $}
      \left( q \cdot K(\theta,\Hat{n}) \cdot  q \right) \ .
 \end{equation}  
 Since $K_{0}(\theta,\Hat{n})$ is a symmetric real matrix, the maximum of $q \cdot K_{0} \cdot q$
 is achieved by picking out the maximal eigenvalue, so the corresponding $q_{\opt}$ is
 the maximal eigenvector.  The eigenvalues of  $K_{0}(\theta,\Hat{n})$ are $\lambda =(3,-1,-1,-1)$,
 so all we need to do is find the eigenvector of $\lambda = 3$.  However, we will first use some
 linear algebra manipulations to simplify the appearance of our equations. We note that
 \begin{itemize}
 \item  Since the characteristic equation we used to solve for the eigenvalues is
 $ \det( K_{0} - \lambda I_{4})=0 $, \emph{adding a constant} to $K_{0}$ simply adds the same constant
 to the eigenvalues, while \emph{scaling} $K_{0}$ simply scales the eigenvalues by  a constant.
 \item  The eigenvectors of any well-behaved eigensystem can be computed from the \emph{adjugate matrix}
  of the vanishing-determinant characteristic equation into which a valid eigenvalue 
 has been substituted.   This follows from the fact that for any matrix,
 $M\cdot \mbox{Adj}(M) = \det M\; I_{4}$; the latter
 equation also indicates that the four columns of $\mbox{Adj}(K_{0})$ are all formally eigenvectors
 of the \emph{same eigenvalue}. 
 \item The eigenvectors themselves can be multiplied by any non-vanishing scale factor without
 changing the eigenvalue equation, since the eigenvector equation is homogenous in the
 eigenvector itself.
 \end{itemize}


 Therefore, we can replace our original matrix $K_{0}(\theta,\Hat{n})$ in \Eqn{BIProfileA.eq} by
 adding one copy of the identity matrix and dividing by 4 to yield a new matrix 
 \begin{equation} \label{BIProfileB.eq}
  K(\theta,\Hat{n})  =\frac{1}{4}\left(K_{0}(\theta,\Hat{n}) + 1\times I_{4} \right)  
 =    \frac{1}{2} \left[ 
\begin{array}{cccc}
  1+ c &   s \,  \hat{n}_{1}  &   s \,  \hat{n}_{2}   &   s \,  \hat{n}_{3}   \\
   s \,  \hat{n}_{1}   &  (1-c) \,  \hat{n}_{1}^2  &  (1-c) \,  \hat{n}_{1} \hat{n}_{2}
   &  (1-c) \,  \hat{n}_{1} \hat{n}_{3} \\
   s \,  \hat{n}_{2}   &  (1-c) \,  \hat{n}_{1} \hat{n}_{2} &  (1-c) \,  \hat{n}_{2}^2 
   &  (1-c) \,  \hat{n}_{2} \hat{n}_{3} \\
   s \,  \hat{n}_{3}   &  (1-c) \,  \hat{n}_{1} \hat{n}_{3} &  (1-c) \,  \hat{n}_{2}
   \hat{n}_{3} &  (1-c) \,  \hat{n}_{3}^2  \\
\end{array}
\right]  \ .
  \end{equation}
 whose eigenvalues are now $(1,0,0,0))$, so the maximal eigenvalue is now $\lambda_{\opt}= 1$,
 but whose normalized eigenvectors are preserved. $K(\theta,\Hat{n}) $ has some interesting
 properties.  First, we see from  \Eqn{qAAform.eq} for $q(\theta,\Hat{n})$ that we have
 simply rediscovered \Eqn{4DQQAA.eq}, except that now we perceive it in a new light,
 as the root of an eigensystem whose maximal eigenvector determines $q_{\opt}$.  Furthermore,
 whichever of the two forms of $K(\theta,\Hat{n})= K(q)$  we use, the eigenvectors
 corresponding to the eigenvalues are just
 \begin{equation}
 \left\{ \begin{array}{cccc}
   \left[  \begin{array}{c}  q_0 \\q_1\\ q_2 \\ q_3 \\ \end{array} \right] &
   \left[  \begin{array}{c}  -q_1 \\q_0\\ 0 \\ 0 \\ \end{array} \right] &
   \left[  \begin{array}{c}  -q_2 \\0\\ q_0 \\ 0 \\ \end{array} \right] &
   \left[  \begin{array}{c}  -q_3 \\0 \\ 0 \\ q_0 \\ \end{array} \right]  
   \end{array} \right\} \ .
 \label{theKeigvecs.eq}
 \end{equation}
and the quaternion itself, $q = (q_0 ,\,q_1,\, q_2 ,\, q_3)$ is trivially the maximal
eigenvector with $\lambda_{\opt} = 1$.  

However, there is a deeper meaning in the eigensystem generated by the Bar-Itzhack
variational method that tells us everything that is important about the non-trivial manifold
in which quaternions live.  \emph{Given the profile matrix $K$}, we can compute the
maximal eigenvector corresponding to $\lambda_{\opt} = 1$ simultaneously in four
different ways by writing down the characteristic equation of $K$ with $\lambda = 1$
and computing the $4\times 4$ \emph{adjugate matrix}.  We can use any equivalent
form we like, but $K(q)$ is particularly simple: first we examine
\begin{align}
\mbox{characteristic equation:\ } = \; \chi \; &= K(q) - 1 \times I_{4} \\
  & =   K(q) - (q \cdot q) \times I_{4} \ ,
\label{KcharEqn.eq}
\end{align}
and then remarkably, when we compute the adjugate of the maximal
eigenvalue's characteristic equation for $K(q)$, we find it is simply
\emph{the negative  of $K(q)$ itself}:
\begin{align}
\mbox{Adjugate}(\chi) =& -  \left[
\begin{array}{cccc}
 {q_0}^2 & q_0 q_1 & q_0 q_2 & q_0 q_3 \\
 q_0 q_1 & {q_1}^2 & q_1 q_2 & q_1 q_3 \\
 q_0 q_2 & q_1 q_2 & {q_2}^2 & q_2 q_3 \\
 q_0 q_3 & q_1 q_3 & q_2 q_3 & {q_3}^2 \\
\end{array}
\right]   = - K(q) \ .
\label{KAdjrEqn.eq}
\end{align}
Recall that the minus sign can be removed and the positive quadratic matrix employed
to represent the family of alternative unnormalized maximal eigenvectors, since the
eigenequation is insensitive to the scale of the eigenvectors.
We already know these solutions for $q_{\opt}$ are correct, since each row (or column, as it is symmetric) is
proportional to the maximal eigenvector $q = (q_0 ,\,q_1,\, q_2 ,\, q_3)$.  However,
in addition we observe a repetition of our observation in Section \ref{3DdirectSoln.sec}
that the four rows of superficially equivalent solutions are \emph{not} equivalent, but indicate that
any of the fourteen combinations of appearances of zeroes in one, two, or three of
the quaternion components $(q_0 ,\,q_1,\, q_2 ,\, q_3)$ renders the entire row useless
for computing the correct quaternion corresponding to the measured rotation matrix,
and another quadratic row with nonsingular normalization must be used for the
calculation.

\qquad

\qquad



\subsection{Bar-Itzhack Variational Approach to 3D Noisy Data}
\label{BarItzh-3D-noise.sec}

We have found the solution for the quaternion manifold's four solutions (eight with signs)
in terms of a perfect orthogonal measurement of the rotation data.  As noted by
\citet{BarItzhack2000},  the same basic
procedure can be used for measured rotation matrices with errors, and the resulting quaternion
produces a perfect orthonormal rotation matrix that is the \emph{optimal approximation}
to the provided errorful  data.   The issue of handling rotation data with wide-ranging errors
has been studied by \citet{SarabandiR2Q2018}; their heuristic approach to determining
an optimal rotation omits critical steps in the noisy-data treatment of Bar-Itzhack that
we handle rigorously here.
Our starting point is a measured $3\times 3$ matrix $R(m)$ that is assumed to originate from a
3D rotation matrix, but cannot be guaranteed to be orthonormal due to measurement
error;   we write the components of this input data matrix as
\begin{equation}
  R(m) = \left[
\begin{array}{ccc}
m_{11} & m_{12} & m_{13} \\
m_{21} & m_{22} & m_{23} \\
m_{31} & m_{32} & m_{33} \\
\end{array}  \label{rotmatMij.eq}
\right] \ .
\end{equation} 
We set up the Bar-Itzhack variational problem starting with a symbolic
rotation in the quadratic quaternion form $R(q)$ given in \Eqn{Rofqq.eq},
and write down the cross-term of the Fr\"{o}benius norm to define a maximization problem
that will be our optimization target:
\begin{equation}
\Delta_\mathbf{F} (q,m)=  \tr R(q) \cdot R(m)^{\t}  =   q \cdot  K_{0}(m) \cdot q \ .
\end{equation}
Our initial profile matrix resulting from rearranging the quadratic quaternion terms
into the form of a scalar-valued symmetric matrix multiplication takes the form
\begin{align}   
K_{0}(m) = &  
\left[             
\begin{array}{cccc}
 m_{11}+m_{22}+m_{33} & m_{32}-m_{23} & m_{13}-m_{31} & m_{21}-m_{12} \\
 m_{32}-m_{23} & m_{11}-m_{22}-m_{33} & m_{12}+m_{21} & m_{13}+m_{31} \\
 m_{13}-m_{31} & m_{12}+m_{21} & -m_{11}+m_{22}-m_{33} & m_{23}+m_{32} \\
 m_{21}-m_{12} & m_{13}+m_{31} & m_{23}+m_{32} & -m_{11}-m_{22}+m_{33} \\
\end{array}
\right]
\label{noisyK0mat.eq}
\end{align}
Note that $K_0(m)$ is traceless, and since $K_0(m)$ is a real symmetric matrix, it will have real eigenvalues,
The  eigenvector  $q_{\opt}$ of  $K_0(m)$'s  maximal eigenvalue $\lambda_{\opt}(m)$   will maximize 
$\Delta_\mathbf{F} (q,m)$.   This maximal eigensystem  will solve the optimization problem  
 \begin{align}  \label{argmaxFrob3DRmat.eq}
q _{\opt}   =\  \raisebox{-.9em}{$\stackrel{\textstyle\argmax}{\textstyle q} $}
     \left(\tr  R(q) \cdot R(m)^{\t}    \right)
      \ = \  \raisebox{-.75em}{$\stackrel{\textstyle\argmax}{\textstyle q} $}
      \left( q \cdot K_{0}(m) \cdot  q \right) \\[0.1in]
    \lambda _{\opt}   = \Delta_\mathbf{F} (q_{\opt},m) = \left( q_{\opt} \cdot K_{0}(m) \cdot  q_{\opt} \right)
    \  .
 \end{align} 
However, to be a proper quaternion, the optimizing value of the eigenvector $q$  of 
$\lambda_{\opt}$ will have to be normalized to become $q_{\opt}$, and we have
argued throughout that this is not always possible, and must be dealt with using the quaternionic
manifold $\Sphere{3}$ with eight covering coordinate patches instead of relying on a single function.

Fortunately, we know from the exact-rotation-data  case in the preceding section how to deal correctly with 
this issue in the Bar-Itzhack context.   We note that while it was useful in the exact case to get a
very clean set of formulas by performing an eigenvector-preserving transformation of the form
\begin{equation}
K(m)  = \mbox{scale} \times \left( K_{0}(m) + \mbox{constant} \times I_{4} \right)
\end{equation}
to adjust our maximal eigenvalue to the identity, in the general case, we cannot find a single
pair of constants that will be all that useful, though one might choose to normalize to obtain
a unit maximal eigenvalue by dividing by the value of $K_{0}$'s maximal eigenvalue
$\lambda_{\opt}(m)$ .  We will assume that if there is some reason to readjust
$K_{0}(m)$ to a form $K(m)$ with the same eigenvectors, up to   scaling,
preserving the corresponding maximal eigenvalue $ \lambda_{\opt}(m)$ up to an additive constant,
we may do so, and thus we will continue with that abstract $K(m)$ to complete our argument.

Obviously what we have to do first is find $\lambda_{\opt}(m)$.  Standard numerical eigensystem
software packages can easily accomplish this, and, for symmetric real matrices up to $4 \times 4$ 
in size, one can even calculate the maximal eigenvalue analytically using Cardano's solution of   4th degree
polynomials (see, e.g, \cite{Hanson:ib5072} for a review).  The last step is then to
form the characteristic equation's matrix as before, using now the \emph{numerical} eigenvalue, giving
\begin{align}\label{KofmCharEqn.eq}
\mbox{characteristic matrix:\ } =  \chi(m) = & K(m) - \lambda_{\opt}(m) \times I_{4} \\
  = &  K(m) - (q_{\opt} \cdot K(m) \cdot q_{\opt}) \times I_{4} \nonumber \\[.0in]
  \mbox {where} &  \nonumber\\
   \det \chi(m) \equiv& 0  \nonumber\  .
\end{align}
Finally, our full quaternion-space covering manifold solution is
determined from the adjugate of \Eqn{KofmCharEqn.eq},
which is an entirely numerical matrix having the following (possibly
scaled) relation to the sign-ambiguous set of eight possible quaternion
formulas:
\begin{align}  
 \left[  \begin{array}{cccc}
 {q_0}^2 & q_0 q_1 & q_0 q_2 & q_0 q_3 \\
 q_0 q_1 & {q_1}^2 & q_1 q_2 & q_1 q_3 \\
 q_0 q_2 & q_1 q_2 & {q_2}^2 & q_2 q_3 \\ 
 q_0 q_3 & q_1 q_3 & q_2 q_3 & {q_3}^2 \\
\end{array}  \right]  = & \mbox{Adjugate}(\chi(m))  \ .
\label{KAdjrEqn2.eq}
\end{align}
Recall that the adjugate is determined only up to a nonvanishing scale of either sign,
and that one picks the ordinal index $k$ with the largest diagonal value ${q_{k}}^{2}$
 in the adjugate matrix, and normalizes that row to obtain $q_{\opt}$. Finally, one 
 calculates the \emph{optimal pure rotation approximation} to the numerically
 measured $R(m)$ using this quaternion selected from the adjugate matrix:
 \begin{equation}
 R_{\opt}(m) = R(q_{\opt})
 \end{equation}
 and our treatment of how to compute a quaternion from any ideal or noisy rotation
 matrix data is done.
 
 \qquad

\subsection{Graphical Illustration of the 3D Rotation Case}

We can get an intuitive feeling for what is going on with the multiple valid regions
for the quaternion solution representations be drawing representative spaces corresponding
to the unnormalized (and potentially singular) solutions, and then making pointwise
normalization maps from those spaces to the actual quaternion subspaces, and noting
where validity of the normalization map fails.

\begin{figure}[h!]
\vspace{-0.5in}
\figurecontent{
\centering
   \includegraphics[width=6.25in]{figspdf/Figure3-v1} }
   \vspace{-0.2in}
  \caption[]{\ifnum\ShowFiles=1 {\bf 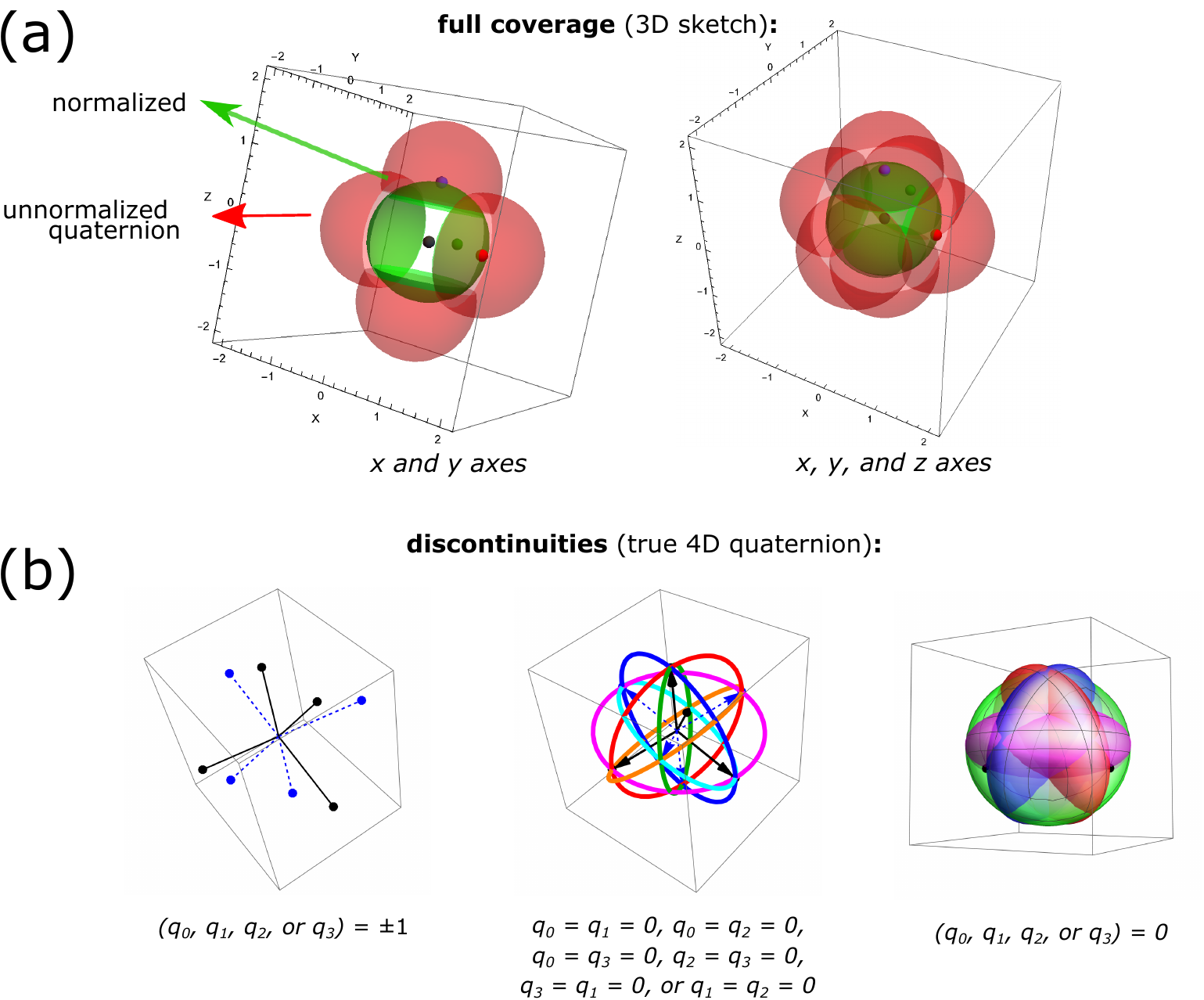}\fi 
  \ifnum\ShowFiles=1 {\it newer: new-figures-mar18-fig3.pdf}\fi 
    \ifnum\ShowFiles=1 {\it original: S2-projection-4.eps, S2-projection-6.eps, }\fi    
   \ifnum\ShowFiles=1 {\it original: threeZeroAdjSing.eps,  \ twoZeroAdjSing.eps, 
                                justOneZeroAdjSing.eps, oneZeroAdjSing.eps. }\fi
 \footnotesize 
  {\bf   (a)}   {\bf 3D subspace showing three axes of singularities.}
  In this 3D subspace of quaternion space, there are partial spheres instead
  of partial circles as in 2D, but the singularity occurs in the same way: as the sphere
  closes in on the origin, normalization is impossible.  At left,  the $x$ and $y$
  axes coincide with the four red spheres that produce partial coverings for the inner green
  quaternion-subspace spheres.  At the right, we include the $z$ axis  component
  and show all six singularity-limited quaternion-subspace normalized hemispheres in green,
  giving the 3D analog of the 2D case \Fig{2Dquaternionsingularities.fig}. \\
  {\bf (b)}
    {\bf  Visualizing the fourteen quaternion singular-normalization domains.}
  (Left)     The allowed quaternions if any three components vanish are the  four $\Sphere{0}$
  subspaces  at the  ends of the 4D axes, $q_{0}=\pm 1$, $q_{1}=\pm 1$, $q_{2}=\pm 1$, 
  and $q_{3}=\pm 1$.
  (Middle) Allowed quaternions with two zeroes are six topological circles
 $\Sphere{1}$.    A single circle projected to 3D from the unit quaternion 
 with $q_{0}=q_{1}=0$ is just the curve  $(0,0,x,y)$ with $x^2+y^2 = 1$.
 The union of all six curves matches the six edges of a complex tetrahedron.
  (Right) Regions of singularity for quaternions with a single zero are topological spheres
 $\Sphere{2}$ that correspond to four faces of a complex tetrahedron, with the allowed $q_{0}=0$ 
 subspace, for example, being the spherical surface   $(0, x,y,z )$ with $x^2+y^2 +z^2= 1$.
    }
 \label  {ZeroAdjSingAB.fig} 
 \end{figure}

In 2D, these maps were fairly simple to see in Figure  \ref{2Dquaternionsingularities.fig},
 where a simple unnormalized circle mapped to a half-circle
in the 2D reduced quaternion plane.  Going one step higher in complexity, we
can produce images in 2D that correspond to a quaternion map, but one dimension
lower.  The top part of Figure \ref{ZeroAdjSingAB.fig} illustrates pairs of 2D 
spheres embedded in $\R{3}$
centered at $(\pm 1, 0,0)$, $(0,\pm 1,0)$, and $( 0,0,\pm1)$, shown in red.
Taking each point on the red spheres and applying the normalization operation,
we obtain the green spheres centered at the origin $(0,0,0)$.  As the red spheres'
points approach the origin, the normalization approaches a divide-by-zero at the
equator of each green sphere, and the map can go no farther.

The bottom part of Figure \ref{ZeroAdjSingAB.fig} illustrates the geometry in 
quaternion space of the 14 singular subspaces, where one or more normalizations
of quaternion parameterizations in terms of rotation matrices will fail, and a test
must be made to identify a legal normalization step.  When one of the four sets
of three vanishing quaternions occurs, only the point pair at the tip of a 4D axis
is permitted, as shown in the left bottom part of the figure.  If one of six pairs
of zeros occurs, the six circles in the middle part denote the allowed subspace
of quaternions, and if only one quaternion vanishes, there are four allowed
spherical subspaces as shown at the left.  These fourteen collections of
4, 6, and 4 subspaces correspond roughly to the vertices, edges, and faces of
a complex tetrahedron, as a tetrahedron has 4 vertices, 6 edges, and 4 faces.
Additional illustrations of the higher dimensional properties of these singularities
are given in Appendix \ref{AdjZeroMatrices.app}, Figures (\ref{obliqueS2Axes.fig})
 and (\ref{4D8arcs.fig}).



  \clearpage

\comment{ 
 \begin{figure}[h!]
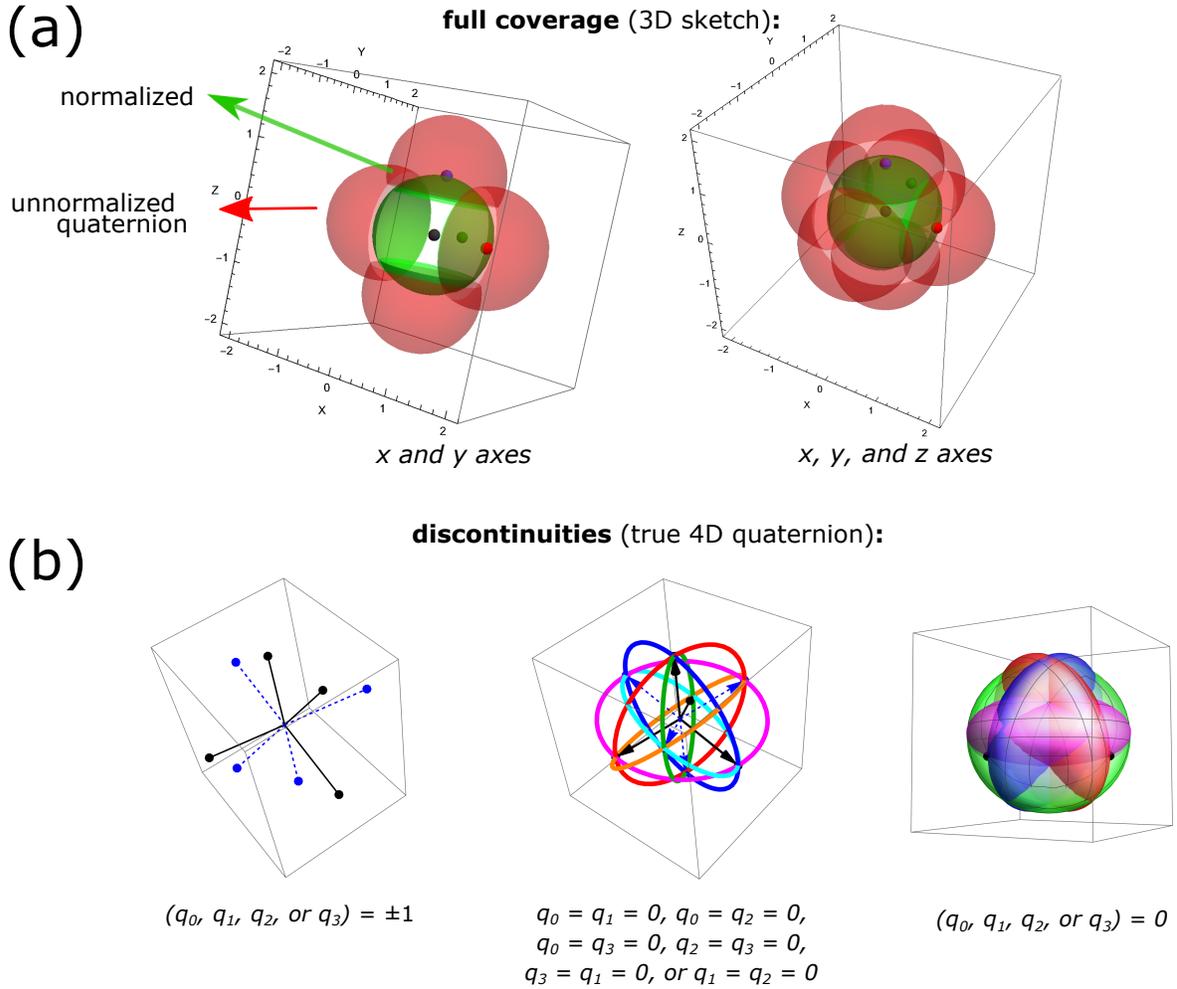

\vspace{0in}
\figurecontent{
\centering
 \includegraphics[width=2.7 in ]{S2-projection-4.eps} \hspace{.1in}
    \includegraphics[width=2.7 in ]{S2-projection-6.eps}  \\
    \hspace*{1.2in}  (a) \hfill (b)   \hspace*{1.2in} \\
    } 
\caption[]{\ifnum\ShowFiles=1 {\bf S2-projection-4.eps, S2-projection-6.eps. }\fi
 \footnotesize  
 {\bf 3D subspace showing three axes of singularities.}
  In this 3D subspace of quaternion space, there are partial spheres instead
  of partial circles, but the singularity occurs in the same way, as the sphere
  closes in on the origin, normalization is impossible.  (a) The $x$ and $y$
  axes coincide with the four red spheres that produce partial coverings for the inner green
  quaternion-subspace spheres.  (b) Adding in the $z$ axis  to show the full
  story of how this subspace of quaternions is covered in a nonsingular fashion
  be three pairs of partial spheres analogous to the $(1+c,s)$ circular arcs
  in \Fig{2D4arcs.fig}. 
  }
 \label{3D6arcs.fig}
\end{figure} 
    } 

\section{Applications of the Adjugate Representation}

We now apply the quaternion adjugate variable representation to the classic problems
of estimating the optimal rotations needed to align sets of pointwise matched
data. The basic appearance of the data for each of the   problems we shall treat
is summarized visually in   \Fig{2D3DRMSDPose.fig} in the Introduction. 
 We  first examine in sections 5.1 and 5.2 the simple but pedagogically instructive 2D cloud
problems, namely matching rotation-related pairs of 2D clouds, and then
pose discovery given a 2D cloud and a 1D point-matched image derived
from that cloud.  The more relevant 3D cloud matching case has been solved
in many ways, so again our adjugate variable approach in section 5.3 is  mainly of pedagogical
interest.  The reader will find the most interesting results in section 5.4, in which
we exploit the adjugate reduction in the power of the least squares loss function
for orthographic 3D-cloud-to-2D-image matching, discovering a new closed form
solution to that least squares problem. Finally, in section 5.5, we apply the methods
of our orthographic pose estimation solution to produce a novel and highly accurate
three-step procedure to solve the 3D-to-2D perspective projection pose estimation
problem.  Comparisons with frequently cited standard results for this problem show
that our sparse adjugate-based procedure matches or exceeds the loss profiles for random
simulated data sets of other known methods, which rely on  more
complicated  iterative  procedures.


\mypar{Universal Least-Squares Loss Function Framework.}
We choose as our framework the basic least squares formulas arising when we require a
set of template reference data to be rotated to agree with another set of 
(assumed noisy) measured data.   The universal least squares loss function applicable
to all of our matching problems applies an unknown rotation to the fixed reference set 
and compares it to  a  jittered test data set, and takes the following form:
\begin{equation}  
\begin{aligned}  
 {\mathbf S}_{\mbox{\footnotesize (2D,3D : Match,Pose)}} & =
              \sum_{k=1}^{K} \left\| R(\mbox{quaternion variables}) \cdot  \Vec{x}_{k} - \Vec{u}_{k} \right\| ^2 \\
     &  =\sum_{k=1}^{K} \left( \Vec{u}_{k}\cdot \Vec{u}_{k} - 
          2 \Vec{u}_{k}  \cdot  R(\mbox{vars}) \cdot  \Vec{x}_{k}     
     \:  + \;  \Vec{x}_{k} \cdot R^{\t}(\mbox{vars}) \cdot R(\mbox{vars}) \cdot  \Vec{x}_{k} \right) \\
        & = \tr( {U} \cdot  {U}^{\t})  - 2 \tr \left(  R(\mbox{vars}) \cdot     {X} \cdot  {U}^{\t}  \right)
               + \; \tr \left( R^{\t}(\mbox{vars}) \cdot R(\mbox{vars}) \cdot   {X}  \cdot  {X}^{\t}   \right) \ .
 \end{aligned} 
 \label{UniversalLoss.eq}
\end{equation}
Here $\{\Vec{x}_{k}\}$ is the set of $K$ reference points describing a cloud in 
dimension $D=2$ or $D=3$,
with $ {X}$ denoting a $D \times K$  matrix allowing us to absorb the sums over $k$.
We write  $\{\Vec{u}_{k}\}$ or  $\ {U}$  for the set of $K$ test points.
For the cloud-to-cloud matching
problem,  ${X}$ and  ${U}$  denote  reference data and rotated  data of dimension $D=2$ or $D=3$,
 and for these same-dimension matching   problems, \Eqn{UniversalLoss.eq} greatly simplifies
 due to $R^{\t} \cdot R = \mbox{Identity Matrix}$, reducing the least squares minimization
 problem to the equivalent maximization problem for the cross term 
 \begin{equation}
 \Delta = \tr ( R  \cdot     {X} \cdot  {U}^{\t})
 \label{UniversalCrossTerm.eq}
 \end{equation}
 after eliminating all constant terms.  For the pose-estimation problem, these become a (corresponding-point-matched) projected image cloud of dimension $D=1$ or $D=2$, and
 the necessarily incomplete rotation matrices in the projection process must remain,
 so the full  form of \Eqn{UniversalLoss.eq} must be retained.
 
 \comment{. 
   For the cloud-cloud matching problems,
we take the rotation $R$ to be a quaternion-based paramaterization, either
the 2D rotation $R(a,b)$ or the 3D rotation $R(q)$; for these full
rotations, $R^{\t} \cdot R = (q\cdot q)^{2} \times I_{D} =\mbox{(Identity Matrix)}$, 
and thus the last term reduces
to a constant $ {X}\cdot  {X}^{\t}$;  only the term \emph{linear} in $R$ appears,
so the problem is \emph{quadratic} in $q$, which makes it much easier to solve.
However, for the pose-estimation problem, although obtaining  the full rotation $R$ is
still our goal,  the rotation matrix in \Eqn{UniversalLoss.eq} must be reduced to
a projection by dropping the last line, that is,  reducing to
either a $1\times2$ projection matrix $P(a,b)$ from 2D to 1D, or a $2\times3$ 
projection matrix $P(q)$ from 3D to 2D.  

Thus for pose estimation,  the last term   in \Eqn{UniversalLoss.eq}
 \emph{does not  reduce to a constant},
but in fact re-introduces   \emph{quartic quaternion terms}  into the problem, so the
quadratic quaternion matrix eigensystem methods frequently exploited in cloud-cloud 
matching problems no longer apply.  However, we emphasize that the pose problem 
is still solvable, as the orthonormality conditions on $R(q)$ permit the missing
bottom row to be reliably reconstructed from  the cross-product of the first two rows,
or, indeed, any five elements of the first two rows. 
}  

There are a variety of approaches to solving each of these problems to obtain the
rotation matrix that optimally aligns the two data sets $ {X}$ and $ {U}$.  We will
pay particular attention to the behavior of the least squares optimization problem
represented by the loss function \Eqn{UniversalLoss.eq} when the quaternion parameters of
the rotation matrices and projections are expressed in terms of the non-singular quadratic
 parameterizations that we have been exploring.
We will hereafter define the variables resulting from the quadratic quaternion forms
as  the \emph{adjugate variables}, which we now define explicitly as:
\begin{equation}
\label{substAdj.eq} 
\begin{aligned}
\mbox{2D:\ \ } &R(a,b) \!\! &  \to   R(a^2,b^2,a b)  \to \, &R(\alpha,\beta,\gamma)&\ \ \ \ &
    P(a,b) \!\! &  \to   P(a^2,b^2,a b)  \to \, & P(\alpha,\beta,\gamma) \\
\mbox{3D:\ \ } &R(q) \!\!&  \to  \ \ \  R(q_{i} q_{j}) \ \ \   \to \,  &     R(q_{ij}) &\ \ \ \ &
 P(q) \! \!&  \to  \ \ \  P(q_{i} q_{j}) \ \ \  \to \, &   P(q_{ij})   \ .
\end{aligned}
\end{equation} %
The adjugate matrices themselves are thus given by \Eqn{2DadjugateNab.eq} and \Eqn{KAdjrEqn.eq}, 
reiterated here for convenient reference:
\begin{equation} \label{AdjugateSummary,eq}
\left. \begin{aligned}
A(a,b) & =
 \left[ \begin{array}{cc} a^2 & a b \\ a b & b^2 \\
\end{array} \right]  \, \equiv \, \left[ \begin{array}{cc}
             \alpha &  \gamma\\ \gamma & \beta\\
     \end{array} \right]  \\[0.2 in]
A(q) &=  \left[
\begin{array}{cccc}
 {q_0}^2 & q_0 q_1 & q_0 q_2 & q_0 q_3 \\
 q_0 q_1 & {q_1}^2 & q_1 q_2 & q_1 q_3 \\
 q_0 q_2 & q_1 q_2 & {q_2}^2 & q_2 q_3 \\
 q_0 q_3 & q_1 q_3 & q_2 q_3 & {q_3}^2 \\
\end{array} \right] \, \equiv \,  \left[
\begin{array}{cccc}
   q_{00}  & q_{01} & q_{02} & q_{03} \\
  q_{01} &  q_{11} & q_{12} & q_{13} \\
 q_{02}  & q_{12} & q_{11} & q_{23} \\
  q_{03} & q_{13} & q_{23} & q_{33} \\
\end{array} \right]
\end{aligned} \right\}  \ .
\end{equation}

While the basic mathematics is of course unchanged, the conversion from
formulas quadratic or quartic in $q$ to formulas linear or quadratic,
respectively, in $q_{ij}$
leads to some additional insights, not to speak of being explicitly nonsingular,
based on the arguments in the preceding sections.


 We consider applying the loss \Eqn{UniversalLoss.eq} here in two ways.
 One canonical standard for obtaining a numerical quaternion from
 the loss function is the \emph{argmin} function, which takes as its
 input the measured data points and the unknown rotation parameters,
 with their constraints, which are the single equation $ q \cdot q = 1 $
 for a pure quaternion formulation.
 
 However, our purpose here is to add the quaternion adjugate variables
 to the arsenal of our analysis tools.   We already are familiar with
 changing variables from the  four unit-quaternion
 elements $q = (q_{0},\, q_{1}, \, q_{2},\, q_{3})$ to their
 ten possible quadratic expressions   written as $q_{ij}$,
 denoting variables whose behavior corresponds to $q_{i} \times q_{j}$.
 Since the unit-length constraint $q\cdot q = 1$ implies there are only
 three independent variables available, there must be seven  constraints
 on the ten quaternion adjugate variables $q_{ij}$ derived from their definition.
 For example, $q_{00}=q_{0} q_{0}$,   $ q_{11}=q_{1} q_{1}$, and
$q_{01}=q_{0} q_{1}$  impose the constraint $q_{00} \, q_{11}  =  {q_{01}}^2$.
While there are a number of alternative ways of extracting our needed
seven constraints, we have found the following to be a universally
useful combination: 
 \begin{equation} \label{TheqqConstraints.Eq}
\begin{aligned}
         q_{00} + q_{11}  + q_{22}+ q_{33}   &=   1      \\
q_{00} \, q_{11} &=  {q_{01}}^2 , &&  q_{00} \, q_{22} &= {q_{02}}^2, &&
        q_{00} \, q_{33} & = {q_{03}}^2 \\
  q_{22} \, q_{33} &= {q_{23}}^2, &&   q_{11} \, q_{33} &= {q_{13}}^2,  &&
  q_{11}\,  q_{22}  &=    {q_{12}}^2 \  .  \\   
\end{aligned}
\end{equation}
This set of constraints reduces the ten adjugate variables to three
independent rotation parameters, and using standard numerical
constraint solution software provided in systems like
Matlab or Mathematica, we get reasonable quaternion
answers for noisy data in all but very unusual cases using \emph{argmin}
on the least squares loss function combined with constraints on the
manifold of legal variations, e.g., 
%
 \begin{align*} \label{UniversalArgmin.eq}
 \mbox{\rm argmin}& [ \,\sum_{\mbox{\small points}} \{\,\left\|\, R[q]\cdot \mbox{\rm reference}[x,y,z] -
                          \mbox{\rm test}[u,v[,w]]\, \right\|^2,\\
         & {q_{0}}^{2} + {q_{1}}^{2} + {q_{2}}^{2}+  {q_{3}}^{2} = 1 \},\\
         &\{ q_{0}, q_{1}, q_{2}, q_{3}\} \ ]\\[0.25in]
\mbox{\rm argmin}& [ \,\sum_{\mbox{\small points}} \{\,\left\|\, R[q_{ij}]\cdot \mbox{\rm reference}[x,y,z] -
                          \mbox{\rm test}[u,v[,w]]\, \right\|^2,\\
&\{q_{00} + q_{11} + q_{22} +  q_{33} = 1, q_{00} q_{11} = {q_{01}}^2, q_{00} q_{22} = {q_{02}}^2, \\
&   q_{00}  q_{33} = {q_{03}}^2, 
    q_{22}  q_{33} = {q_{23}}^2,  q_{11} q_{33} = {q_{13}}^2, 
  q_{11} q_{22} =    {q_{12}}^2 \}\}, \\
  &\{ q_{00}, q_{01}, q_{02}, q_{03}, q_{11}, q_{12}, q_{13}, q_{22}, q_{23}, q_{33} \} \ ] \ .
   \end{align*}
 Appropriately configured machine learning applications presumably should
 function identically to \emph{argmin}.  However,  there are many choices
 to be made in designing neural networks for such purposes, and it may,
 for example, be challenging to reliably design an implementation,
 possibly leading to the observed misbehavior of such networks. 
 We hope to deal with these issues more thoroughly in a separate venue.
 
 Here, we will focus on a top-level overview of the ways in which quaternion
 adjugate variables can provide insights not only into the relationship between
 quaternions and rotations in numerical settings, but also into the 
 least squares equations for point-cloud matching and their algebraic solutions.  
 In the remaining parts of this Section,  we will show how these two contexts
 are intimately related.
  
\subsection{2D Point Cloud Orientation Matching}

We begin with the simplest example, which we will label as
``2D Matching,'' namely the problem of aligning a pair of 2D point clouds, 
where  $\{\Vec{x}_{k}\} = \{[x,y]_{k}\}$ is a set of $K$ 2D column vectors
describing points in a reference set,  and $\{\Vec{u}_{k}\} = \{[u,v]_{k}\}$  describes
a set of 2D test points.  Following our conventions in \Eqn{UniversalLoss.eq},
the least squares loss function then takes the form
\begin{align}
\mathbf{S}_{\mbox{\small (2D Match)}} =& \sum_{k=1}^{K} 
\left\| R \cdot \Vec{x}_{k} -  \Vec{u}_{k} \right\|^2  \ .
\end{align}
The corresponding optimization problem is easily solved in closed form in terms of the $2\times 2$
cross-covariance matrix 
\[ E_{ab}= \sum_{k}{x_{a}}^{k} {u_{b}}^{k} =
\left[ \begin{array}{cc}   x\! \cdot \! u & x\! \cdot \! v \\  y\! \cdot \! u &  y\!  \cdot \! v \\ \end{array}  \right]  \]
 using a wide variety
of methods, including quaternion forms exploiting the parameterization $R(a,b)$
  \citep[see, e.g.,][supporting information, Sec.~5]{Hanson:ib5072}.
We now examine what happens if we parameterize $R$ not by the 2D quaternion
itself, $q = [a,b]$, but by the adjugate-inspired variables $\{\alpha,\beta,\gamma\}$
replacing $\{a^2,b^2, a b\}$, subject to the constraints  $\alpha+\beta = 1$ 
and $\alpha \beta = \gamma^{2}$.  With $R(\alpha,\beta,\gamma)$, we thus find the loss function %
%
\begin{align} \label{S2DCloudPair.eq}
\mathbf{S}_{\mbox{\small \rm (2D Match) }} =& 
\, u\!\cdot\! u + v\!\cdot\! v - 2 \alpha \, x\!\cdot\! u + 2 \beta \,x\!\cdot\!  u 
   - 4 \gamma\, x\!\cdot\! v + {\alpha}^2 x\!\cdot\! x + 4 {\gamma}^2 x\!\cdot\! x - 
 2 \alpha \beta \, x\!\cdot\! x +  \nonumber\\
  & {\beta}^2 x^2 + 4 \gamma\, y\!\cdot\! u- 2 \alpha \,y\!\cdot\! v 
  + 2 \beta \, y\!\cdot\! v + {\alpha}^2 y\!\cdot\! y + 
 4 {\gamma}^2 \,y\!\cdot\! y - 2\alpha \beta \,y\!\cdot\! y +{\beta}^2 y\!\cdot\! y  \nonumber\\
  \MoveEqLeft[6]{=  
  \, u\!\cdot\! u + v\!\cdot\! v - 2 \alpha \, x\!\cdot\! u + 2 \beta \,x\!\cdot\!  u 
   - 4 \gamma\, x\!\cdot\! v    +   4 \gamma\, y\!\cdot\! u- 2 \alpha \,y\!\cdot\! v 
  + 2 \beta \, y\!\cdot\! v +  x\!\cdot\! x  +   y\!\cdot\! y }\ ,
\end{align}
where we used both constraints to remove the quaternion dependence of the  $x\!\cdot\! x  +   y\!\cdot\! y$ terms  (this
is equivalent to exploiting $R\cdot R^{\t} = I_{2}$ in \Eqn{UniversalLoss.eq}).
The least squares solution minimizing $\mathbf{S}$ in \Eqn{S2DCloudPair.eq} is easy to
find in the adjugate variable framework by using the constraints to eliminate $\beta$ and
$\gamma$ in terms of $\alpha$, requiring the derivative with respect to $\alpha$ to
vanish, and solving the resulting quadratic equation for $\alpha$.  There is one subtle 
point, which is  that, because of the form of the constraints,  the relative sign of $a$ and $b$ 
(the sign of $\gamma$) can be indeterminate.   Both signs give an adjugate matrix  that handles the possible singularities at. $[a,b]=[0,1]$ and $[a,b]=[1,0]$, and in fact
we can determine the appropriate  sign from the data,
yielding a result identical with the sign-resolved quadratic
products of the results from the 2D quaternion eigensystem methods noted 
in earlier sections.

The solution for the $2\times 2$ adjugate matrix can be written as
\begin{equation}
A_{R\cdot x \to  u} =  \left[ \begin{array}{cc}
             \alpha &  \gamma\\ \gamma & \beta\\
     \end{array} \right] \; = \; \left[ \begin{array}{cc} a^2 & a b \\ a b & b^2 \\
\end{array} \right] = \left[
\begin{array}{cc}
 \frac{1}{2} \left(1+ \frac{\textstyle{x\cdot u}+{y\cdot v}}
            {\textstyle  \lambda(x,y,u,v)} \right) & 
            \frac{1}{2} \;\frac{\textstyle{x\cdot v}-{y\cdot u}} 
                      {\textstyle  \lambda(x,y,u,v)  } \\
 \frac{1}{2}\; \frac{\textstyle{x\cdot v}-{y\cdot u}}
                      {\textstyle  \lambda(x,y,u,v)  } &
    \frac{1}{2} \left(1-\frac{\textstyle{x\cdot u}+{y\cdot v}}
                   {\textstyle  \lambda(x,y,u,v)   } \right) \\
\end{array}
\right] \ .  \label{SolveMatch2D.eq}
\end{equation}
where $ \lambda(x,y,u,v) = \sqrt{(x \cdot u + y \cdot v)^2 + ( x \cdot v - y \cdot u)^2 }$
actually turns out to be the maximal eigenvalue appearing naturally also in the
quaternion matrix approach.
One can easily verify that $\alpha + \beta =1$ and $\alpha \beta = \gamma^2$.
The  optimal aligning 2D quaternion $[a_{\opt},b_{\opt}]$
 is found as usual by identifying
 the maximum of $\alpha$ and $\beta$ and normalizing its row, and the 
 corresponding rotation matrix is $R_{\opt} = R(a_{\opt},b_{\opt})$.   We can write
 the latter also in closed form as
  \begin{equation}
  R_{\opt} (X,U )\, = \,\left[ \begin{array}{cc}
 \frac{\textstyle{x\cdot u}+{y\cdot v}} {\textstyle  \lambda(x,y,u,v)}   & 
                  -   \frac{\textstyle{x\cdot v}-{y\cdot u}} {\textstyle  \lambda(x,y,u,v)  } \\[0.2in]
  \frac{\textstyle{x\cdot v}-{y\cdot u}}   {\textstyle  \lambda(x,y,u,v)  } &
    \ \frac{\textstyle{x\cdot u}+{y\cdot v}}  {\textstyle  \lambda(x,y,u,v)   }  \\
\end{array}
\right] \ .  \label{SolveMatch2DRot.eq}
\end{equation}
Equations for this problem equivalent to \Eqn{SolveMatch2DRot.eq} are well-known
\citep[see, e.g.,][]{Haralick-pose-1989}, but our adjugate-based derivation is novel.
\qquad


\subsection{2D Pose Estimation}

Our next topic is the 2D pose estimation problem, ``2D Pose," which is actually a more
complicated orientation problem than the 2D matching problem,  even though
its least squares loss function  is one-dimensional and has only a single  term.
The problem studies the action of  projections  acting on  2D cloud points  to obtain
an image that is a line of matched points.   The projection can be 
obtained by truncating the $2\times 2$ rotation matrix \Eqn{2DabRotN2.eq}
to the first line, so $P(a,b)= [ a^2 - b^2, -2\, ab]$  or 
$P(\alpha,\beta,\gamma)= [\alpha  - \beta, -2 \gamma]$ ,
with the adjugate matrix coordinates  $\alpha = a^2,\,\beta = b^2,\, \gamma = a b$. 
In principle we should enforce the constraints $\alpha + \beta =1$ and $\alpha \beta = \gamma^2$  to guarantee compatibility with the quaternion roots of our task, but
it turns out that the reduced dimension of the pose estimation problem allows us
some interesting additional freedom.  Our least squares optimization problem  
takes this  explicit form:
\begin{align}  \label{LSQ2DPose.eq}
 {\mathbf S}_{\mbox{\footnotesize (2D Pose)}}  =& 
              \sum_{k=1}^{K} \left( P( \alpha,\beta,\gamma) \cdot  [x_{k},\,y_{k}]^{\t} - u_{k} \right)^2  
\,  = \, \sum_{k=1}^{K} \left( (\alpha -\beta)\, x_{k} -2 \gamma \, y_{k} - u_{k} \right)^2 \\
  \MoveEqLeft[4] {=   ( \alpha -\beta)^2\,  x \! \cdot \!  x -4 \gamma( \alpha -\beta) \, x  \! \cdot \!  y 
     + 4 {\gamma}^{2} \, y  \! \cdot \!  y
        -2 (\alpha -\beta) \, x \! \cdot \!  u  + 4 \gamma \, y \! \cdot \!  u \  +    u  \! \cdot \!  u 
          \label{LSQ2DPoseExpr.eq}  }  \ .  
\end{align} 
Already we see something that might be interesting: while this equation is \emph{quartic} in the quaternion  $(a,b)$ variables, making it unapproachable by the matrix methods 
applicable for the ``2D Match''
problem,  it is only \emph{quadratic} in the adjugate $(\alpha,\beta,\gamma)$ variables.  Might it
be possible to complete the squares, and get an elegant system solvable as a transformed
``2D Match'' problem?   Unfortunately,  this fails because  the square  completion transformation
 requires that the quadratic terms be represented by a symmetric
nonsingular matrix (as its inverse is required), and the corresponding matrix 
following from  \Eqn{LSQ2DPoseExpr.eq} has \emph{vanishing determinant}. 


To exploit the reduced dimension of the adjugate parameters in the loss function 
 \Eqn{LSQ2DPoseExpr.eq}, we proceed by requiring the vanishing of the derivatives of the loss
 function with respect to each variable, ignoring the constraints for the moment. 
 We obtain three equations, but only the following two linear equations for our three variables
 are independent:
\begin{equation} 
\left. \begin{aligned}
\frac{d S}{d \alpha} &= x \! \cdot \!  u - \alpha \, x \! \cdot \!  x + \beta  \,   x \! \cdot \!  x
    + 2 \gamma \, x \! \cdot \!  y =0 \\
\frac{d S}{d \gamma} &=  y\! \cdot \!  u  - \alpha \,   x\! \cdot \!  y\   + \beta \,  x\!\cdot \! y
     + 2 \gamma \, y \! \cdot \!y =  0
\end{aligned} \right\} \ ,
\label{2DPosedervs.eq}
\end{equation}
We find the partial solutions
\begin{equation}   
\left. \begin{aligned}
\alpha &= \beta  + \frac{x\! \cdot \!u\  y\! \cdot \!y  -   y\! \cdot \!u \ x\! \cdot \!y    }
    {x\! \cdot \!x \   y\! \cdot \! y \ - \ (x\! \cdot \!y)^2}\\
\gamma &= \frac{1}{2} \left(  \frac{ x\! \cdot \!u \ x\! \cdot \!y -  y\! \cdot \!u \ x\! \cdot \!x }
  { x\! \cdot \!x \  y \! \cdot \! y \ - \ (x\! \cdot \!y)^2}  \right)
\end{aligned} \right\} \ ,
\label{2DPoseaa:ab.eq}
\end{equation}
At this point we are ready to use one constraint, $\alpha = 1-\beta$, inserted into
the first line of \Eqn{2DPoseaa:ab.eq} to solve for $\beta$; inserting \emph{that} back
into our equation for $\alpha$, we have a complete solution in terms of only the 
cross-covariances: introducing the notation $\sum x_k x_k \to \text{xx}$, $\sum x_k y_k \to \text{xy}$,  
$\sum y_k y_k \to \text{yy}$,   $\sum u_k x_k \to \text{ux}$,  $\sum u_k y_k \to \text{uy}$,
and $\sum u_k u_k \to \text{uu}$
for the cross-covariance sums over $k$, we end up with
\begin{align} 
 \alpha & =  \frac{1}{2} \left( 1 + \frac{ \text{ux} \ \text{yy} \ - \ \text{uy}\    \text{xy} }
                                 { \text{xx}\  \text{yy}-\text{xy}^2}  \right)  \\
       \beta & =  \frac{1}{2}
    \left(1  -  \frac{ \text{ux} \ \text{yy} \ - \ \text{uy}\    \text{xy} } { \text{xx}\  \text{yy}-\text{xy}^2}\right)\\
    \gamma & =  \frac{1}{2}   \frac{\text{ux}\ \text{xy} - \text{uy}\ \text{xx}}    
                                                     { \text{xx}\  \text{yy} - \text{xy}^2   } \ . 
   \label{lsq2DPoseSoln.eq}
   \end{align}

We will find it useful to 
employ a notation using the $3\times  3$ matrix of all term-by-term cross-covariance 
elements, including the self-covariance elements, of $[x,y]$ and $[u ]$
summed over $k$.  Accordingly, we define
\begin{align}
C &= \left[ \begin{array}{ccccc} 
\text{xx} & \text{xy}  &  \text{ux}  \\
\text{xy} & \text{yy}  & \text{uy}   \\
\text{ux} & \text{uy}  &\text{uu}    \\
\end{array} \right] \ .
\label{3by3ccarray.eq}
\end{align}
Our algebraic solutions reduce to ratios of
 $2 \times 2$ subdeterminants of \Eqn{3by3ccarray.eq}, which we choose 
to write using the notation
\begin{equation} \label{CC2Dsubdets.eq}
\begin{aligned}
    d_{1}\to \left[
\begin{array}{ccc}
 \text{xx} & \text{xy}  \\
 \text{xy} & \text{yy}   \\
\end{array} \right] & \ \ 
d_{2}\to \left[ \begin{array}{ccc}
 \text{ux} & \text{xy}  \\
 \text{uy} & \text{yy}  \\
 \end{array} \right] & 
d_{3} \to \left[ \begin{array}{ccc}
 \text{ux} &  \text{xx} \\
 \text{uy} & \text{xy}   \\
\end{array} \right]\\
    \end{aligned}
  \end{equation}
Thus
\begin{equation}\label{lsqDet2DPoseSoln.eq}
\begin{array} {c@{\ \ \ }c@{\ \ \ }c} 
     \alpha  \  = \  \frac{\textstyle 1}{\textstyle 2} 
     \left( 1 + \frac{\textstyle \mktall{\!}{d_{2}}}{\textstyle d_{1}}  \right)  &
     \beta \ =  \ \frac{\textstyle 1}{\textstyle 2} \left( 1 - \frac{\textstyle \mktall{\!}d_{2}}{\textstyle d_{1}}  \right)  &
     \gamma \  =  \  \frac{\textstyle 1}{\textstyle 2}  \, \frac{\textstyle \mktall{\!}d_{3}} {\textstyle d_{1}} 
      \end{array}
   \end{equation}

To complete our solution of the 2D Pose problem, we note that while
Eqs.~(\ref{lsqDet2DPoseSoln.eq}) satisfy the constraints $\alpha +\beta = 1$,
they do not satisfy our second constraint $\alpha \beta = \gamma^{2}$.  However,
it is important to remember that there is one less matrix element
in the loss function \Eqn{LSQ2DPose.eq} than a full rotation matrix loss.  
If we attempt to construct the projection, that is the top line of the 2D rotation
matrix, and insert the solutions \Eqn{lsqDet2DPoseSoln.eq}, we find the following
approximate first step, which in fact gives perfect solutions for noise-free test data:
\begin{equation} \label{lsqDetSoln2DRot.eq}
\left. \begin{aligned}
\tilde{P}(\alpha, \beta, \gamma) & = \left[ \alpha - \beta, \ \  -2 \gamma \right] \\[0.05in]
 & =  \left[  \frac{(u\! \cdot \!x \ y\! \cdot \!y - u\! \cdot \!y \  x\! \cdot \!y}
                                { x\! \cdot \!x \  y \! \cdot \! y \ - \ (x\! \cdot \!y)^2},   \ \ 
                  \frac{(u\! \cdot \!y \ x\! \cdot \! x - u\! \cdot \! x \  x\! \cdot \!y}
                                { x\! \cdot \!x \  y \! \cdot \! y \ - \ (x\! \cdot \!y)^2} \right] \\
   & = \left[ \frac{d_{2}}{d_{1}} ,  -\frac{ d_{3}}{d_{1}} \right] 
\end{aligned} \right\}  \\[0.05in]
\end{equation}
 Evaluating this tentatively as the projection   in the 2D Pose loss 
 function, \Eqn{LSQ2DPose.eq}, against an arbitrary
 list of pure or noisy data sets,  we find that even though its scale 
 varies through a range near unity, unlike a rotation matrix row, it
 still scores very  well as a target for a minimizer of  \Eqn{LSQ2DPose.eq}.
 Remarkably, without doing any further computation, but simply normalizing 
  \Eqn{lsqDetSoln2DRot.eq}  to produce a legal partial rotation matrix element,
 results in mean losses of $\approx 10^{-30}$ for pure data,
 and mean losses smaller than those using the \emph{known} initial value for
 the 2D projection \Eqn{lsqDetSoln2DRot.eq}.  Combining the projection matrix element
  \Eqn{lsqDetSoln2DRot.eq} with its orthogonal partner (the 2D cross-product)
  and normalizing produces a perfect orthogonal rotation matrix, which is the
  solution to the 2D Pose least squares problem:
 \begin{equation} \label{lsqDetSoln2DRotFull.eq}
\left. \begin{aligned}
 {R}(\alpha, \beta, \gamma) 
 & =  \frac{ \left[  \begin{array}{cc}
     (u\! \cdot \!x \ y\! \cdot \!y - u\! \cdot \!y \  x\! \cdot \!y ) &
                                                    (u\! \cdot \! y \ x\! \cdot \! x - u\! \cdot \! x \  x\! \cdot \!y)\\
         ( u\! \cdot \! x \  x\! \cdot \!y - u\! \cdot \!y \ x\! \cdot \! x  ) &    
                                      (u\! \cdot \!x \ y\! \cdot \!y - u\! \cdot \!y \  x\! \cdot \!y )   \\         
                            \end{array}\right] } 
           {\left( (u\! \cdot \!x \ y\! \cdot \!y - u\! \cdot \!y \  x\! \cdot \!y )^2 +                  
                            (u\! \cdot \! y \ x\! \cdot \! x - u\! \cdot \! x \  x\! \cdot \!y)^2 \right)^{1/2} }\\
   & =  \frac{1}{\sqrt{{d_{2}}^2 + {d_{3}}^{2}}}
   \left[  \begin{array}{cc}  d_{2} & - d_{3} \\  d_{3} & \ d_{2} \\ \end{array}  \right] \ .
\end{aligned} \right\}  \\[0.05in]
\end{equation}
 
 \qquad
 
 \noindent{\bf Remark:} We will see in the 3D pose problem that, to handle noisy data that
 move the least squares solution away from a pure rotation, we will take one more step:
 the transition from the projection-matrix solution that works for noise-free data to a full
 rotation matrix that preserves its form for noisy data will require applying  the Bar-Itzhack
 procedure to find the \emph{optimal} pure rotation matrix given  an approximate candidate.
 The 2D case has much less structure, and can avoid that complication.  One can check
 that applying the Bar-Itzhack process to the numerator of \Eqn{lsqDetSoln2DRotFull.eq}
 produces exactly the same rotation matrix.

\begin{figure}[t]
\vspace{-0.5in}
\figurecontent{
\centering
 \includegraphics[width=6.1in]{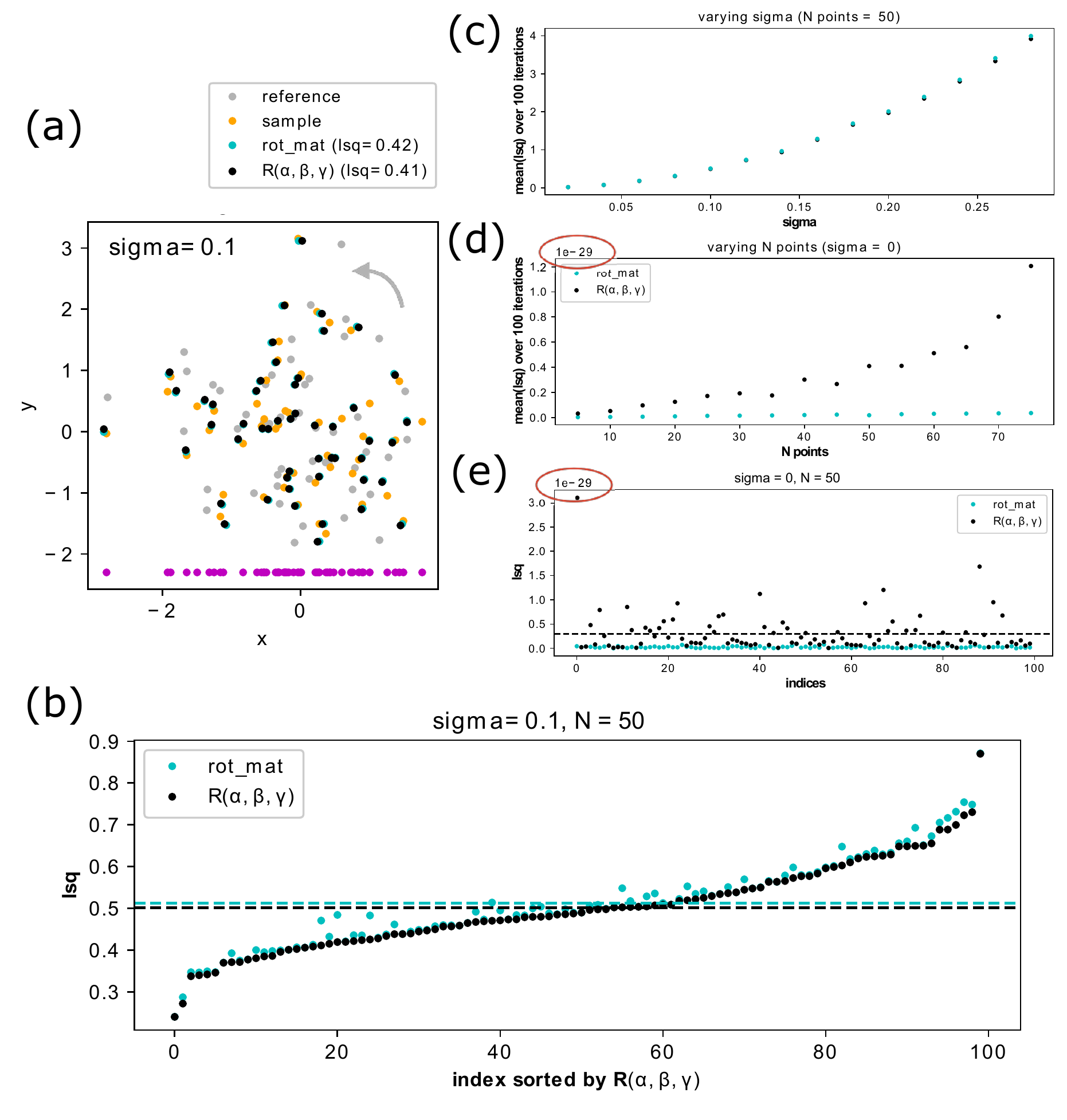}}   
 \caption[]{\ifnum\ShowFiles=1 {\bf PointClouds-dir/figspdf/Figure4-v2.pdf. }\fi
 \ifnum\ShowFiles=1 {\it recent: PointClouds-dir/figspdf/Fig4-2D.pdf. }\fi 
 \ifnum\ShowFiles=1 {\it originals:  lsq2DSVDPoseLosses.eps. }\fi  
 \footnotesize 
 {\bf Results using Analytical Solution to the 2D Point-Cloud Projection Problem.}
 (a) Example data for a small rotation with noise $\sigma=0.1$: original reference 
 data in grey, rotated sample points in orange, points using rotation matrix 
 without noise in cyan (mostly hidden), results using 
 our analytical solution $R(\alpha,\beta,\gamma)$   in black,
 and projected points in magenta.
 (b) Comparison of least squared errors between our analytical solution 
 $R(\alpha,\beta,\gamma)$ and the original \emph{rot\_mat} for 100 random
  2D point clouds with $N=50$  and $\sigma=0.1$. Data  are sorted by the 
  $R(\alpha,\beta,\gamma)$ results, and dashed lines indicate the mean.
  (c) Exploration of the dependency of the  least squared errors 
     on $\sigma$, here with $N= 50$, and we plot the mean results over 100 iterations.
      The coloring is as in  (b).
  (d,e) Exploring the no-noise case for $R(\alpha,\beta,\gamma)$: The original \emph{rot\_mat} 
  performs consistently better, especially as the number of points $N$ in the cloud increases;
   one set of 100 iterations with $N=50$ is shown for clarity. In both cases note 
   that the scale of the $y$ axis is $10^{-29}$. }
      
\label{2DPoseLosses.fig} 
\end{figure}

  In \Fig{2DPoseLosses.fig},  we show that $R(\alpha,\beta,\gamma)$ in  \Eqn{lsqDetSoln2DRotFull.eq}
   is indeed  an exact rotation matrix giving the relation between an initial point cloud
  $[x,y]$ and its projection $[u]$ after a rotation and added noise.  
 We show that  it outperforms the rotation that  was used to simulate the pose data (referred
 to as \emph{rot\_mat});  this should be very good, but  somewhat random and less 
 deterministic than the least squares solution, since
  it has \emph{no way of knowing} what the least squares formula is.

\mypar{Final remarks on the 2D Pose least squares problem.}  The fact that
\Eqn{lsqDetSoln2DRot.eq} satisfies both constraints and achieves basically
vanishing loss for perfect data was somewhat unexpected, but is very likely
related to the fact that the eigenvalues for the quaternion profile matrix approach
to solving the ``Match'' class of problems are rotation-invariant;  this feature is 
maintained in all the numerical experiments we have done, and presumably there
is a direct proof related to our invariance proof for the ``Match'' problem in
Appendix \ref{RMSDRotInvariance.app}.
 The literature on the subject  of pose estimation contains numerous papers mentioning closed form
 solutions and good approximation methods, but typically the exact solutions are lists
  of alternative roots of equations 
 that must be evaluated one by one against the loss function, and, in the end, numerical
 optimization methods such as Newton's method  or neural networks
 are often preferred to achieve experimental results
  \cite[see, e.g.,][]{Haralick-pose-1989,OlssonCVPR2006,WientapperACCV2016,WientapperCVIU2018,WientapperACCV2016,ZhouWangKaess-ICRA2020}. 
   In 2D, the simple least squares solution \Eqn{lsqDetSoln2DRotFull.eq} that we found 
   can undoubtedly be obtained by any of many equivalent methods; however, our derivation based on the
   quaternion adjugate matrix provides a clear picture of what is happening.
   The key is that rotation parameterizations are subtle structures, even in 2D, as we saw in Section
    \ref{2DRotations.sec}.   There are several  fundamental insights that appear:  one
     is that if you drop one line of a rotation matrix to produce a projection matrix, there is still
     enough information present to reproduce the full $N \times N$ rotation matrix by simply
     taking the cross-product of the lines of the projection matrix to find the missing line.  
     Next is that both the algebraic
     methods to extract a least-squares solution from a squared-difference loss function such
     as \Eqn{LSQ2DPose.eq} and the numerical methods such as \emph{argmin} methods
     produce answers without anomalies \emph{only if the correct, reduced, number of
     constraints is imposed}; what we found was that for the 2D Match problem, both the constraints
     must be applied, but for the 2D Pose problem, only the first constraint is necessary,  in addition to
     normalization, and in
     fact one gets excessive ambiguous branched solutions otherwise (we will see a similar  thing
     in the 3D Pose problem later). Finally, the particular constraints that are successful in guiding
     \emph{argmin} include, for example,  the \emph{topological constraints on orthonormality of the
     rows of the projection matrix}, which for 2D are simply $\alpha +\beta = 1$ plus the
     one-line normalization.  These are the
     key observations; we conjecture that any successful pose estimation problem has these 
     properties, which are clarified by formulating the problem using quaternion adjugate variables to
     parameterize the rotation, which is computed \emph{directly}, without going through an
     isolated quaternion stage.  We conclude be noting that \emph{if} a quaternion is desired, which
     can often be the case if one wants to visualize global features of families of rotations,
     the quaternion can be at once extracted using the fundamental methods such as the
     Bar-Itzhack optimization described in Sections \ref{2DRotations.sec} and 
     \ref{3DRotations.sec}.

\comment{ 
\begin{figure}[h!]
\vspace{0.0in}
\figurecontent{
\centering
 \includegraphics[width=4.5 in ]{lsq2DSVDPoseLosses.eps}}
 \caption[]{\ifnum\ShowFiles=1 {\bf lsq2DSVDPoseLosses.eps. }\fi
 \footnotesize  The ranges of the 2D pose problem losses from
 \Eqn{LSQ2DPose.eq}  using noisy data
 with projection matrices derived from alternate formulas.
 [Green] original, simulation-creating, rotation; [Black] the bare least squares
 solution that is perfect for noise-free data, but veers away from a good rotation
 for noisy data; [Magenta] altering the least squares form by normalizing by
 the square root of the determinant; [Blue] the Bar-Izhack optimal rotation
 approximating the least squares solution $\to$ this is seen to be
 \emph{identical}  to the normalized least squares matrix.  This is unusual,
 and the 3D case does not have this simple relation.}
\label{2DPoseAdjArgQ.fig}
\end{figure}
    }   


\clearpage

\subsection{3D  Point-Cloud Matching}
\label{3DMatch.sec}

Moving on to 3D data, we consider first the classic ``3D Match'' problem, also known as the RMSD or 
``Generalized Procrustes'' problem, 
whose task is to  find the 3D rotation best aligning a possibly noisy test cloud with a reference cloud
to which it corresponds.   As noted, here we choose a least squares 
loss function for our purposes that
reverses the common order and applies the rotation matrix 
to align the reference cloud with the test cloud. 
We assume we have two 3D point clouds of size $K$,
 a reference set $\{\Vec{x}_{k}\}=\{[x,y,z]_{k}\}$ that
 we consider as a list of $K$ columns of 3D points, and a test set of noisy measured
 3D points $\{\Vec{u}_{k}\}=\{[u,v,w]_{k}\}$
 that is believed to be related, pointwise, to the reference set  by an unknown rotation. 
 If we choose to express that rotation using
 \Eqn{Rofqq.eq} as $R(q)$, then we can write the least-squares optimization
 target as \cite[see, e.g.,][]{Horn1987,Hanson:ib5072}.
 \begin{equation} 
 {\mathbf S}_{ \mbox{\footnotesize(3D Match) }} (q)= \sum_{k=1}^{K} \| R(q) \cdot \Vec{x}_{k} - \Vec{u}_{k} \| ^{2}\ .
 \label{RMSDeqn.eq}
 \end{equation}
 There are well-known procedures using quaternion eigensystems to solve
 this problem in closed form either from the least squares functional ${\mathbf S}(q)$ 
 (see \citet{FaugerasHebert1983,FaugerasHebert1986}), or from the
 cross-term, which reduces to a trace over the rotated $3\times 3$ cross-covariance matrix $E_{ab}$
  \cite[see, e.g.,][]{Horn1987},
 \begin{equation}
 \Delta(q) = \tr R(q) \cdot \left( {X} \cdot  {U}^{\t} \right) = \tr R(q) \cdot E
 \label{basic3DCross.eq} \ . 
 \end{equation} 
  Because the $R^{\t}\cdot R$  factor in \Eqn{RMSDeqn.eq} disappears from
this least squares optimization function, we can always express the problem of
 finding the optimal quaternion using a   \emph{quadratic} function
 of quaternions, and the problem is  solvable using standard linear algebra.
  (Non-quaternion methods  such as SVD are also widely used.)   We reiterate that in the
 3D$\to$2D pose-estimation problem to be dealt with in the next subsection, those terms no
 longer cancel, and the  expression for $\mathbf {S}(q)$ becomes \emph{quartic} in $q$,
 and \Eqn{basic3DCross.eq} is no longer applicable.

 \mypar{The Adjugate.} We now explore how the adjugate variables can be incorporated into this problem.
 Starting from the standard framework of \Eqn{RMSDeqn.eq}, we see there is little
 motivation to use the full $\mathbf S(q)$ form, since all the quartic terms disappear,
 and only the non-constant cross-term $\Delta(q)$ shown in \Eqn{basic3DCross.eq} is relevant.   
In the standard quaternion solution of the optimal rotation problem,  we rearrange
the cross-term into a form that is optimized by the maximal eigenvector of the
profile matrix, $M(E)$  (see, e.g., \citet{Hanson:ib5072} for further references and a review).
We find
\begin{equation}
 \Delta(q)\,  = \,  \tr R(q) \cdot E  \, = \, (q_0,q_1,q_2,q_3) \cdot M(E) \cdot
 (q_0,q_1,q_2,q_3)^{\t} \equiv q \cdot M(E) \cdot q \ ,
\label{qM3qnn.eq}
\end{equation}
 where $M(E)$ is  the traceless, symmetric  $4\times 4$ matrix  that is composed of
 linear functions of the elements of the
  $3\times 3$ cross-covariance matrix $E={X} \cdot  {U}^{\t}$  of the data:
\begin{equation} 
     M(E) \!  = \!
\left[ \begin{array}{cccc}
 \!\!   E_{xx} + E_{yy} + E_{zz} & E_{yz} - E_{zy} & E_{zx} - E_{xz} &
                     E_{xy} - E_{yx} \!  \\
 \!   E_{yz} - E_{zy} &  E_{xx} - E_{yy} - E_{zz} & E_{xy} + E_{yx} & 
                     E_{zx} +E_{xz}  \! \\
 \!   E_{zx} - E_{xz} &  E_{xy} + E_{yx} & - E_{xx} + E_{yy} - E_{zz} & 
          E_{yz} + E_{zy}  \!  \\
  \!   E_{xy} - E_{yx} &  E_{zx} +E_{xz}  &  E_{yz} + E_{zy} & 
              - E_{xx} - E_{yy} + E_{zz}  \!\!
                     \end{array} \right] \ .
       \label{basicHornnn.eq}              
\end{equation} 

In the usual method, the transformed loss function \Eqn{qM3qnn.eq}, now a 
maximization problem, is solved by computing the maximal eigenvalue  
$\lambda{\opt}$ of $M(E)$, and identifying its normalized eigenvector as
exactly $q_{\opt}$,   with $R_{\opt} = R(q_{\opt})$ solving the matching problem.
  However, in every such calculation, there is an often-hidden
step that relates precisely to one of our main points in this paper:  there are
\emph{always} fourteen submanifolds of possible quarternion solutions that
can obstruct obtaining a normalizable quaternion from $M(E)$ and
$\lambda_{\opt}$!   These are avoided by scaling one element of the unknown
eigenvector to unity, and solving the eigenvector equation for the three remaining
elements using Kramer's rule;
if that eigenvector element happens to vanish, this fails, and one sets the \emph{next}
element to unity, repeating until successful.
Whatever methods a library eigensystem program
uses to return a valid eigenvector of this system corresponding to the maximal
eigenvalue of $M(E)$, it will always be exactly equivalent to computing
the characteristic matrix and its adjugate,
\begin{align*}
\chi(E) & =  \left[ M(E) \, - \, \lambda_{\opt}\, I_{4} \right] \\
 A(E) &= \mbox{Adjugate} \, ( \chi(E))  \ , 
 \label{RMSDAdjugate.eq}
 \end{align*}
finding the maximum-magnitude diagonal of $A(E)$, and normalizing that row to guarantee
the computation will not encounter one of the singular domains of unnormalizable
eigenvectors.

 For our purposes, we now
  rephrase the 3D Match problem to employ the quaternion adjugate variables $q_{ij} = q_i q_j$
 instead of the quaternions $q_{i}$ themselves.  Then, with  $R(q)$'s quadratic form
 in $q$ replaced by the adjugate form 
\begin{equation}
    R(q_{ij}) =  \left[ 
 \begin{array}{ccc}
  q_{00} +q_{11}-q_{22} - q_{33} & 2 q_{12}  - 2 q_{03}  & 2 q_{13} + 2 q_{02}   \\
 2 q_{12} + 2 q_{03}  & q_{00} -q_{11} + q_{22} - q_{33}  &  2 q_{23} - 2 q_{01}  \\
  2 q_{13} - 2 q_{02}  &  2 q_{23} + 2 q_{01}  &  q_{00} -  q_{11} - q_{22} + q_{33}
\end{array}  \right]   \  ,
\label{qadjrmsdeq}
\end{equation}
 we find that $\Delta(q_{ij})$ now defines  a superficially \emph{linear} optimization problem,
 in ten dimensions,  that takes the form
 \begin{align}
 \Delta(q_{ij})\,  = &\,  \tr R(q_{ij}) \cdot E  \,  = \, \sum_{i\le j} q_{ij} \, M(E)_{ij}  \ .
\label{M3qadj.eq}
\end{align}
If all the $q_{ij}$ were independent up to an overall scale,  the solution 
$q_{ij} \propto M_{ij}$ would immediately be seen to maximize $\Delta$. 
This opportunity is unfortunately obstructed by the fact that the ten 
adjugate variables are not independent, but are related by the
  seven constraints of  \Eqn{TheqqConstraints.Eq}
 that reduce the superficial problem
 in ten free variables down to the required three parameters of 3D rotations.
%

 We can now recast the problem traditionally solved using the maximal eigenvalue of $M(E)$
 and its associated normalized eigenvector, which is just $q_{\opt}$,  in several ways.  First,
 we can simply use \Eqn{TheqqConstraints.Eq} to reduce all the ten $q_{ij}$ to functions
 of just three independent adjugate variables such as $(q_{11},\, q_{22},\, q_{33})$ , and
 require that all three corresponding derivatives of \Eqn{M3qadj.eq} vanish.  This is an
 effective solution with the drawback that there are eight alternative sign-permuted solutions
 due to the square roots in the constraint equations, and the correct values of 
 $(q_{11},\, q_{22},\, q_{33})$  appear in only one of these terms for any data set, and
 which term that is appears to be indeterminate.  In addition, the signs of the remaining
 seven terms in the list of adjugate variables are indeterminate as well.  The correct 
 adjugate matrix can always be found be checking all permutations substituted into
 the optimization function \Eqn{M3qadj.eq} and using the choices giving the maximal
 value, but that is an awkward algorithm compared to the maximal quaternion eigensystem
 method.
   
    \qquad
   
   We note one other alternative approach that can be used, driven by the proof in Appendix
   \ref{RMSDRotInvariance.app} that for \emph{error-free data}, the maximal eigenvalue of
   $M(E)$ is independent of $R(q)$.  Thus  the maximal eigenvalue 
   is the same as the maximal eigenvalue of
   $M(E_{0})$, where $E_{0} = X \cdot X^{\t}$ is the self-covariance of the reference data,
   and that value is simply
   \begin{equation}
   \lambda_{\opt} = \tr E_{0} \ .
   \end{equation}
   A candidate quaternion applicable to error-free data thus emerges from the hybrid
   characteristic equation
   \[ \chi(E, E_{0}) = \left[ M(E) - \tr(E_{0}) I_{4} \right] \]
   upon computing the adjugate,
   \[ A(\chi) = \mbox{Adjugate} \left(\chi(E,E_{0})\right) \ , \]
   and computing the optimal quaternion $q_{\opt}$ from the largest-magnitude row of $A(\chi)$.
   For noise-containing data, the procedure becomes somewhat circular for the particular task
   of 3D cloud-matching, as the rotation $R_{\opt} = R(q_{\opt})$ becomes inexact, and one
   unavoidably has to compute a more complicated maximal eigenvalue to solve the Bar-Itzhack
   problem, producing an optimal exact rotation matrix $R_{\mbox {\tiny BI}} \approx R_(q_{\opt})$ that
   produces an acceptable solution to the 3D cloud matching problem.



\qquad

 \subsection{3D to 2D Pose Estimation }
 \label{3DPose.sec}
 
Finally, we turn our attention to the ``3D Pose'' pose-estimation problem that corresponds
most closely  to the classic 3D RMSD squared-difference optimization.  
 We suppose we have  a 3D point cloud reference set ${X}$ that
 we consider as a list of $K$ columns of 3D points $\{\Vec{x}_{k}\}$, and 
 a 2D  test set of points  ${U}$, with 2D image-plane components $\{\Vec{u}_{k}\}$ 
 that are considered as paired images of each point in the 3D cloud projected
in parallel from some to-be-determined camera orientation.  Here we study  only
the idealized case of the parallel projection described by a $2\times 3$ projection
matrix $P(q)$ that is extracted from the top two rows of a 3D rotation matrix;  this
is of course quite relevant for applications like microscopy for which the effective
focal length relative to the scale of the data is infinite.

       If we choose to express our rotation using \Eqn{Rofqq.eq} as $R(q)$, then
 we may write the projection as
   \begin{equation}
    P(q) =  \left[ 
 \begin{array}{ccc}
 {q_0}^2+{q_1}^2-{q_2}^2 - {q_3}^2 & 2 q_1 q_2  -2 q_0 q_3 
& 2 q_1 q_3 +2 q_0 q_2   \\
 2 q_1 q_2 + 2 q_0 q_3  &  {q_0}^2-{q_1}^2 + {q_2}^2 - {q_3}^2   
      &  2 q_2 q_3 - 2 q_0 q_1  \\
\end{array}  \right]   \  ,
\label{qrotproj.eq}
\end{equation}
 and the least-squares optimization target can  be written
 \begin{align} \label{3D2DPoseLSQ.eq}  
 \mathbf{S}_{\mbox{\small 3D Pose}} =&
 \sum_{k=1}^{K} \| P(q) \cdot \Vec{x}_{k} -\Vec{u}_{k} \| ^{2}\ .
 \end{align}
 While the 3D point cloud matching loss function in Section \ref{3DMatch.sec}
 can be reduced to the quadratic cross-term $\Delta$ and solved using
  an optimal quaternion eigenvector,  \emph{this approach fails for pose estimation}.
  In the pose-estimation problem we can no longer
  eliminate the quartic quaternion part of the optimization and the problem
  becomes potentially much more complex.  

The adjugate formalism now comes into play: we replace the individual quaternions
in \Eqn{3D2DPoseLSQ.eq}, as they appear in  \Eqn{qrotproj.eq},
by their adjugate quadratic forms, $q_{i}q_{j} \to q_{ij}$, so our adjugate-valued
projection matrix becomes
\begin{equation}
    P(q_{ij}) =  \left[ 
 \begin{array}{ccc}
  q_{00} +q_{11}-q_{22} - q_{33} & 2 q_{12}  - 2 q_{03}  & 2 q_{13} + 2 q_{02}   \\
 2 q_{12} + 2 q_{03}  & q_{00} -q_{11} + q_{22} - q_{33}  &  2 q_{23} - 2 q_{01}  \\
\end{array}  \right]   \  .
\label{qadjproj.eq}      
\end{equation}
We note the projection matrix  is lacking these matrix elements,
  $q_{13} -q_{02}$, $q_{23} + q_{01}$, and $q_{00} -q_{11} - q_{22} + q_{33}$, 
  so the number of constraints needed can in principle be reduced.  
  Secondly, those variables are in a specific sense 
recoverable because, since $P(q_{ij})$ is part of an orthonormal $3\times 3$ matrix, the
missing bottom row can be computed, if the first two rows have been
determined,  by taking the cross-product of the two rows of ${P(q_{ij})}_{\opt}$ and
normalizing the result (if necessary) to get the missing last row of a full orthonormal rotation matrix.
The resulting form of \Eqn{3D2DPoseLSQ.eq}
  now becomes a quadratic function in the adjugate variables $q_{ij}$ that can be
solved, in principle, using least squares algebraic
methods, resulting in a computable adjugate matrix  from
which a guaranteed non-singular quaternion can be extracted.  
Our loss function, written in terms of the measured data components summed
over $K$,  is complicated by the appearance of both linear and quadratic terms
in $q_{ij}$, and takes the form:
\begin{align} 
\MoveEqLeft{ 
\mathbf{S}_{\mbox{\small 3D Pose}}(q_{ij},x,y,z,u,v) =
               } \nonumber \\ 
&{q_{00}}^2 \,{x\!\cdot \!x}   +{q_{11}}^2 {x\! \cdot \!  x}+{q_{22}}^2 {x\! \cdot \!  x} +{q_{33}}^2 {x\! \cdot \!  x}  
+{q_{00}}^2\, {y\!\cdot \!y} +{q_{11}}^2 {y\! \cdot \!  y}+{q_{22}}^2 {y\! \cdot \!  y} 
  +{q_{33}}^2 {y\! \cdot \!  y} \nonumber \\ &
-4 {q_{00}} {q_{01}} \, {y\! \cdot \!  z} +4 {q_{00}} {q_{02}}\, {x\! \cdot \!  z}+2 {q_{00}} {q_{11}} \,{x\! \cdot \!  x}  
-2 {q_{00}} {q_{11}}\, {y\! \cdot \!  y}+8 {q_{00}} {q_{12}}\, {x\! \cdot \!  y} \nonumber \\ &
+4 {q_{00}} {q_{13}}\, {x\! \cdot \!  z}-2 {q_{00}} {q_{22}}\, {x\! \cdot \!  x}    
+2 {q_{00}} {q_{22}}\, {y\! \cdot \!  y}+4 {q_{00}} {q_{23}}\, {y\! \cdot \!  z}    
-2 {q_{00}} {q_{33}}\,  {x\! \cdot \!  x}-2 {q_{00}} {q_{33}}\,  {y\! \cdot \!  y}  \nonumber \\ &
-2 {q_{00}}\,  {u\! \cdot \!  x}-2 {q_{00}}\,  {v\! \cdot \!  y}  
+4 {q_{01}}^2 {z\! \cdot \!  z}-8 {q_{01}} {q_{03}}\,  {x\! \cdot \!  z}  
+4 {q_{01}} {q_{11}}\,  {y\! \cdot \!  z}-8 {q_{01}} {q_{12}}\,  {x\! \cdot \!  z}    \nonumber \\ &
-4 {q_{01}} {q_{22}}\,  {y\! \cdot \!  z}  
 -8 {q_{01}} {q_{23}}\,  {z\! \cdot \!  z}  +4 {q_{01}} {q_{33}}\,  {y\! \cdot \!  z}   
+4 {q_{01}}\,  {v\! \cdot \!  z}+4 {q_{02}}^2 {z\! \cdot \!  z}-8 {q_{02}} {q_{03}}\,  {y\! \cdot \!  z} 
 \nonumber  \\ &
+4 {q_{02}} {q_{11}}\,  {x\! \cdot \!  z} +8 {q_{02}} {q_{12}}\,  {y\! \cdot \!  z}  
+8 {q_{02}} {q_{13}}\,  {z\! \cdot \!  z}-4 {q_{02}} {q_{22}}\,  {x\! \cdot \!  z}
-4 {q_{02}} {q_{33}}\,  {x\! \cdot \!  z}-4 {q_{02}}\,  {u\! \cdot \!  z}  \nonumber \\ &
+4 {q_{03}}^2 {x\! \cdot \!  x}
+4 {q_{03}}^2 {y\! \cdot \!  y}     -8 {q_{03}} {q_{11}}\,  {x\! \cdot \!  y}  
+8 {q_{03}} {q_{12}}\,  {x\! \cdot \!  x}  -8 {q_{03}} {q_{12}}\,  {y\! \cdot \!  y}   
-8 {q_{03}} {q_{13}}\,  {y\! \cdot \!  z}  \nonumber \\ &
+8 {q_{03}} {q_{22}}\,  {x\! \cdot \!  y}+8 {q_{03}} {q_{23}}\,  {x\! \cdot \!  z} 
  +4 {q_{03}}\,  {u\! \cdot \!  y} -4 {q_{03}}\,  {v\! \cdot \!  x} 
+4 {q_{11}} {q_{13}}\,  {x\! \cdot \!  z}-2 {q_{11}} {q_{22}}\,  {x\! \cdot \!  x}  \nonumber \\ &
  -2 {q_{11}} {q_{22}}\,  {y\! \cdot \!  y} 
-4 {q_{11}} {q_{23}}\,  {y\! \cdot \!  z}-2 {q_{11}} {q_{33}}\,  {x\! \cdot \!  x} 
+2 {q_{11}} {q_{33}}\,  {y\! \cdot \!  y}  
-2 {q_{11}}\,  {u\! \cdot \!  x}+2 {q_{11}}\,  {v\! \cdot \!  y}  \nonumber \\ &
  +4 {q_{12}}^2 {x\! \cdot \!  x}+4 {q_{12}}^2 {y\! \cdot \!  y} 
+8 {q_{12}} {q_{13}}\,  {y\! \cdot \!  z}+8 {q_{12}} {q_{23}}\,  {x\! \cdot \!  z} 
-8 {q_{12}} {q_{33}}\,  {x\! \cdot \!  y}   -4 {q_{12}}\,  {u\! \cdot \!  y}  \nonumber \\ &
  -4 {q_{12}}\,  {v\! \cdot \!  x}
+4 {q_{13}}^2 {z\! \cdot \!  z}-4 {q_{13}} {q_{22}}\,  {x\! \cdot \!  z}  
-4 {q_{13}} {q_{33}}\,  {x\! \cdot \!  z}   -4 {q_{13}}\,  {u\! \cdot \!  z} 
 +4 {q_{22}} {q_{23}}\,  {y \! \cdot \!  z}  \nonumber \\ & 
 +2 {q_{22}} {q_{33}}\,  {x\! \cdot \!  x}-2 {q_{22}} {q_{33}}\,  {y\! \cdot \!  y} 
+2 {q_{22}}\,  {u\! \cdot \!  x}-2 {q_{22}}\,  {v\! \cdot \!  y}  
+4 {q_{23}}^2 {z\! \cdot \!  z}-4 {q_{23}} {q_{33}}\,  {y\! \cdot \!  z}   \nonumber  \\ &
-4 {q_{23}}\,  {v\! \cdot \!  z}
 +2 {q_{33}}\,  {u\! \cdot \!  x}  +2 {q_{33}}\,  {v\! \cdot \!  y}+{u\! \cdot \!  u}+{v\! \cdot \!  v}   \label{3DposeLossFcn.eq} \ .
\end{align}
(Note that it is the sum of $q_{ii}$ that sums to unity,  not the sum of  ${q_{ii}}^2$, so there
is no simplification in the first line.)
Parallel to the 2D case,  one cannot complete the squares to recover a simpler quadratic form
in a transformed variable set because the $10 \times 10$ matrix incorporating
the quadratic products of the adjugate variables is singular.
As usual, the redundancy of the adjugate variables has to be reduced by the
imposition of  constraints  such as those in \Eqn{TheqqConstraints.Eq}.  
However,  we are potentially lacking some degrees of freedom
in the projection-matrix adjugate variables,  so if we try to constrain all the
variables, we may come up with no solutions.  Thus it appears possible
that we do not need (or cannot utilize) all seven adjugate constraints
in order to obtain the canonical three rotational degrees of freedom in a rotation.

\qquad

\subsubsection{Solving the 3D Pose Least-Squares Loss Function Algebraically}

We can get  full solutions of the least squares problem
defined by the  general form of  \Eqn{UniversalArgmin.eq}
using a specific choice of  the adjugate constraints in \Eqn{TheqqConstraints.Eq}.
 When we impose the four constraints containing
the adjugate variable $q_{00}$,  with \verb|lossRPose3DAdj| denoting
the algebraic expression \Eqn{3DposeLossFcn.eq} and symbols for the
cross-covariance terms $x\cdot y$ and etc., this Mathematica  
expression yields a list of eight candidate solutions:

   \begin{quote}
\begin{small}
\begin{verbatim}
the3D2DAdjSolns = 
 Module[{eqn = lossPose3DAdj}, 
  Solve[ { D[eqn, q00] == 0, 
      D[eqn, q11] == 0, D[eqn, q22] == 0,  D[eqn, q33] == 0,
      D[eqn, q01] == 0, D[eqn, q02] == 0,  D[eqn, q03] == 0,
      D[eqn, q23] == 0, D[eqn, q13] == 0, D[eqn, q12] == 0,
      q00 + q11 + q22 + q33 == 1, 
      q00 q11 == q01 q01,  q00 q22 == q02 q02,  q00 q33 == q03 q03
 (* q22 q33==q23 q23,q11 q33==q13 q13,q11 q22==q12 q12 *)   }, 
    {q00,  q11, q22, q33, q01, q02, q03, q23, q13, q12}]]. .
  \end{verbatim}\end{small}   \end{quote}  
%
%

(The three unused constraints are commented out, retained for later reference.)
The resulting set of eight algebraic expressions can be tested  by substituting
randomly generated rotations applied to a cloud of points, and adding noise 
 to generate a 2D projected data image.  We test each of the list of 8 against
 100 data sets to see, first, 
whether they produce an adjugate matrix that  provides a solution,
 and then to see whether the resulting solutions 
obey all seven constraints in \Eqn{TheqqConstraints.Eq}.  For exact data,
four of the solutions are usually complex, and thus unusable.  Four of the
solutions are always real, and, strangely, exactly \emph{one} of them
always produces the quaternion (via the adjugate procedure) that was used
to generate the data.   However, we have more work to do to achieve a
 deterministic algorithm.  For pure data, all the constraints are in fact obeyed,
 while for errorful data, the constraint identities that
were \emph{enforced} are always maintained, while those that were not
enforced (commented out in the \verb|the3D2DAdjSolns| expression)
 are no longer valid.  We can do better, but first we need some notation,
 as the immediate algebraic solutions in some cases are ten megabytes
 in length.

We have found a useful symbolic representation of the first four solutions,
one of which always  gives the right quaternion for error-free data sets.  We
introduce first the $5\times 5$ matrix of all term-by-term cross-covariance 
elements, including the self-covariance elements, of $[x,y,z]$ and $[u,v]$
summed over $k$, denoting $\sum x_k x_k \to \text{xx}$, $\sum x_k y_k \to \text{xy}$, \ldots,
$\sum z_k v_k \to \text{vz}$, so we have
\begin{align}
C &= \left[ \begin{array}{ccccc} 
\text{xx}& \text{xy} & \text{xz}& \text{ux} & \text{vx} \\
\text{xy}& \text{yy} & \text{yz}& \text{uy} & \text{vy} \\
\text{xz}& \text{yz} &\text{ zz}&\text{ uz} & \text{vz} \\
\text{ux}& \text{uy} & \text{uz} &\text{uu} &\text{uv}  \\
\text{vx}& \text{vy} & \text{vz} & \text{uv} &\text{vv} \\
\end{array} \right] \ .
\label{5by5ccarray.eq}
\end{align}
All of the algebraic solutions reduce to ratios of order 3 products of the elements
of cross-covariances \Eqn{5by5ccarray.eq} or square roots of appropriate powers
of such elements.  We define the $3 \times 3$ subdeterminants of \Eqn{5by5ccarray.eq}
using the notation
\begin{equation} \label{CCsubdets.eq}
\begin{aligned}
   & \ \  d_{1}\to \left[
\begin{array}{ccc}
 \text{xx} & \text{xy} & \text{xz} \\
 \text{xy} & \text{yy} & \text{yz} \\
 \text{xz} & \text{yz} & \text{zz} \\
\end{array} \right]     \\ 
d_{2}\to \left[ \begin{array}{ccc}
 \text{xx} & \text{xy} & \text{ux} \\
 \text{xy} & \text{yy} & \text{uy} \\
 \text{xz} & \text{yz} & \text{uz} \\
\end{array} \right] &  \ \ 
d_{3} \to \left[ \begin{array}{ccc}
 \text{xx} & \text{xy} & \text{vx} \\
 \text{xy} & \text{yy} & \text{vy} \\
 \text{xz} & \text{yz} & \text{vz} \\
\end{array} \right] &    
d_{4}\to \left[ \begin{array}{ccc}
 \text{xx} & \text{xz} & \text{ux} \\
 \text{xy} & \text{yz} & \text{uy} \\
 \text{xz} & \text{zz} & \text{uz} \\
\end{array} \right] \ \ \   \\ 
d_{5}\to \left[ \begin{array}{ccc}
 \text{xx} & \text{xz} & \text{vx} \\
 \text{xy} & \text{yz} & \text{vy} \\
 \text{xz} & \text{zz} & \text{vz} \\
\end{array} \right] &  \ \ 
d_{6}\to \left[ \begin{array}{ccc}
 \text{xx} & \text{ux} & \text{vx} \\
 \text{xy} & \text{uy} & \text{vy} \\
 \text{xz} & \text{uz} & \text{vz} \\
\end{array} \right]      &
 d_{7} \to \left[ \begin{array}{ccc}
 \text{xy} & \text{xz} & \text{ux} \\
 \text{yy} & \text{yz} & \text{uy} \\
 \text{yz} & \text{zz} & \text{uz} \\
\end{array} \right] \ \ \  \\ 
 d_{8} \to \left[ \begin{array}{ccc}
 \text{xy} & \text{xz} & \text{vx} \\
 \text{yy} & \text{yz} & \text{vy} \\
 \text{yz} & \text{zz} & \text{vz} \\
\end{array} \right] & \ \
d_{9}\to \left[ \begin{array}{ccc}
 \text{xy} & \text{ux} & \text{vx} \\
 \text{yy} & \text{uy} & \text{vy} \\
 \text{yz} & \text{uz} & \text{vz} \\
\end{array} \right]  
&   d_{10}\to \left[  \begin{array}{ccc}
 \text{xz} & \text{ux} & \text{vx} \\
 \text{yz} & \text{uy} & \text{vy} \\
 \text{zz} & \text{uz} & \text{vz} \\
\end{array} \right]  \ ,   \\
    \end{aligned}
  \end{equation}
  where $d_{1}$ plays a special role as the \emph{self-covariance} of the reference
  cloud's point values.
The four usable versions of the least-squares solutions differ by pairs of signs 
of square roots in the expressions for $(q_{01},q_{02},q_{23},q_{13})$,
so we  can write the  3D pose least squares solutions as a function of
their corresponding square root signs,  which we denote by $s_{ij}$.
We can then express the four solutions in terms of 
the combinations of signs that distinguish one 
from another as $\omega(s_{01},s_{02},s_{23},s_{13})$, where
\begin{equation}
\left. \begin{aligned}
\mbox{\rm soln}(1) &= \omega(+1,+1; \, +1,+1) \\
\mbox{\rm soln}(2) &= \omega(+1,-1; \, +1,-1) \\
\mbox{\rm soln}(3) &= \omega(-1,+1; \, -1,+1) \\
\mbox{\rm soln}(4) &= \omega(-1, -1; \, -1,-1) 
\end{aligned}  \ \ \right\} \ .
\end{equation}
Then a more explicit form of the solutions  terms of the cross-covariance
determinants of \Eqn{CCsubdets.eq} takes the form
\begin{equation}
   \label{pose3Dsoln.eq}
      \begin{aligned}
\begin{array}{lr}
 \hspace*{-5.25in}{\omega(s_{01},s_{02},s_{23},s_{13}) = } & \   \\[.05in]
 \end{array}\\
 \left[ \begin{array}{l}
 {q_{00}}\to \frac{\sqrt{{d_{1}}^2
   \left(-{d_4}+({d_5}-{d_7})^2-{d_8}\right)}+{d_1}
   ({d_7}-{d_5})}{4 {d_1}^2} \\[0.1in]
 {q_{11}}\to \frac{{d_1} (2 {d_1}+{d_5}+{d_7})-\sqrt{{d_1}^2
   \left(-{d_4}+({d_5}-{d_7})^2-{d_8}\right)}}{4 {d_1}^2} \\[0.1in]
 {q_{22}}\to -\frac{\sqrt{{d_1}^2
   \left(-{d_4}+({d_5}-{d_7})^2-{d_8}\right)}+{d_1} (-2
   {d_1}+{d_5}+{d_7})}{4 {d_1}^2} \\[0.1in]
 {q_{33}}\to \frac{\sqrt{{d_1}^2
   \left(-{d_4}+({d_5}-{d_7})^2-{d_8}\right)}+{d_1}
   ({d_5}-{d_7})}{4 {d_1}^2} \\[0.1in]
 {q_{01}}\to -\frac{{s_{01}} \sqrt{2 ({d_1}+{d_5}) \left(\sqrt{{d_1}^2
   \left(-{d_4}+({d_5}-{d_7})^2-{d_8}\right)}+{d_1}
   ({d_7}-{d_5})\right)-{d_1} ({d_4}+{d_8})^2}}{4 \sqrt{{d_1}^3}} \\[0.15in]
 {q_{02}}\to -\frac{{s_{02}} \sqrt{2 ({d_1}-{d_7}) \left(\sqrt{{d_1}^2
   \left(-{d_4}+({d_5}-{d_7})^2-{d_8}\right)}+{d_1}
   ({d_7}-{d_5})\right)-{d_1} ({d_4}+{d_8})^2}}{4 \sqrt{{d_1}^3}} \\[0.15in]
 {q_{03}}\to \frac{{d_4}+{d_8}}{4 {d_1}} \\[0.1in]
 {q_{23}}\to \frac{{d_3}}{2 {d_1}}-\frac{{s_{23}} \sqrt{2 ({d_1}+{d_5})
   \left(\sqrt{{d_1}^2
   \left(-{d_4}+({d_5}-{d_7})^2-{d_8}\right)}+{d_1}
   ({d_7}-{d_5})\right)-{d_1} ({d_4}+{d_8})^2}}{4 \sqrt{{d_1}^3}} \\[0.15in]
 {q_{13}}\to   \frac{{d_2}}{2 {d_1}} + \frac{{s_{13}} \sqrt{2 ({d_1}-{d_7}) \left(\sqrt{{d_1}^2
   \left(-{d_4}+({d_5}-{d_7})^2-{d_8}\right)}+{d_1}
   ({d_7}-{d_5})\right)-{d_1} ({d_4}+{d_8})^2}}{4 \sqrt{{d_1}^3}} \\[0.1in]
 {q_{12}}\to \frac{{d_8}-{d_4}}{4 {d_1}} \\
\end{array} \right] \ .
\end{aligned}
 \end{equation}

With careful examination and experimentation, the daunting form of \Eqn{pose3Dsoln.eq} reveals
some remarkable structure.  If one takes a list of error-free data sets, and, 
evaluates all four functions in \Eqn{pose3Dsoln.eq}  \emph{against the pose loss function} \Eqn{3DposeLossFcn.eq}, all four least-squares alternate solutions produce a \emph{perfect
match}. 
This seems impossible until one carefully checks the steps, and discovers that, although the elements
$(q_{01},\,q_{02},\,q_{23},\,q_{13})$ differ among the four functions, when all are combined
together to form the $2 \times 3$ \emph{projection matrix} $P(q_{ij})$, \emph{the differences cancel
and all four produce the same projection with no square roots}.

  Since only the top two lines enter into the least squares loss formula, this is completely logical:
 our solution only asks to minimize those two lines, and the third line does not appear
 at all,  and in fact if we substitute the four versions of $q_{ij}$ solutions in \Eqn{pose3Dsoln.eq}
 into the third line of the adjugate-parameterized rotation matrix \Eqn{qadjrmsdeq}, they are
 \emph{all different}.   Now we come to a procedure that remarkably takes us full circle
 back to the calculations for optimal matches to noisy rotation matrices in Section
 \ref{BarItzh-3D-noise.sec}.  First, since we get a perfect first two lines of the rotation matrix
 for all four solutions, we can simply construct the third line by taking the cross-product of
 the two projection-matrix components.   In the second step, we observe that for noisy
 test data, the perfect success of \Eqn{pose3Dsoln.eq} is deformed and, in general, we
 do not know exactly what process is going on.  However, the basic fact is that what
 we need is an optimal rotation matrix that is the \emph{ideal} approximation, with perfect
 orthonormality, to our solution that (so far) behaves well only for perfect data.  We already know  how to do that, from our work  in previous sections on extracting adjugate vectors,
 and hence quaternions, using methods like those in the introductory sections.
 
  For pedagogical reasons that will become clear, we first write
 the  initial form of the projection-matrix solution in terms of the cross-covariance
 determinants in \Eqn{CCsubdets.eq} as follows:
\begin{equation}
\tilde{P}(x,y,z;u,v)  = \left[ \begin{array}{ccc}
 \displaystyle\frac{\textstyle d_{7}}{\textstyle d_{1}} &     
              -   \displaystyle\frac{\textstyle d_{4}}{\textstyle d_{1}}& 
                        \displaystyle    \frac{\textstyle d_{2}}{\textstyle d_{1}}\\[0.15in]
 \displaystyle \frac{\textstyle d_{8}}{\textstyle d_{1}} &
    -   \displaystyle \frac{\textstyle d_{5}}{\textstyle d_{1}}&
          \displaystyle \frac{\textstyle  d_{3}}{\textstyle  d_{1}}\\
\end{array} \right] \ .
\label{PoseSolnPmatUN.eq}
\end{equation}
On any error-free data set, this projection remarkably is a perfect least-squares solution,
is orthonormal, and produces a vanishing loss function to 30 orders of magnitude accuracy.
If we ignore the disagreements with the form of the third rotation-matrix line coming from
the four solutions for $q_{ij}$, we can simply take the cross-product of the two lines,
that is $P_{1} \times P_{2}$, and that will give a unique answer for
a third line that will also be orthonormal on pure data, establishing
our initial form for the full 3D Pose rotation matrix solution of the form
\begin{equation}
\tilde{R}(x,y,z;u,v)  = \left[ \begin{array}{ccc}
 \displaystyle\frac{\textstyle d_{7}}{\textstyle d_{1}} &     
              -   \displaystyle\frac{\textstyle d_{4}}{\textstyle d_{1}}& 
                        \displaystyle    \frac{\textstyle d_{2}}{\textstyle d_{1}}\\[0.15in]
 \displaystyle \frac{\textstyle d_{8}}{\textstyle d_{1}} &
    -   \displaystyle \frac{\textstyle d_{5}}{\textstyle d_{1}}&
             \displaystyle \frac{\textstyle  d_{3}}{\textstyle  d_{1}}\\[0.15in]
         \displaystyle \frac{\textstyle d_{6}}{\textstyle d_{1}} &
       \displaystyle \frac{\textstyle d_{9}}{\textstyle d_{1}}&
          \displaystyle \frac{\textstyle  d_{10}}{\textstyle  d_{1}}\\ 
          \end{array} \right] \ .
\label{PoseRotSolnPmatUN.eq}
\end{equation}
Why must we call this our ``initial form'' instead of our final solution?
Our issue here is virtually identical to the distinctions we found in 
Sections \ref{3DdirectSoln.sec}, \ref{3DVariationalSoln.sec},
 and \ref{BarItzh-3D-noise.sec}  in the treatment of exact rotation
 matrices vs error-containing measurements of rotation matrices.
 When we use error-containing data for the pose problem, the perfect
 match of  \Eqn{PoseSolnPmatUN.eq}  and its extension to
 an actual $3\times 3$ camera model matrix in  \Eqn{PoseRotSolnPmatUN.eq}
 breaks down, just as it did when we introduced the data-generic Bar-Itzhack
 method in Section \ref{BarItzh-3D-noise.sec}.  As soon as we insert data
 with errors in these equations, the different components are \emph{not even
 normalized to unity}, much less orthogonal.  This cannot be the optimal answer
 for a rotation matrix placing a noisy 2D point image into a corresponding 
 3D cloud scene.   It will still be a least-squares solution minimizing the cost
 function \Eqn{3D2DPoseLSQ.eq}, but since it does not preserve the properties
 of a rotation matrix, it will not actually correspond to an optimal \emph{rotation},
 which is what we require of the pose estimation problem.

 Finally, we can \emph{reverse-engineer} a new version of the adjugate variables
 in \Eqn{pose3Dsoln.eq} found by hand-solving the least-squares optimization.  We know that \Eqn{PoseRotSolnPmatUN.eq}
 corresponds with the adjugate variables via \Eqn{qadjrmsdeq}, so if we simply solve that for
 the $q_{ij}$, we find the adjugate variables in directly in terms of the cross-covariance determinants
 that produce  \Eqn{PoseRotSolnPmatUN.eq}:
 \begin{equation} \label{cleanPose3Dsoln.eq}
 \left.  \begin{array}{l}
 {q_{00}}\to \frac{\textstyle 1}{\textstyle 4 d_1} ({d_1}+{d_{10}}-{d_5}+{d_7}) \\ [0.1in]
 {q_{11}}\to \frac{\textstyle 1}{\textstyle 4 d_1} ({d_1}-{d_{10}}+{d_5}+{d_7}) \\   [0.1in]
 {q_{22}}\to \frac{\textstyle 1}{\textstyle 4 d_1} ({d_1}-{d_{10}}-{d_5}-{d_7}) \\   [0.1in]
 {q_{33}}\to \frac{\textstyle 1}{\textstyle 4 d_1} ({d_1}+{d_{10}}+{d_5}-{d_7}) \\ [0.1in]
 {q_{01}}\to \frac{\textstyle 1}{\textstyle 4 d_1} ({d_9}-{d_3}) \\ [0.1in]
 {q_{02}}\to \frac{\textstyle 1}{\textstyle 4 d_1} ({d_2}-{d_6}) \\ [0.1in]
 {q_{03}}\to \frac{\textstyle 1}{\textstyle 4 d_1} ({d_4}+{d_8}) \\ [0.1in]
 {q_{23}}\to \frac{\textstyle 1}{\textstyle 4 d_1} ({d_3}+{d_9}) \\ [0.1in]
 {q_{13}}\to \frac{\textstyle 1}{\textstyle 4 d_1} ({d_2}+{d_6}) \\ [0.1in]
 {q_{12}}\to \frac{\textstyle 1}{\textstyle 4 d_1} ({d_8}-{d_4})  \\
  \end{array}   \right\}  \ .
  \end{equation}

 \mypar{The Adjugate and the Solution to the 3D Pose Estimation Problem.} \nopagebreak
 Let us state our problem clearly: we have a least squares solution that works on all data,
 but that solution corresponds to a rotation matrix only for perfect data.  This latter behavior
 is almost certainly a manifestation of the rotation-invariant eigenvalues that occur for
 perfect data in the 3D cloud alignment problem of Section \ref{3DMatch.sec}, outlined
 in Appendix \ref{RMSDRotInvariance.app}.   To solve the problem, we must
 \emph{restrict the subspace of solutions to pure rotations}.   But we know \emph{exactly}
 how to accomplish that!  We simply take our ``good approximation,'' namely our
 initial solution $\tilde{R}(\Vec{x};\Vec{u})$ given in \Eqn{PoseRotSolnPmatUN.eq},
  consider it as an \emph{error-containing rotation matrix}, and apply the Bar-Itzhack
  optimization as a \emph{second iteration optimization} to produce a new quaternion
  adjugate corresponding to the perfect rotation matrix \emph{best approximating}
  our initial  formula \Eqn{PoseRotSolnPmatUN.eq}; this rotation will then apply
  in both the case of perfect data and in the more general case when, as we can see,
  the data themselves produce a completely reasonable example of an inexact
  rotation matrix that must be optimized.  We would argue that no more accurate
  least-squares-related pure rotation matrix solving the pose estimation problem can
  be found; without question this solves the perfect-data pose problem, and very
  plausibly is the best solution to the errorful-data pose estimation problem.

   For completeness, we review the actual steps producing a closed form algebraic
  solution to the 3D Pose task.  First, we examine the matrix \Eqn{PoseRotSolnPmatUN.eq}, 
  which in the errorful-data   case effectively is the cross-covariance matrix 
  appearing in the 3D Match task  \Eqn{basic3DCross.eq}.  While the expression $\tilde{R}$  of
   \Eqn{PoseRotSolnPmatUN.eq} is a rotation matrix for perfect data, and continues
   to give a least squares solution minimizing \Eqn{3D2DPoseLSQ.eq} for noisy data,
   it no longer obeys all the constraints needed to make it a valid rotation matrix.
   We can thus apply the Bar-Itzhack optimization we explored earlier to find the
   \emph{nearest} rotation matrix to  \Eqn{PoseRotSolnPmatUN.eq}.   Since the
   Bar-Itzhack loss function needs the \emph{inverse} of the desired approximate
   matrix to find the quaternion  of the closest pure rotation matrix, we now compute
   the profile matrix of the \emph{transpose}  of $\tilde{R}$,  \Eqn{PoseRotSolnPmatUN.eq},
   which, from \Eqn{basicHornnn.eq}, now becomes:
   \begin{equation} 
     M(x,y,z; u,v) \!  = \!
\left[ \begin{array}{cccc}
 \!\!   d_{7}  - d_{5} + d_{10} &-( d_{9} - d_{3} )&   -(  d_{2} - d_{6})&   -( d_{8} + d_{4}) \!  \\
 \!      -(  d_{9} - d_{3} )&  d_{7} + d_{5} - d_{10} &     d_{8} - d_{4}  &            d_{6} +d_{2}  \! \\
 \!     -(  d_{2} - d_{6}) &         d_{8} - d_{4}  & - d_{7}  - d_{5} - d_{10} &    d_{3} + d_{9}  \!  \\
  \!     -(  d_{8} +d_{4}) &        d_{6} +d_{2}  &          d_{3} + d_{9} &     - d_{7} + d_{5} + d_{10}  \!\!
                     \end{array} \right] \ ,
       \label{poseHornrot.eq}              
\end{equation} 
that is, the last three rows of the first column, and the last three columns of the first
row are negated to correspond to the matrix ${\tilde{R}}^{\t}$.
We then compute the maximal eigenvalue using any method we like, but we note that
it is not hard to compute the analytic algebraic formula using the Cardano equations \citep{Hanson:ib5072}.  Given that eigenvalue
\begin{equation} \label{maxeigMxyzyv.eq}
\lambda_{\mbox{\footnotesize max}} = \mbox{\bf  (Maximal Eigenvalue) }\left( M(x,y,z; u,v)\right) \ ,
\end{equation}
we form the characteristic matrix $\chi$ with vanishing determinant by subtracting
 $\lambda_{\mbox{\footnotesize max}}$,
\begin{equation} \label{3DposeCharMat.eq}
\chi(x,y,z;u,v) = \left[ M(x,y,z;uv) - \lambda_{\mbox{\footnotesize max}} \; I_{4} \right] \ .
\end{equation}
Recall that the critical feature is the maximal eigenvector,  whose normalized value is the
quaternion giving the optimal solution for the sought-for rotation matrix.   As usual, we now
just compute the adjugate, which, up to a normalization, will now always be four copies
of the needed optimal quaternion,
\begin{equation} \label{poseHornrotAdj.eq}
A(x,y,z;u,v) \, = \, \mbox{Adjugate}(\chi(x,y,z;u,v)) = 
  \left[  \begin{array}{cccc}
 {q_0}^2 & q_0 q_1 & q_0 q_2 & q_0 q_3 \\
 q_0 q_1 & {q_1}^2 & q_1 q_2 & q_1 q_3 \\
 q_0 q_2 & q_1 q_2 & {q_2}^2 & q_2 q_3 \\ 
 q_0 q_3 & q_1 q_3 & q_2 q_3 & {q_3}^2 \\
\end{array}  \right]  \ .
\end{equation}
The final answer is found by choosing a nonsingular
row from the adjugate $A(x,y,z;u,v)$ for normalization to determine $q_{\opt}$:
\begin{equation} \label{maxeigvecQ.eq}
q_{\opt} =   \mbox{\bf (Normalize Row with Largest Diagonal)}\left( A(x,y,z; u,v)\right) \ .
\end{equation}
 We noted in previous simpler examples that, while $q_{\opt}$ can be very complicated, 
 reassembling the quaternions
into the rotation matrix itself, $R_{\opt} = R(q_{\opt})$,  can result in a simplified final form; 
we have not yet accomplished that explicitly for the 3D problem, but a differentiable
final form of the optimal quaternion can be obtained using the noted algebraic form
of the solution.

\qquad

{\bf Note: }  One finds experimentally an odd feature of this optimization process:  because
our fundamental standard for optimization is the Fr\"{o}benius norm of the $3\times 3$ rotation
matrix differences, the appropriate test of differences is based on the \emph{rotation matrices}.
If one compares the \emph{quaternions} resulting from \Eqn{maxeigvecQ.eq} using a
\emph{quaternion distance measure}, one occasionally finds quaternions that are nearly
inverses of one another, and thus very far apart.  However, examining the resulting rotation
matrices, which will also be close to inverses of the expected matrix, one finds that, indeed,
the inverse-like rotation matrix will correspond to a \emph{smaller}  Fr\"{o}benius norm, and
so is technically correct.   This   phenomenon, which seems to appear for $|q_0| \ll 1$,
is one of many odd features appearing when we add substantial noise to rotation matrices.

\qquad

%
   
\comment{  
\begin{figure}[h!]
\vspace{0.0in}
\figurecontent{
\centering
 \includegraphics[width=4.5in]{lsq3DSVDPoseLosses.eps}}
\caption[]{\ifnum\ShowFiles=1 {\bf lsq3DSVDPoseLosses.eps. }\fi
  \footnotesize
Plotting the values of the 3D Pose loss function \Eqn{3D2DPoseLSQ.eq}
for 100 randomly chosen, noise injected, reference and test pose data sets.
The projection values are chosen from [Green] the original rotation used
in the creation of the simulated data; [Black] the least squares solution
that is perfect for perfect data, but does not exactly give a rotation matrix
for noisy data; [Magenta] the Bar-Itzhack optimal pure rotation \emph{closest}
to the not-quite-a-rotation least squares solution.  The differences between
these losses for various permutations at the bottom show that indeed the
least squares solution can be less than the optimal rotation solution, and
that the original rotation is always a poorer choice than the optimal rotation.
}
\label{3DPoseAdjArgQ0.fig}
\end{figure}
  }   

\begin{figure}[t]
\vspace{-0.75in}
\figurecontent{
\centering \hspace*{-0.5in}
 \includegraphics[trim=1in  0 0 0, clip=true,width=6.9in]{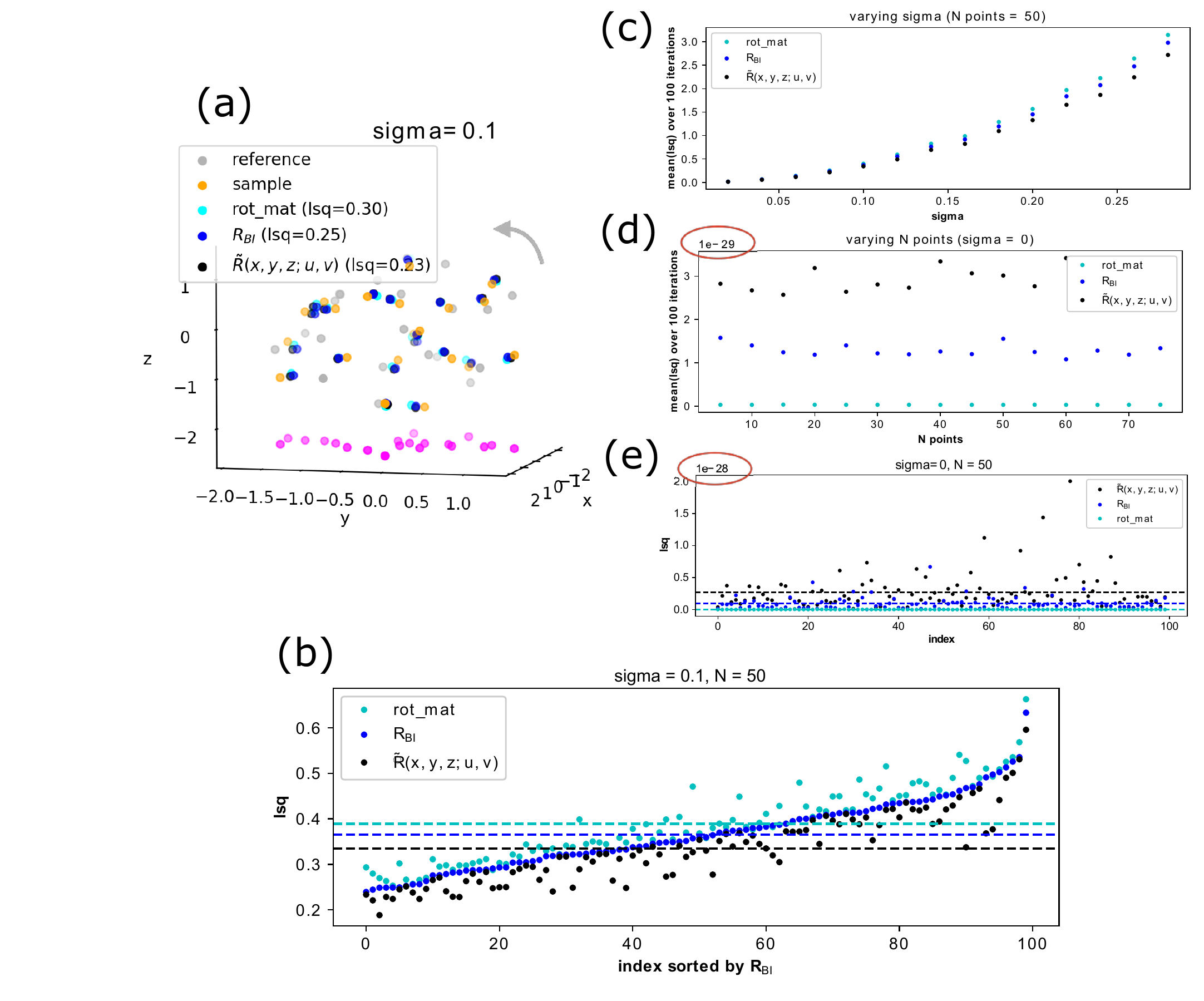}}   
 \caption[]{\ifnum\ShowFiles=1 {\bf PointClouds-dir/figspdf/Figure5-v2.pdf. }\fi
 \ifnum\ShowFiles=1 {\it newer: PointClouds-dir/figspdf/Fig4-3D.pdf. }\fi 
\ifnum\ShowFiles=1 {\it originals: lsq3DSVDPoseLosses.eps. }\fi  
 \footnotesize 
 {\bf Results using Analytical Solution to the 3D Point-Cloud Projection Problem.}
 (a) Example data for a small rotation with noise $\sigma=0.1$: original reference 
 data in grey, rotated sample points in orange, points using rotation matrix 
 without noise in cyan, 
 our analytical solution  $\tilde{R}(x,y,z:u,v)$  in blue, the rotation-corrected
 solution $R_{\mbox{\tiny BI}}$  in black, and projected points in magenta.
 (b) Comparison of least squared errors between our analytical solution 
 $\tilde{R}(x,y,z:u,v)$  in blue, the rotation-corrected
 solution $R_{\mbox{\tiny BI}}$  in black, and
   the original \emph{rot\_mat}  in cyan for 100 random
  3D point clouds with $N=50$  and $\sigma=0.1$. Data  are sorted by the 
  $R_{\mbox{\tiny BI}}$ results, and dashed lines indicate the mean.
  (c) Exploration of the dependency of the  least squared errors 
     on $\sigma$, here with $N= 50$, and we plot the mean results over 100 iterations.
      The coloring is as in  (b).
  (d,e) Exploring the no-noise case for $\tilde{R}(x,y,z:u,v)$ and  $R_{\mbox{\tiny BI}}$:
    the original \emph{rot\_mat} 
  performs consistently better, but without any dependency on the number of points;
   one set of 100 iterations with $N=50$ is shown for clarity.  In both cases note 
   that the scale of the $y$ axis is $10^{-29}$.
  }
\label{3DPoseLosses.fig} 
\end{figure}


To bring this to a close, we test our new solutions using randomly generated 3D point clouds 
and corresponding 2D projections after a rotation and added noise  (\Fig{3DPoseLosses.fig}). 
For each data set, we consider both the exact solution \Eqn{PoseRotSolnPmatUN.eq}, which 
gives a valid rotation matrix only for noise-free data but always minimizes the least-squares, 
as well as the solution after applying Bar-Itzhack via the profile matrix \Eqn{poseHornrot.eq}.
  When we compute these and compare the list of 
losses to those of the original rotation used to simulate the
pose data, we find we outperform this original rotation matrix (referred to as \emph{rot\_mat}) 
and our results improve as more noise is added (\Fig{3DPoseLosses.fig}).  The least squares
solution, which is not a rotation for noisy data, can be even better, but of
course those results are not useful.   We also checked these
results for zero noise, and our optimal rotation and the original
rotation substituted into the loss function are uniformly zero to machine
accuracy  $\approx 10^{-30}$, consistent with an exact least-squares  
loss minimizing solution to the pose estimation optimization problem.


\clearpage

\subsection{3D Pose Estimation with Perspective Projection}

We have assumed that orthographic projection could be used in
our 2D and 3D Pose estimation exercises so far in order to find closed-form 
least squares solutions for   optimal adjugate matrices of unnormalized
quaternions and their corresponding rotation matrices.   Pieces of our
solution methods can be applied also to the more difficult problem
of perspective projection, with finite focal length.  For our final
application, we now explore an approach to perspective  3D pose
estimation that exploits the same methods  as the previous parts
of this Section.  

   We will make no attempt to review the vast literature on this subject,
but it is appropriate to mention a few of the influential  developments
along with more current pieces of literature,
starting with the classic work of Haralick et al. \cite{Haralick-pose-1989}, which
defines the problem for the case of corresponding points, which has been our
context throughout.   Weintapper et al. and Zhou et al.
\cite{WientapperACCV2016,WientapperCVIU2018,ZhouWangKaess-ICRA2020} 
continue with some recent developments of the classic methods.
   \cite{LuHagerMj-FastPose-2000} invoke an approach
similar to ours, utilizing multiple stages, but without the closed-form aspects that
we are available to us,   \cite{GuoEtAl-CamOrientwFocal-2021} 
study other approaches, while related quaternion methods are employed by 
 \cite{ForbesPoseEst2011}.   We focus exclusively on the rotation aspects, but a
number of authors examine full 6 degree-of-freedom methods, most recently making 
heavy use of machine learning,  such as, e.g.,  
\cite{XiangY2018PoseCNN,FuaEtAl-6DPoseEst-CVPR2020}.

We start as  usual with a loss function based on least squares
that needs to be minimized in order to find the rotation matrix that rotates
a 3D reference cloud to its best possible alignment with a planar 2D image.
While we were able
to get away with just the top two lines of the rotation matrix in the orthographic
projection squared-error function, now we need the entire matrix because the
bottom row determines the depth coordinate that implements the perspective
division.   Thus we start with  the two-row projection \Eqn{qadjproj.eq} determining
the numerator of the rotated 3D cloud, and adjoin to that the 3rd line in
quaternion adjugate coordinates
\[ R_{3} = D(q)= \left[
\begin{array}{ccc}
 2 q_{13} - 2 q_{02}  &  2 q_{23} + 2 q_{01}  & q_{00} - q_{11} - q_{22} + q_{33} 
    \end{array} \right] \]
to produce the relevant depth element $z' = D(q)\cdot [x,y,z]$. 

\begin{figure}[h!]
\vspace{0.0in}
\figurecontent{
\centerline{
 \includegraphics[width=6.5in]{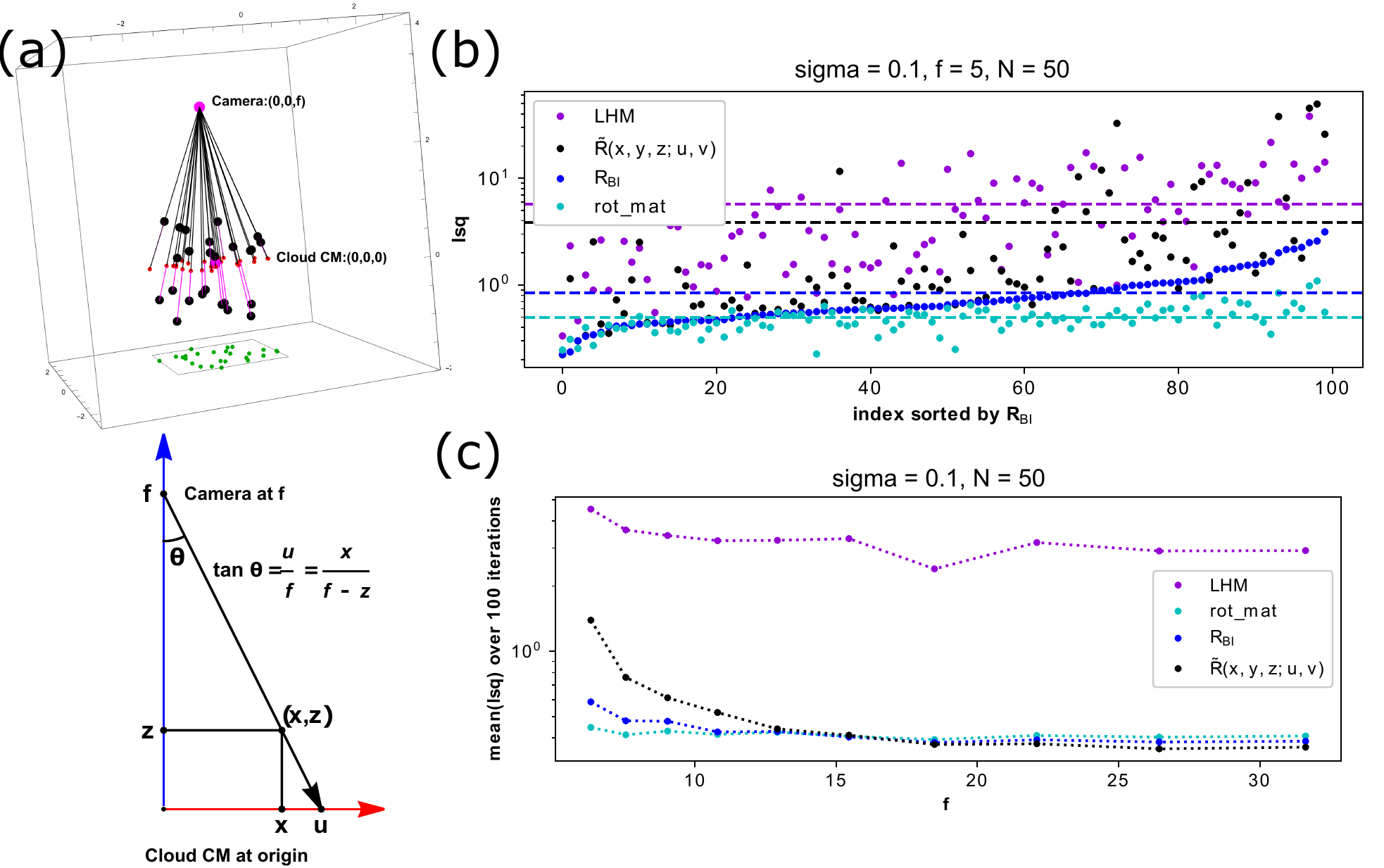} }}
\caption[]{\ifnum\ShowFiles=1 {\bf Figure6-v4.eps} \fi    \footnotesize
 {\bf 3D Perspective Pose Problem and Loss Spectra: Camera at $\mathbf{ z=f}$.}
   (a) Geometry of perspective projection from camera at $(0,0,f)$ from
   the 3D reference data to centered at the origin noisy 2D image, with $u/f=x/(f-z)$ defining
   the projected coordinates $(u,v)$.
(b) Plotting the values of the 3D perspective loss function of \Eqn{3D2DFPoseLSQ.eq}
  with focal length $f$ for the 3D Pose estimation problem,
comparing  the original data-generating rotation \emph{rot\_mat}, 
the LHM method's results \citep{LuHagerMj-FastPose-2000}, our closed-form least 
squares solution  $\tilde{R}$,
and the corrected exact rotation $R_{\mbox{\tiny BI}}$.
An error distribution with $\sigma = 0.1$ is used for the simulated projection data with 
50 points.
(c)  The mean least-squares error averaged over 100 sample data sets of size 50 with
normal error $\sigma=0.1$ as a function of the focal length $f$, on a logarithmic scale.
The  $R_{\mbox{\tiny BI}}$ rotation solution does very well at smaller focal lengths,
while $\tilde{R}$ (not a rotation) least-squares solution gets better at large camera 
distances, and $R_{\mbox{\tiny BI}}$ appears to outperform the LHM. } 
\label{3DPoseGraphs.fig}
\end{figure}

\begin{figure}[h!]
\vspace{0.0in}
\figurecontent{
\centering \hspace{-0.5in}
 \includegraphics[width= 6.25in]{figspdf/Figure7-v3}}
  \caption[]{\ifnum\ShowFiles=1 {\bf 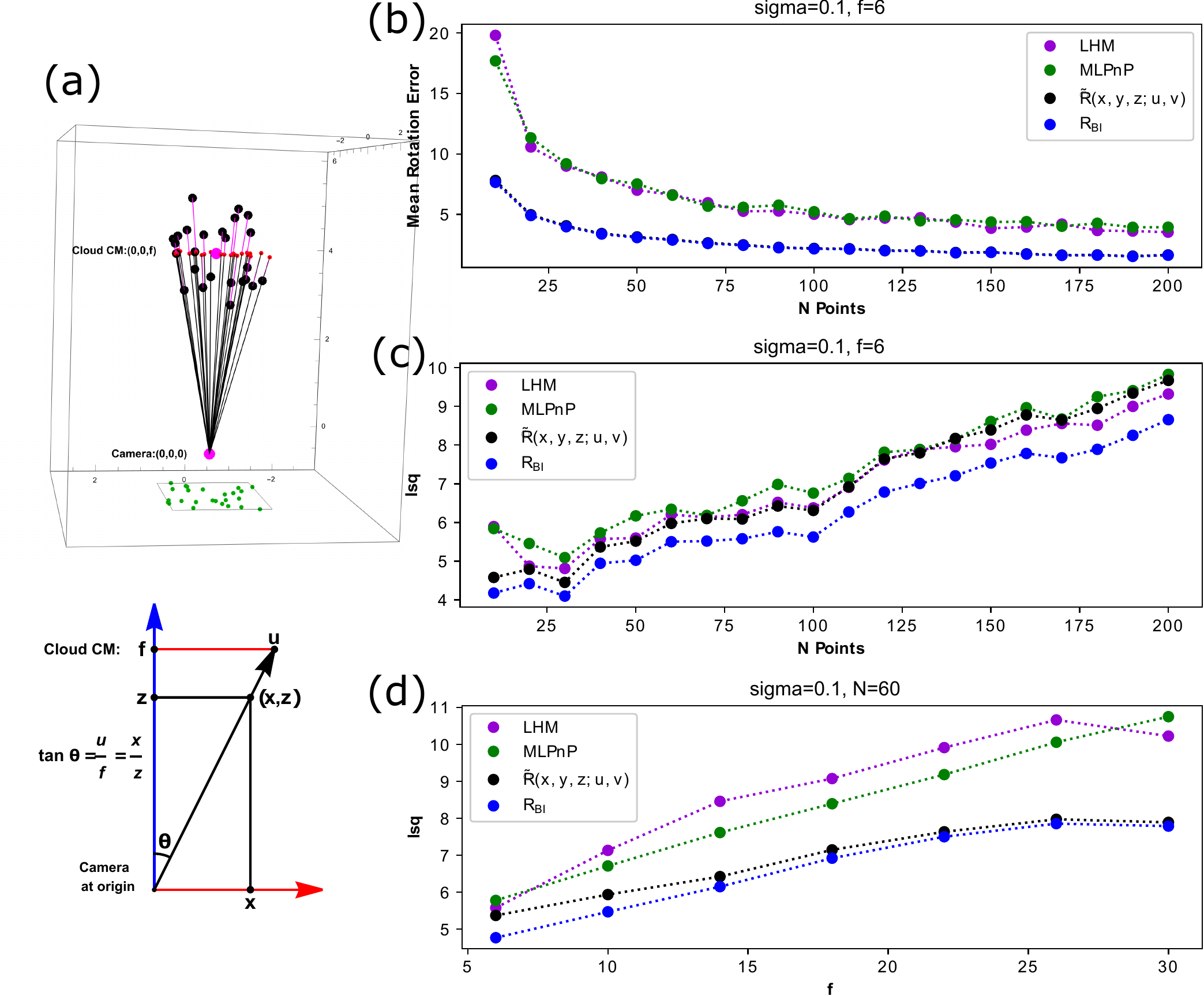 }\fi
  \footnotesize
  {\bf 3D Perspective Pose Problem and Loss Spectra: Camera at Origin.}
   (a) Geometry of perspective projection from camera at $(0,0,0)$ to
    3D reference data to centered at $(0,0,f)$, producing a noisy 2D image,
    with $u/f=x/f$ defining  the projected coordinates $(u,v)$. 
    (b) For 3D point cloud  sizes ranging from 10 to 200,  random quaternion
rotations were applied to produce noisy 2D projected images with  standard
error $\sigma = 0.1$ and focal length 6. We plot the
relative performances of the LHM and MLPnP algorithms compared to
ours using  our version of the ``Mean (Quaternion) Rotation Error'' in  
\Eqn{qoptvsqinput.eq}.  In this context, our
idealized least-squares solution $\tilde{R}$ and our corrected-to-robust-rotation
solution $R_{\mbox{\tiny BI}}$ are basically indistinguishable, with $\tilde{R}$'s
black dots hidden behind the blue dots of $R_{\mbox{\tiny BI}}$.
(c) For the same collection of point-cloud sizes and parameters, we plot
the  corresponding 3D pose least-squares measure \Eqn{3D2DFPoseLSQAlt.eq},
and find similar results, with the optimal-rotation solution   $R_{\mbox{\tiny BI}}$ 
responding best to this measure. 
(d) Here we fix the cloud sample size at 60 and plot mean least-squares measure
across a spectrum of different focal lengths.  Again, $R_{\mbox{\tiny BI}}$  shows
a good response.}
\label{QmeanRotErr-vs-Npoints.fig}
\end{figure}

  {\bf Alternate Choices of Camera Location.}
   We can choose from two alternative ways of looking at the least squares loss
function for perspective projection, illustrated in \Fig{3DPoseGraphs.fig}
and \Fig{QmeanRotErr-vs-Npoints.fig}
noting that a perspective
projection loss function analogous to the orthographic loss \Eqn{3D2DPoseLSQ.eq}
must  incorporate an additional division by the depth.  One traditional approach
 places the  point cloud's center of mass at the origin and the image plane at $z=0$,
 with the camera looking down from a pinhole camera at focal distance $f$,
 so  $x_{\mbox{cam}} = (0,0,f)$.  This has the advantage that the similar-triangles
 perspective formula contains only the inverse focal length $\bar{f} = 1/(\mbox{focal length})$
 multiplying the depth:
\begin{align} \label{3D2DFPoseLSQ.eq} 
 \mathbf{S}_{\mbox{\small 3D Pose  $\bar{f}$}} &= \ \mathbf{S}(\bar{f} ) \ = \ 
      \sum_{k=1}^{K} \left\| \frac{\textstyle P(q) \cdot \Vec{x}_{k}}
               {\textstyle (1 - \bar{f} D(q)\cdot\Vec{x}_{k})}   - \Vec{u}_{k} \right\| ^{2}\ .
 \end{align}
This form allows one to easily take the orthographic  limit $\bar{f} \to 0$ without needing infinite numbers.   An alternative perspective formula commonly used in machine vision
reverses the camera and the point cloud, so the camera is at the origin,  the image
is at $z = f$, and the cloud center of mass is off-center at  $(0,0,f)$, so the similar-triangles
computation yields the least squares expression involving only $f$, as opposed to only
$\bar{f}= 1/f$:
\begin{align} \label{3D2DFPoseLSQAlt.eq} 
 \mathbf{S}_{\mbox{\small 3D Pose  $f$}} &=   \  \mathbf{S}(f)  \ = \ 
 \sum_{k=1}^{K} \left\| \frac{\textstyle f \,P(q) \cdot \Vec{x}_{k}}
               {\textstyle  D(q)\cdot\Vec{x}_{k}}   - \Vec{u}_{k} \right\| ^{2}\ .
 \end{align}
The cloud-driven geometry of these two approaches is shown in  \Fig{3DPoseGraphs.fig}
and \Fig{QmeanRotErr-vs-Npoints.fig}.
In either case,  we will assume that
the focal point of the pinhole camera appears well outside the cloud so that the optimization problem has smooth mathematical behavior.

{\bf Remark on focal length determination.}  The focal length can be determined
from the data at any stage by setting the derivative of $\mathbf{S}(f)$ or
 $\mathbf{S}(\bar{f})$ to zero and solving
for $f$ or $\bar{f}$, provided  one has a candidate for the rotation matrix
 $R(q_{\opt}) =  \left[P(q_{\opt}),D(q_{\opt})\right]$.  
 However, the $\bar{f}$ version results
upon differentiation in a very high degree formula in $\bar{f}$ whose vanishing 
point needs to be found numerically, though it is typically well-behaved.  In contrast,
 given some known $R(q_{\opt})$ that
produces $R(q_{\opt})\cdot \left[x,y,z\right]= \left[x',\,y',\,z'\right]$,
the $f$ version gives a closed form  solution that is simply
\begin{align} \label{3D2DSolveALTforF.eq} 
 f  =&
  \frac{ \textstyle  \sum_{k=1}^{K} \left( (u_{k} x'_{k}+ v_{k} y'_{k})/z'_{k}\right) }
  { \textstyle  \sum_{k=1}^{K} \left(  ({x'_{k}}^2+  {y'_{k}}^2)/{z'_{k}}^2\right) } \ .
     \end{align}
 
 
 \qquad

While we propose an iterative three-step solution in Appendix \ref{perspective_soln.app}, 
here we simply test our orthographic solutions $\tilde{R}$ and $R_{\mbox{\tiny BI}}$, described 
above by \Eqn{PoseRotSolnPmatUN.eq}, and 
\Eqn{poseHornrot.eq}, against existing solutions for the perspective pose estimation problem, 
with the difference that the least-squares loss is measured as described in \Eqn{3D2DFPoseLSQ.eq} 
or \Eqn{3D2DFPoseLSQAlt.eq}, appropriately.  In \Fig{3DPoseGraphs.fig} we study the case in which
the point cloud is at the origin, and compare to the iterative LHM method \citep{LuHagerMj-FastPose-2000}.
In  \Fig{QmeanRotErr-vs-Npoints.fig}, we have adapted the available MatLab code for the MLPnP
method as in \citep{MLPnP-Urban-2016} to study the case where the camera is at the origin,
and compare our orthographic results to both LHM and MLPnP.
In all cases, we employ noisy ($\sigma = 0.1$) data, comparing different focal lengths, and in the case of
the MLPnP tests, different numbers of points N.  For the MLPnP tests, we supplement our least-squares measure with
a  quaternion-quaternion distance measure, to be more similar to the existing literature (e.g., \citet{MLPnP-Urban-2016}).
Here we use the quaternion-quaternion angle measure, 
\begin{equation}
\mbox{Rotation Error}(q_{\opt}, q_{0}) =  2 \arccos (q_{\opt} \cdot q_{0}) \ ,
\label{qoptvsqinput.eq}
\end{equation}
where $q_{0}$ is chosen to be the RMSD solution between the reference 3D point cloud and
the sample 3D point cloud after rotation and added noise, but before projection.  

It is plain to see that in both \Fig{3DPoseGraphs.fig} and \Fig{QmeanRotErr-vs-Npoints.fig}, 
our $R_{\mbox{\tiny BI}}$ rotation matrix outperforms both LHM and MLPnP, with improved performance
at longer focal lengths, as would be expected given our orthographic assumption.

 \clearpage
 
\subsection{Remarks}

 {\bf Properties of variational approaches:} The algebraic solution methods we have outlined here are not necessarily the most
effective approach.  In particular, numerical approximation methods involving
\emph{argmin} or \emph{argmax} numerical searches for the numerical optimization
of the loss functions that we have considered here can be quite effective, and
can efficiently enforce constraints on either the quaternion variables themselves,
or on the adjugate variables.  The latter have the advantage of course that the
singular domains of the quaternion determination can be explicilty avoided,
and the constraints essentially function as Lagrange multipliers if the
optimization is considered as a dynamical system.  Similarly, trainable
neural networks can function more or less equivalently to numerical search
optimization, with the particular advantage that, if successfully trained, the
expensive search process in an \emph{argmin} implementation, which is repeated
for each and  every new data instance, is skipped in every later application of a neural network.
Using the loss functions that we have presented to implicitly guide the training
target \emph{without   explicit training data} is basically the same context as
\emph{argmin}, except that the resulting successful search path can be efficiently
encoded for arbitrary future exploitation using one-time trained weights. 
 Strategies parallel to the \emph{argmin} method using the adjugate variable
 approach to the loss function can clearly be implemented using
neural networks, and such approaches should be effective for pose estimation.
Constraints can supplement the loss function used to update the weights (to function
much the same as Lagrange multipliers) in a neural network's virtual
dynamical system.  We intend to  treat these issues in detail elsewhere.

\qquad 

{\bf Possible applications to multiple datasets, bundle adjustment, and cryo-EM:} Here we have studied optimal alignment of single datasets with a single rigid point cloud and a single camera model. Many important applications examine collections of camera models providing an assembly of
data imposing restrictions on one or more imprecisely known point clouds. 
For example, the bundle adjustment problem \citep{schoenberger2016sfm,schoenberger2016mvs,TriggsEtAl:BundleAdj2000,ChenChenWanArxiv2016,Remondino-CIPA-2017} examines a collection of camera models, optimizes them individually, and then optimizes them collectively in combination with the optimization of the candidate point cloud coordinates.  This problem has many properties in parallel with the field of cryo-EM single particle analysis
 \citep[see, e.g.][]{relion-2012,cryosparc-2017,relion-2018, singer-singlecryo-2020}
 which determines the 3D structure of a molecule from a collection of 2D images of that molecule. The techniques that we have introduced here may also be able to contribute to the multiple dataset problem.

\qquad


\clearpage
%
%
%

%


\section{Conclusion}


Our objective in this paper has been to establish a clear framework for
understanding how quaternions must be treated in the context of measurable rotation
matrices, whatever the source.  There are many domains in which the use of
quaternions for representing rotations is attractive, and some recent papers \citep{zhou2019continuity,Peretroukhin2020,zhao2020quaternion,xiang2020revisiting}
have cast doubt on the validity of quaternions as an output parameterization
for automatic learning of the quaternion corresponding to implicit or explicit rotation matrices.
We have shown that a variational method based on the work of  \citet{BarItzhack2000}
recasts both ideal and noisy rotation measurements into the framework of
an adjugate matrix; this matrix contains four separate algebraic formulas for the
same quaternion corresponding to a given rotation matrix.  Each expression
is valid in a certain region of the quaternion manifold $\Sphere{3}$, but breaks
down with a singularity in the normalization outside its own region.  Combined,
however, these four
formulas, actually eight if we include their opposite signs, completely
cover the quaternion manifold with nonsingular patches.   The natural
occurrence of these singular regions and the ways to escape them by crossing
between formulas to cover the whole manifold then allow us to understand
quaternions from a consistent mathematical viewpoint.

Having established the importance of the adjugate, we adopted the quaternion adjugate variables, substituting single
adjugate variables for all possible quadratic quaternion forms, as a framework 
for treating matching and pose estimation problems.  This is of interest
because algebraic problems using quaternion variables are reduced in degree
by a factor of two in the adjugate variables.  The cost of this transformation is
the introduction of additional constraints, but in certain cases  the advantage
of using the adjugate variables instead of bare quaternions can be significant. 
Using this framework, we were able to solve the pose estimation problem with orthographic
projection, resulting, resulting in closed form least squares solutions valid
for perfect data, and correctable to optimal rotations for noisy data using a second-stage
Bar-Itzhack optimization.  Furthermore, we applied this result successfully to the pose estimation
problem with perspective projection, and found that even with the imperfect orthographic approximation
our results outperformed those of existing methods. Thus we argue
that the adjugate variables not only solve the question of how to understand the
quaternion manifold in rotation-determination tasks, but have applications of
their own in simplifying certain least squares  problems for optimal rotations.

\newpage

\section*{Acknowledgments}
We are indebted to B.K.P.~Horn for his crucial role in introducing
us to this problem, and for continuing support and encouragement,
and to Pascal Fua and Yinlin Hu for their generous advice and assistance.
SMH acknowledges the support of the Flatiron Institute and 
many helpful interactions with her colleagues there.
  
  
  \bibliography{QRotML}   

     \clearpage
 \appendix
 \section{Example Code for Rotation to Quaternion Algorithm}
  
 \label{RotToQuatCode.app}

 \subsection {Pseudocode}

  \begin{tabular}{rl} 
            & Compute the trace of $R$: $\tr = \mbox{\rm Trace}(R)$, noting that $\tr = 1 + 2 \cos \theta$.\\
             (if) & \emph{if}  $\tr>0$, then set $q_0 = \frac{\sqrt{ \tr + 1}}{2} = \cos(\theta/2)$.\\[0.1in]
            & Compute $s=(2 \sqrt{ \tr + 1})^{-1}= (4 \cos(\theta/2))^{-1}$,  then\\
             & set  $q_{1} = s*(m_{32} - m_{23})$, $q_{2} = s*(m_{13} - m_{31})$,   $q_{3} = s*(m_{21} - m_{12})$ \\[0.05in]
          (else) & \emph{else} check $m_{ii}$:\\
                 &  \emph{if}  $m_{11}$ is largest,  set $s = \sqrt{m_{11} -m_{22} - m_{33}+1}$\\
                 &\ \  \ set $q_{1} = s/2$,   set $s = 1/(2s)$ \\ 
                 & \ \ \ set $q_{0} = s*(m_{3,2} - m_{2,3})$,   $q_{2} = s*(m_{2,1} - m_{1,2})$, $q_{3} = s*(m_{1,3} - m_{3,1} )$\\
                 &   \emph{if}  $m_{22}$ is largest,  set $s = \sqrt{m_{22} -m_{33} - m_{11}+1}$\\
                  & \ \ \ set $q_{2} = s/2$,   set $s = 1/(2s)$ \\ 
                  & \ \ \  set $q_{0} = s*(m_{1,3} - m_{3,1})$,
                    $q_{3} = s*(m_{3,2} - m_{2,3})$, $q_{1} = s*(m_{2,1} - m_{1,2})$\\
                 &  \emph{if}  $m_{33}$ is largest,  set $s = \sqrt{m_{33} -m_{11} - m_{22}+1}$\\
                 & \ \ \  set $q_{3} = s/2$,   set $s = 1/(2s)$  \\ 
                 & \ \ \  set $q_{0} = s*(m_{2,1} - m_{1,2})$,
                    $q_{1} = s*(m_{1,3} - m_{3,1})$, $q_{2} = s*(m_{3,2} - m_{2,3})$\\
                    & Normalize to unity.
      \end{tabular}\\
 
 \subsection {C code}

 \begin{tabular}{|p{\ourfigwidth}|} 
\hline
\begin{footnotesize}
\begin{verbatim}
typedef struct tag_Quat {double w, x, y, z;} Quat;

/* quat->w is the scalar component, translated here
   internally as q[3] to facilitate the manipulation of
   the vector, or imaginary component, using indices 0,1,2 */
   
MatToQuat(double m[4][4], Quat * quat)
{ double  tr, s, q[4];    int  i, j, k;    int nxt[3] = {1, 2, 0};

  tr = m[0][0] + m[1][1] + m[2][2];

  /* check the diagonal */
  if (tr > 0.0) {
    s = sqrt (tr + 1.0);
    quat->w = s / 2.0;
    s = 0.5 / s;
    quat->x = (m[2][1] - m[1][2]) * s;
    quat->y = (m[0][2] - m[2][0]) * s;
    quat->z = (m[1][0] - m[0][1]) * s;
  } else {                
    /* diagonal is negative */
    i = 0;
    if (m[1][1] > m[0][0]) i = 1;
    if (m[2][2] > m[i][i]) i = 2;
    j = nxt[i];
    k = nxt[j];

    s = sqrt ((m[i][i] - (m[j][j] + m[k][k])) + 1.0);
                      
    q[i] = s * 0.5;
                            
    if (s != 0.0) s = 0.5 / s;

    q[3] = (m[k][j] - m[j][k]) * s;
    q[j] = (m[i][j] + m[j][i]) * s;
    q[k] = (m[i][k] + m[k][i]) * s;

    quat->x = q[0];
    quat->y = q[1];
    quat->z = q[2];
    quat->w = q[3];
  }
}
\end{verbatim}
\end{footnotesize}
\\
\hline
\end{tabular}

 \qquad
 
 \subsection{Mathematica Code}
 
 \begin{tabular}{|p{\ourfigwidth}|} 
\hline
 \begin{tabbing}
 (* This code randomizes the sign of q0 in the final step.  This is optional. *)\\[0.25in]
 
 RotToQuat[mat] := \ Module[\{qinit, q0, q1, q2, q3, trace, s, t1, t2, t3\},\\
  xxx  \= xxxx \= \kill
   \> trace = Sum [mat[[i, i]], \{i,1,3\}];\\
    \>  If[trace   $> $ 0, \\
   \> \> s = Sqrt[trace + 1]; \\
   \> \>  q0 = s/2; s = 1/(2 s); \\
   \> \>  q1 = (mat[[3, 2]] - mat[[2, 3]]) s; \\
  \> \>   q2 = (mat[[1, 3]] - mat[[3, 1]]) s; \\
  \> \>   q3 = (mat[[2, 1]] - mat[[1, 2]]) s, \\
 \>   If[mat[[1, 1]] $\geq$ mat[[2, 2]] \&\& mat[[1, 1]] $\geq$ mat[[3, 3]], \\
 \> \>   s = Sqrt[mat[[1, 1]] - mat[[2, 2]] - mat[[3, 3]] + 1];  \\
    \> \>  q1 = s/2;\\
    \> \>   s = 1/(2 s); \\
    \> \>  q0 = (mat[[3, 2]] - mat[[2, 3]]) s;\\
     \> \>      q2 = (mat[[2, 1]] + mat[[1, 2]]) s; \\
   \> \>  q3 = (mat[[1, 3]] + mat[[3, 1]]) s, \\
   \> If[mat[[1, 1]] $<$ mat[[2, 2]] \&\& mat[[1, 1]]  $\geq$   mat[[3, 3]], \\
    \> \>    s = Sqrt[mat[[2, 2]] - mat[[3, 3]] - mat[[1, 1]] + 1 ];  \\
     \> \>   q2 = s/2; s = 1/(2 s); \\
      \> \>   q0 = (mat[[1, 3]] - mat[[3, 1]]) s;  \\
    \> \>  q3 = (mat[[3, 2]] + mat[[2, 3]]) s; \\
   \> \>    q1 = (mat[[2, 1]] + mat[[1, 2]]) s,  \\
    \> \>    s = Sqrt[mat[[3, 3]] - mat[[1, 1]] - mat[[2, 2]] + 1 ];  \\
     \> \>   q3 = s/2; s = 1/(2 s); \\
     \> \>  q0 = (mat[[2, 1]] - mat[[1, 2]]) s; \\
     \> \>  q1 = (mat[[1, 3]] + mat[[3, 1]]) s;  \\
    \> \>    q2 = (mat[[3, 2]] + mat[[2, 3]]) s]]];  \\
  \> qinit = N[\{q0, q1, q2, q3\}]; \\
 \>  qinit = If[Abs[q0] $<   10^{-10}$, qinit, Sign[q0] qinit]; \\
 \>  qinit = (1 - 2 RandomInteger[\{0, 1\}]) qinit; \\
  \> Normalize[qinit]]
  \end{tabbing}
\\
\hline
\end{tabular}
  
  
  \section{Invariance of the Matching Problem Eigenvalues for Error-Free Data}
 \label{RMSDRotInvariance.app}
 
   In most realistic cases of the 3D Match problem, attempting to find the
   best rotation to align a cloud that is misoriented with respect to a the 
   reference cloud of its origin, 
the sample data occur  multiple times, with each instance related to the
reference data by a rotation, \emph{and} the sample data are somewhat
noisy, so that they do not correspond exactly to a pure rotation applied
to the reference data.  However, a particularly interesting feature appears
if there is no noise, or if the noise is effectively negligible.  In this case,
the self-covariance of the reference data completely determines the
eigenvalue of the profile matrix for each sample, and, since the loss function
for each sample is exactly the eigenvalue, they cannot be distinguished.
The \emph{eigenvectors} corresponding to that one single eigenvalue
are of course distinct, and, treated as quaternions, they determine which
rotation  must be applied to align each distinct sample with the
reference data.

  This feature of the  ''ideal`` noise-free situation can be proven as follows:
  first we take a reference data set of $K$ 3D points $Y = \{y_{k}\}$ and
  a given pairwise-matched sample  data set $X = \{x_{k}\}$, with each such
  point having a 3D index $a\in \{1,2,3\}$, e.g., $[x_{k}]_{a}$.  Then we deal with the
  cross-covariance matrix,
  \begin{equation} 
E_{ab} = \sum_{k=1}^{N} \, [x_{k}]_{\textstyle_a} \:[y_{k}]_{\textstyle_b} 
     = \left[ \Vec{X}   \cdot \Vec{Y}^{\t} \right]_{ab} 
\label{Edefnn.eq}
\end{equation}
where  $[x_{k}]$ denotes the $k$th column of $\Vec{X}$,
and the range of the indices $(a,b)$  is the spatial dimension $D=3$, 
and we examine the cross-term of the least-squares loss for the 
matching problem
\begin{equation}
 \Delta(q)\,  = \,  \tr R(q) \cdot E  \, = \, (q_0,q_1,q_2,q_3) \cdot M(E) \cdot
 (q_0,q_1,q_2,q_3)^{\t} \equiv q \cdot M(E) \cdot q \ .
\label{qM3qnnx.eq}
\end{equation}
 Here $M(E)$ is  the traceless, symmetric  $4\times 4$ matrix  
\begin{equation} 
     M(E) \!  = \!
\left[ \begin{array}{cccc}
 \!\!   E_{xx} + E_{yy} + E_{zz} & E_{yz} - E_{zy} & E_{zx} - E_{xz} &
                     E_{xy} - E_{yx} \!  \\
 \!   E_{yz} - E_{zy} &  E_{xx} - E_{yy} - E_{zz} & E_{xy} + E_{yx} & 
                     E_{zx} +E_{xz}  \! \\
 \!   E_{zx} - E_{xz} &  E_{xy} + E_{yx} & - E_{xx} + E_{yy} - E_{zz} & 
          E_{yz} + E_{zy}  \!  \\
  \!   E_{xy} - E_{yx} &  E_{zx} +E_{xz}  &  E_{yz} + E_{zy} & 
              - E_{xx} - E_{yy} + E_{zz}  \!\!
                     \end{array} \right] \ .
       \label{basicHornn.eq}              
\end{equation} 
built from our original $3\times 3$ cross-covariance matrix $E$ defined by \Eqn{Edefnn.eq}. 
We will refer to $M(E)$ from here on as the \emph{profile matrix}.

\qquad

We begin by writing down the eigenvalue expansion of the profile matrix,
\begin{equation}
\det [M - e I_{4}]  \; = \; 
     e^4 + e^3 p_1 + e^2 p_2 + e p_3 + p_4 \; =0\;  \label{3DeigEqnenn.eq}\ ,
\end{equation}
where $e$ denotes a generic eigenvalue,  $I_{4}$ is the 4D
identity matrix, and  the $p_{k}$ are homogeneous polynomials of degree $k$ in the elements of $M$.
For the special case  of a traceless, symmetric profile matrix $M(E)$ defined by \Eqn{basicHornn.eq},
the $p_{k}(E)$ coefficients simplify and can be expressed  numerically as the following 
 functions  either of $M$ or of $E$:
\begin{align}
p_{1}(E) & =  - \tr [ M ] \  = 0   \label{p1-3Dn.eq} \\
p_{2}(E) &   = -\frac{1}{2} \tr [ M \cdot M ]  \ = \   -2 \tr [ E \cdot  {E}^{\t} ] \nonumber \\
 & =  -2
   \left(E_{xx}^2+E_{xy}^2+E_{xz}^2+E_{yx}^2+E_{yy}^2+E_{yz}^2+E_{zx}^2
     +E_{zy}^2+E_{zz}^2\right) 
          \label{p2-3Dn.eq}     \\
p_{3}(E) &   =   -\frac{1}{3} \tr \left[ M\cdot M \cdot M\right] 
\  = \  -8 \det [ E ] \nonumber \\ 
  &  =    8 \left(E_{xx} \,E_{yz} \,E_{zy} +  E_{yy}\, E_{xz} \,E_{zx}+   E_{zz} \,E_{xy} \,E_{yx}   \right)  \nonumber  \\
   & \hspace{3em}   - 8 \left( E_{xx}\, E_{yy}\, E_{zz}+E_{xy}\,  E_{yz}\, E_{zx} + E_{xz}\, E_{zy}\, E_{yx}  \right)   \label{p3-3Dn.eq}  \\
p_{4}(E) & = \det [ M ]   \ = \    
                         2  \tr [  E \cdot  {E}^{\t} \cdot  E \cdot  {E}^{\t} ] - \left( \tr [  E \cdot  {E}^{\t}  ]\right)^2 
     \label{p4-3Dn.eq}  \ .
\end{align}
Interestingly, the polynomial $M(E)$ is arranged so that $-p_{2}(E)/2$ is the (squared) 
Fr\"{o}benius norm of $E$, and $-p_{3}(E)/8$ is its determinant.
Our task now is to express the four eigenvalues   $e = \epsilon_{k}(p_1,p_2,p_3,p_4)$,
 $k=1,\ldots,4$, usefully in 
 terms of the matrix  elements, and also to find their eigenvectors;  we are of course particularly  interested in the  maximal eigenvalue $\epsilon_{\opt}$.

\qquad

{\bf Invariance.}  If we now write the sample matrix to explicitly show its derivation from
the reference matrix, that is, ${x_{k}}^{a} = R(q)_{ab} {y_{k}}^{b} $ for all $k\in \{1,\ldots K\}$,
then the cross-covariance matrix becomes
\begin{align}
E_{ab}\ &= \ \sum_{k=1}^{N} \, R(q)_{ac}[y_{k}]_{\textstyle_c} \:[y_{k}]_{\textstyle_b} 
    \ \   = \ \   R(q)_{ac}  \left[ \Vec{Y}   \cdot \Vec{Y}^{\t} \right]_{cb}  \ .
\label{EdefRnn.eq}
\end{align}
We now take the three distinct components of Eqs.~(\ref{p2-3Dn.eq}), (\ref{p3-3Dn.eq}),
and (\ref{p4-3Dn.eq})  and write them out as
\begin{gather}
\left\{ \label{pkInvarProofnn.eq}
\begin{align} 
\det [ E ]  & =  \det  \sum_{k } R_{ac} {y_{k}}^{c} {y_{k}}^{a}  \nonumber  \\
               & =  \det  R_{ac} \sum_{k }{y_{k}}^{c} {y_{k}}^{a}  \nonumber \\
                  & = \det E[Y,Y] \, \equiv \det E_{0} \\
\tr [ E \cdot  {E}^{\t} ] & =   \sum_{k,k'} R_{ac} {y_{k}}^{\textstyle_c} {y_{k}}^{\textstyle_b} 
        R_{ad} {y_{k'}}^{\textstyle_d} {y_{k'}}^{\textstyle_b} \nonumber  \\
     & = \sum_{k,k'}   {y_{k}}^{\textstyle_c} {y_{k}}^{\textstyle_b} 
                                {y_{k'}}^{\textstyle_c} {y_{k'}}^{\textstyle_b}  
                                   \ \  =  \ \ \tr  E_{0} \cdot  {E_{0}}^{\t}   \\
 \tr [  E \cdot  {E}^{\t} \cdot  E \cdot  {E}^{\t} ] & =   
 \tr [  E_{0} \cdot  {E_{0}}^{\t} \cdot  E_0 \cdot  {E_{0}}^{\t} ] 
 \end{align} \right. \  .
 \end{gather}
Thus we know that, modulo assuming irrelevant errors in the data, the
entire characteristic equation  \Eqn{3DeigEqnenn.eq} is independent of the
rotation matrix applied to obtain the sample data, and thus the eigenvalues
of the profile matrix $M(E)$ are independent of the applied rotations embodied
in the sample data;  only the self-covariance of the reference data 
\begin{equation}
E_{0} = \Vec{Y}   \cdot \Vec{Y}^{\t}
\end{equation}
enters into the determination of the eigenvalues of $M(E)$, and since $E_{0}$ is
symmetric, the last three elements of the top row and left-hand column of
$M(E_{0})$ \emph{vanish},  making the maximum eigenvalue  just 
\[ \mbox{\rm [max eigenvalue]}(\;M(E_{0})\;) =  \tr[E_{0}] . \]

\qquad

 \section{The Adjugate Matrix}  
  \label{adjugatematrix.app}
  
    Standard methods for finding the rotation aligning a rotated 3D point cloud with its
            reference cloud \citep[see, e.g.,][]{Horn1987,Hanson:ib5072}  determine  the optimal quaternion
             by finding the maximal eigenvalue of a certain $4 \times 4$ matrix.
               The normalized eigenvector of that maximal eigenvalue is $q_{\opt}$,  the
             quaternion determining the optimal aligning rotation matrix $R_{\opt} =
             R(q_{\opt})$.  Buried in the last step of this  routine linear algebra calculation,
             we have in fact a mandatory process that closely parallels the complicated
             axis-angle procedure just described above.  The key observation is that there is an
             ambiguity in the process of going from a symmetric real matrix $M$ and one of its
             eigenvalues $\lambda$ to a well-behaved corresponding eigenvector. 

             One needs a small piece of linear algebra to follow this train of thought.
             First, we recall that the determinant of a square real matrix can be computed
             using the   \emph{Adjugate Matrix}  built from the transposed cofactors of the matrix.
             Dividing the Adjugate by the determinant itself yields the 
             \emph{inverse}, that is, if the matrix $M$ is nonsingular, then
             \[ M^{-1} = \frac{\textstyle\mbox{Adjugate}(M)}{\det M}
             \hspace{.2in} \rightarrow \hspace{0.2in}  
             M \cdot \mbox{Adjugate}(M) = \det M\; I_{4} \ ,\]
             where $I_4$ is the 4D identity matrix.   Here we will be particularly interested in
             the special case where $\det M = 0$, so that the inverse does not exist.
             Nevertheless, the  Adjugate matrix   may \emph{still exist} for singular matrices
              because it is completely linear:
             \[ \mbox{Adjugate}(M)_{ij} =  \mbox{Transpose}\left(\mbox{Cofactor Matrix}](M_{ij} )\right)\ .\]
              However,  another step is needed to understand how we will exploit the
             Adjugate.   All the matrices we will be examining will be real symmetric
             4D matrices,  so we there will exist a \emph{characteristic  equation}
             with real eigenvalues $\lambda$ depending on four polynomials of
             the set of matrix elements $\{ m_{ij} \}$ of $M$, which takes  the form
             \begin{equation} \label{charEqnM.eq}
             \begin{aligned}
             \det \left( M(m) - \lambda \,I_{4} \right) &=  \lambda^4 + \lambda^3 \,p_{1}(m)
             + \lambda^2 \, p_{2}(m)+ \lambda  \, p_{3}(m) + p_{4}(m)\\
                 &=  0 \ .
              \end{aligned}
                          \end{equation}
             Here  the $p_{k}(m)$ are polynomials  of degree $k$ in the matrix elements $m$.
             The characteristic equation    \Eqn{charEqnM.eq} of $M$ has four real
             roots, and in ordinary circumstances these are distinct, each has an
             eigenvector, and there is a unique ``maximal'' eigenvector corresponding
             to the largest eigenvalue, which we now denote   by $\lambda_{\opt}$.
             \emph{By definition} substituting $\lambda_{\opt}$ into the characteristic
             equation solves the eigenvalue problem, so we may denote this special
             context of the characteristic equation by the matrix
              \[     \chi = \left[ M - \lambda_{\opt}\, I_{4}  \right] \ , \]
                   where $\chi$ itself is singular, $\det \chi \equiv 0$.
                    Then the \emph{Adjugate
              matrix}  $A(M,\lambda)$ is defined as the transposed cofactor matrix of $\chi$, 
              constructed precisely
              so that each column of $A$ produces the \emph{determinant}   $\det \chi$.
              Now, since the determinant of $\chi$ vanishes by construction, we can split
              the two terms in the definition of $\chi$ as follows:
                   \[   \chi \cdot A = M \cdot A - \lambda A = \det \chi  \equiv 0 \ .\]
                Thus \emph{all four columns }of the  Adjugate $A$ are in principle eigenvectors
                of  the \emph{same} $\lambda_{\opt}$.    The essential caveat is that
                if  any of the  rows or columns of the (symmetric) matrix
                $A$ have a vanishing or very small norm, that eigenvector will be of
                little practical use, and a \emph{different column} must
                be chosen as the eigenvector corresponding to $\lambda_{\opt}$ for further
                calculations.  Typical linear algebra libraries perform these checks by
                default using a variety of methods; here, we will argue that, due to the
                nontrivial spherical manifold $\Sphere{3}$ 
                on which quaternions live, one must explicitly take into account
                the properties of the  \emph{entire}  Adjugate matrix as it appears in  quaternion
                calculations   that involve an explicit or  implicit eigensystem.  We show 
                explicitly in the main text that
                typical  contexts that extract quaternions from measured  rotation matrices
                meet this criterion.

  {\bf Summary of Results.}  
               Only the \emph{Adjugate matrix} of a particular 
               $4\times 4$ matrix, written in terms of
 unnormalized quadratic pre-quaternion forms, can be expressed in a nonsingular
 fashion in terms of rotation matrix elements.  If we simply set \Eqn{Rofqq.eq}
 equal to \Eqn{R.axisangle.eq},  $R(q) =  R (\theta,\Hat{n})$, assuming that
 $R (\theta,\Hat{n})$ corresponds to an ideal noise-free measurement, we will construct
 a symmetric $4\times 4$ matrix of algebraic expressions for all 10 quadratic quaternion
 terms that is, critically,  square-root-free and division-free:
  \begin{equation}  
  K(q) = \! \left[
\begin{array}{cccc}
 {q_0}^2 & q_0 q_1 & q_0 q_2 & q_0 q_3 \\
 q_0 q_1 & {q_1}^2 & q_1 q_2 & q_1 q_3 \\
 q_0 q_2 & q_1 q_2 & {q_2}^2 & q_2 q_3 \\
 q_0 q_3 & q_1 q_3 & q_2 q_3 & {q_3}^2 \\
\end{array}
\right] 
\! =   
\frac{1}{2} \left[
 \begin{array}{cccc}
 1+c & s  \, \hat{n}_{1}   &  s  \, \hat{n}_{2}    & s  \,  \hat{n}_{3}    \\
 s  \,  \hat{n}_{1}   & (1-c)\,  {\hat{n}_{1}\!}^{2} & (1-c) \, \hat{n}_{1}   \hat{n}_{2}  & (1-c) \, \hat{n}_{1}   \hat{n}_{3}  \\
 s  \,  \hat{n}_{2}    & (1-c)  \,\hat{n}_{1}   \hat{n}_{2}  & (1-c) \, {\hat{n}_{2}\!} ^2 &
      (1-c) \, \hat{n}_{2}   \hat{n}_{3}  \\
 s  \,  \hat{n}_{3}   & (1-c) \, \hat{n}_{1}   \hat{n}_{3}  & (1-c) \, \hat{n}_{2}   \hat{n}_{3}  & (1-c)\, {\hat{n}_{3} \!}^2  
\end{array}
\right] . 
\label{3DAAquatAdjugatesX.eq}
\end{equation}
Note that, in the exact case, the quaternion origins of the polynomials in $K(0)$ imply  the existence
of nontrivial constraints.   For the situation of inexact rotation data $R(m)$ 
replacing $R (\theta,\Hat{n})$, similar expressions hold with the same qualitative features.

In all cases, each row (or column) of the measured data on the right-hand side
  corresponds to a full 4-element quaternion \emph{scaled} by a factor of 
  $q_0$, $q_1$, $q_2$, or $q_3$  as 
the only type of object that can be reliably expressed without encountering normalization singularities.
These can be removed by simple normalization \emph{if} the normalizing divisor
is not too near the floating point precision limit for zero.  
There are in fact 14 different sets of singular quaternion submanifolds, corresponding topologically
to the 4 vertices, 6 edges, and 4 faces of an abstract tetrahedron, that is, quaternions
with zeroes in three, two, or one of their four elements.  The constraint $q\cdot q = 1$
guarantees that  \emph{at least one row} of \Eqn{3DAAquatAdjugatesX.eq}
  can always be normalized to produce a valid quaternion $q_{\opt}$ and its
  pure rotation matrix $R(q_{\opt})$ that is the best
approximation to the numerically measured matrix $R(\theta,\Hat{n})$ or $R(m)$.
The algorithm is, not surprisingly, very close to the classic algorithm of \citet{Shepperd1978}:
find the row whose diagonal element is the largest, and normalize that
row to find a quaternion $q_{\opt}$.  This is also similar, but not identical, to the 
Ansatz of \citet{xiang2020revisiting}, which achieves
a non-singular result heuristically without our  more complete theoretical underpinnings.

\clearpage

\section{ The Fourteen Adjugate Matrices Exhibiting Unnormalizable Quaternion Geometry}
\label{AdjZeroMatrices.app}


  
\begin{figure}[h!]
\vspace{-.4in}
\figurecontent{
\centering
 \includegraphics[width=2.7 in]{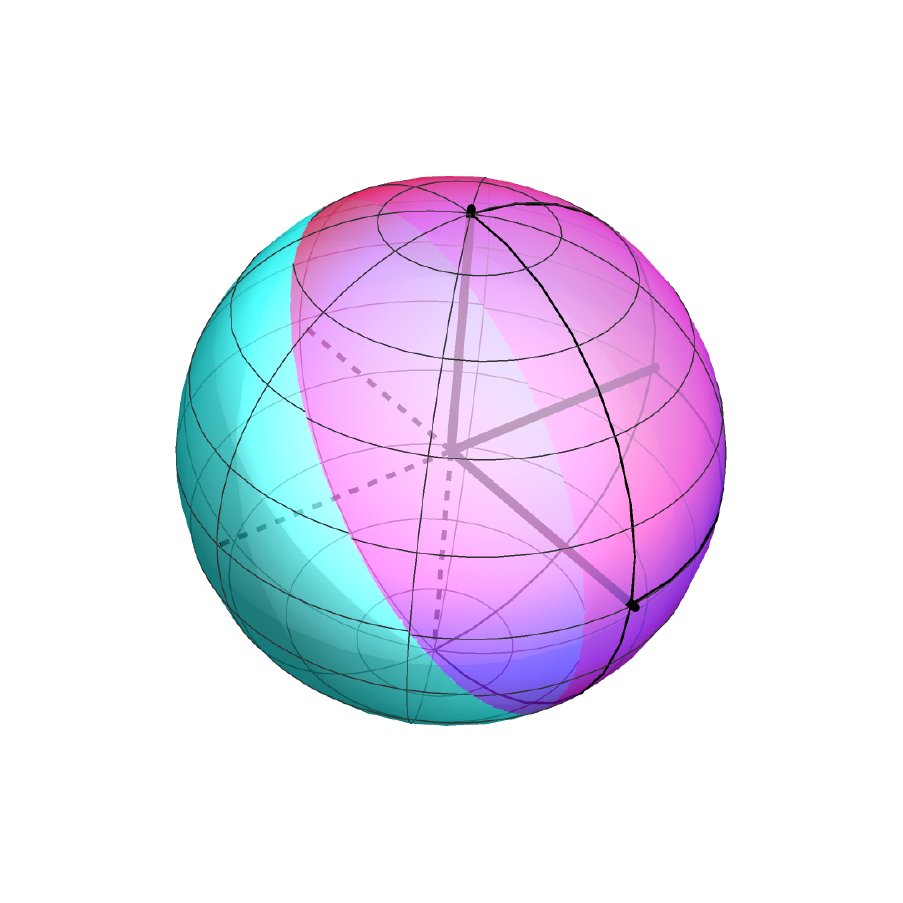} \hspace{.1in}
    \includegraphics[width=2.7 in]{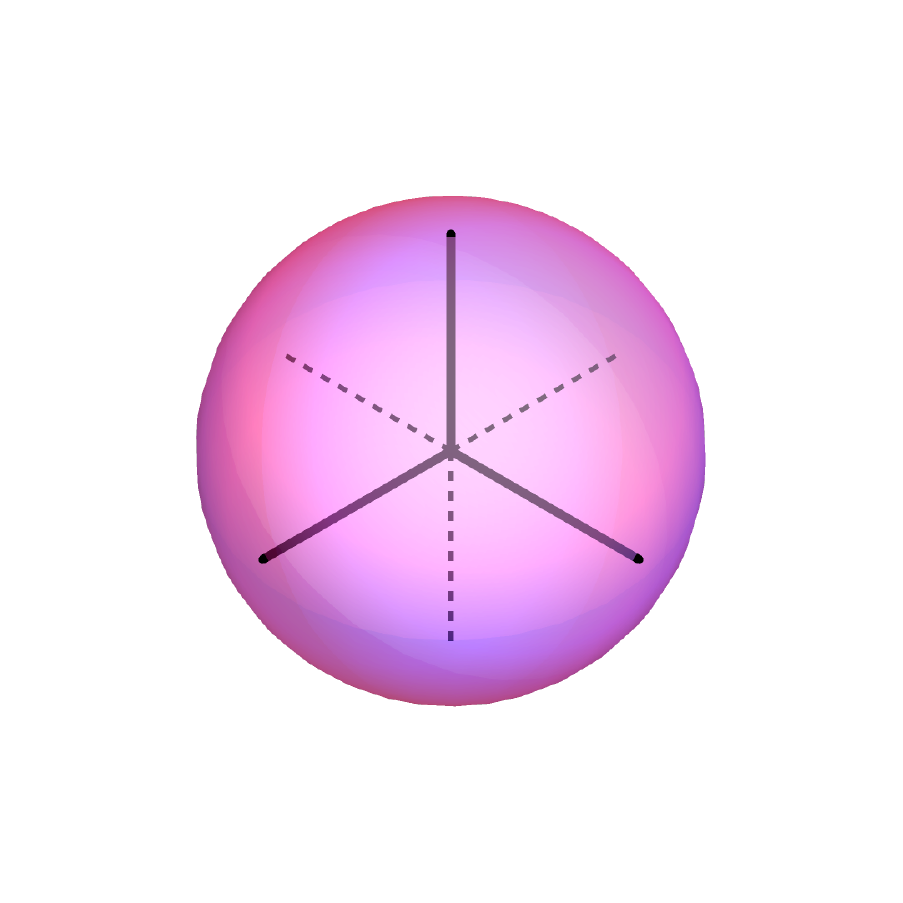}  \\[-.4in]
    \hspace*{1.2in}  (a) \hfill (b)   \hspace*{1.2in} \\[-0.2in]
    \includegraphics[width=2.7 in]{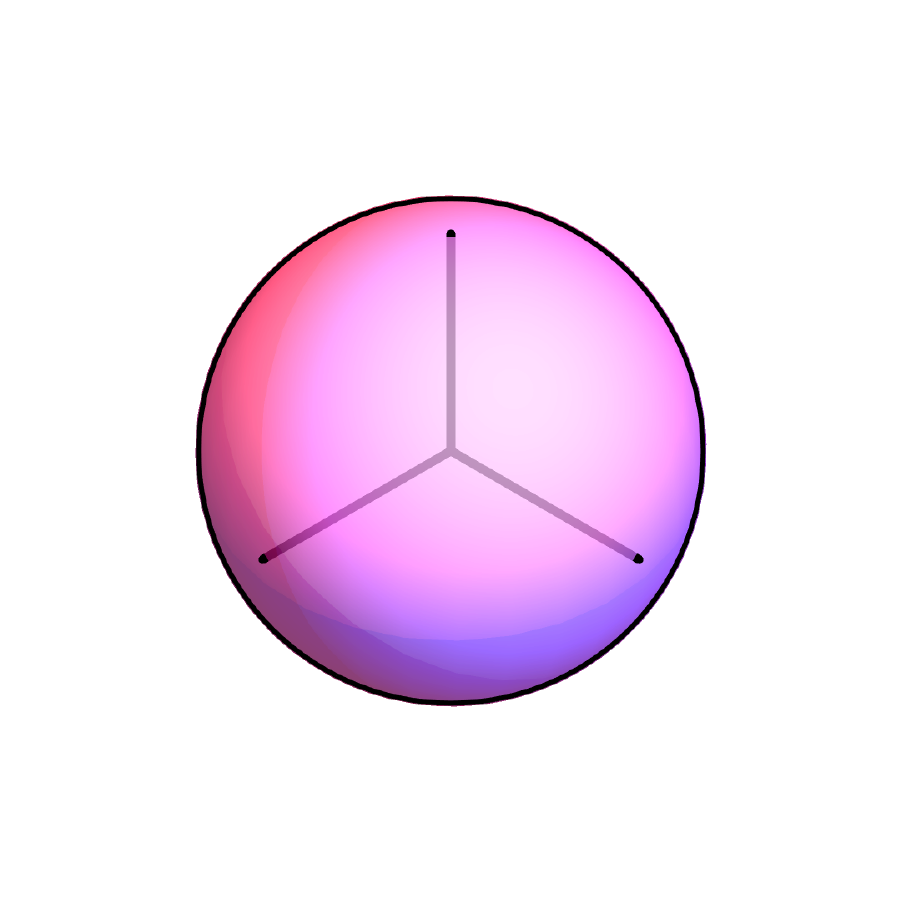} \hspace{.1in}
    \includegraphics[width=2.7 in]{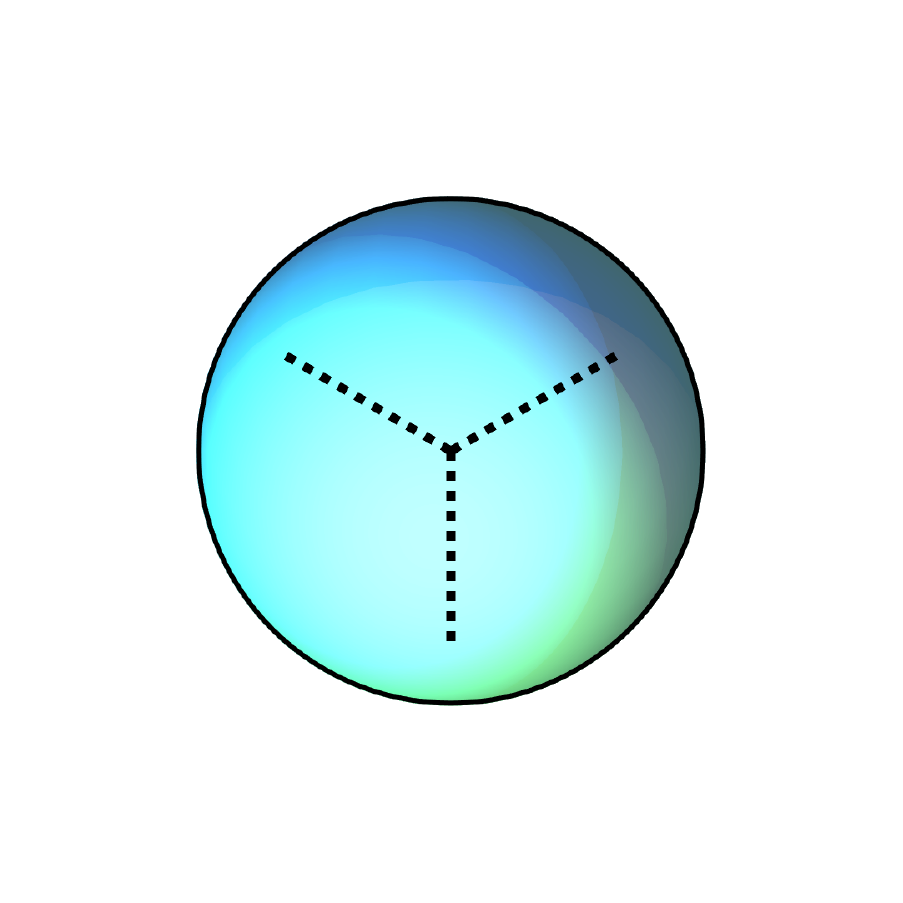}  \\[-.4in]
    \hspace*{1.2in}  (c) \hfill (d)   \hspace*{1.2in} \\[0.4in]
    } 
\caption[]{\ifnum\ShowFiles=1 {\bf figspdf/S2-axes-oblique.pdf, S2-axes-direct.pdf,
S2-disk-plus.pdf, S2-disk-minus.pdf.  }\fi
 \footnotesize  
 {\bf Analog of $\Sphere{3}$ projection with the $\Sphere{2}$ full and hemispherical
 projection.}.  (a) The two-sphere $\Sphere{2}$ contains three orthogonal axes in its
 3D-space projection, shown obliquely here  to expose the positive (solid)  and negative
 (dashed) ends of the coordinate axes using a general viewpoint.
 (b)  If we look straight down the diagonal, the three axes, both positive and negative
 ends of the axes appear in the 2D image to be the vertices of an equilateral
 triangle. (c,d) If we simply display the 2D disk with the positive axes in a planar
 image separately from the 2D disk with the negative axes, we can see a 
 (flattened) depiction in which \emph{every single point} of $\Sphere{2}$ is
 visible and distinct.  In order to make every single point of the quaternion
 hypersphere  $\Sphere{3}$ visible in our images, we will  simply put the \emph{four}
 axes of 4D space at the symmetrical vertices of a tetrahedron, and use a pair 
 of solid  balls (simulating 3D space of course using 3D graphics images) instead
 of a pair of filled disks. \\[0.5in] }
 \label{obliqueS2Axes.fig}
\end{figure}

\mypar{Strategy for depicting the full quaternion map.}  Despite its four-dimensional intrinsic
nature, quaternion geometry can be depicted in a fairly accurate way if we are
willing to follow some analogies between lower dimensional and higher dimensional
spheres.   First, we show in \Fig{obliqueS2Axes.fig}(a) an ordinary sphere $\Sphere{2}$
embedded in 3D Euclidean space $\R{3}$, with the three orthogonal axes $\Hat{x}$,
$\Hat{y}$, and $\Hat{z}$, projected in the familiar way to a 2D image.  Even though
the image is a dimension lower than the actual 3D object being depicted, we are
accustomed to interpreting this image as a 3D object.  Now rotate the sphere
as in  in \Fig{obliqueS2Axes.fig}(b) so that the three axes are projected equally
onto the 2D image, with the ends of the axes forming the vertices of an equilateral
triangle.  Now we see that this projection corresponds to one hemisphere of
$\Sphere{2}$ flattened into a disk containing all three positive axes, and the
back hemisphere as a second disk containing all three negative axes.  It is
clear that if we create two separate images as in \Fig{obliqueS2Axes.fig}(c,d),
\emph{every single point} on the manifold $\Sphere{2}$ can be seen in the
two separate hemispherical images.  We can do the same thing with a full
quaternion map using a \emph{solid ball} containing a 3D quadruple of positive
axes (with the four axis ends being the vertices of a tetrahedron), paired with
a matching solid ball containing the symmetric projections of the four
negative axes.  Every point of the quaternion sphere is visible in the two
solid balls, exactly analogous to the two filled disks for the hemispheres
of $\Sphere{2}$ in \Fig{obliqueS2Axes.fig}(c).

\qquad

\begin{figure}[h!]
\vspace{0in}
\figurecontent{
\centering
 \includegraphics[width=2.7in]{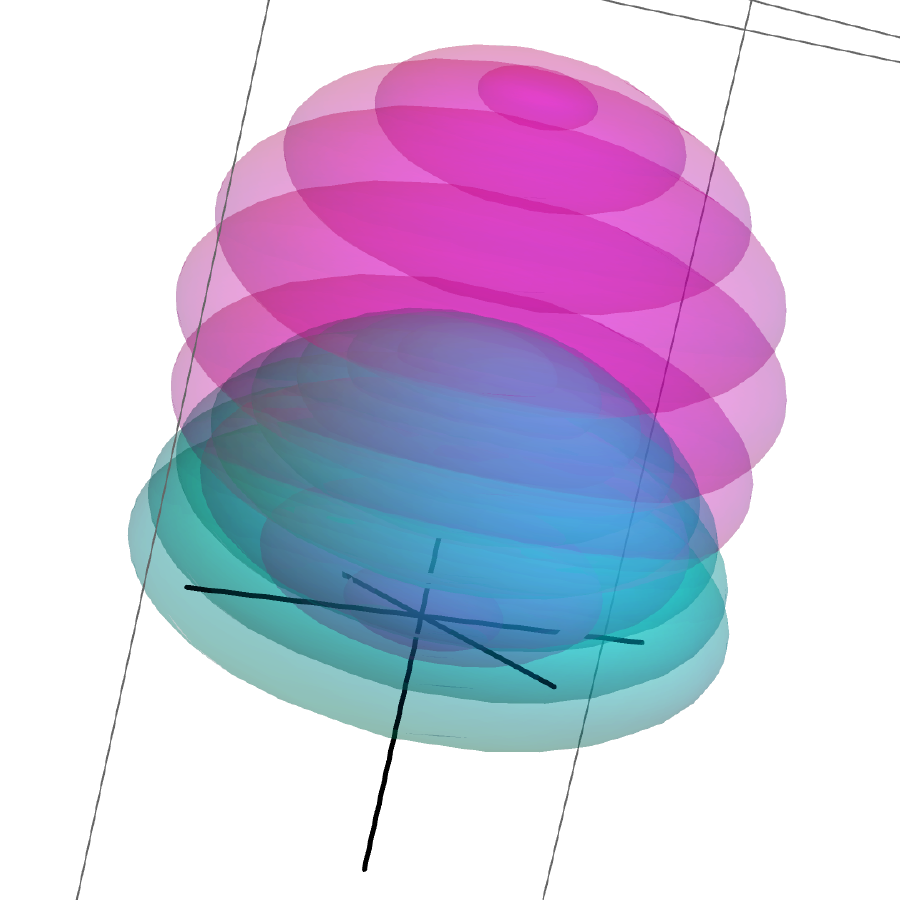} \hspace{.1in}
    \includegraphics[width=2.7in]{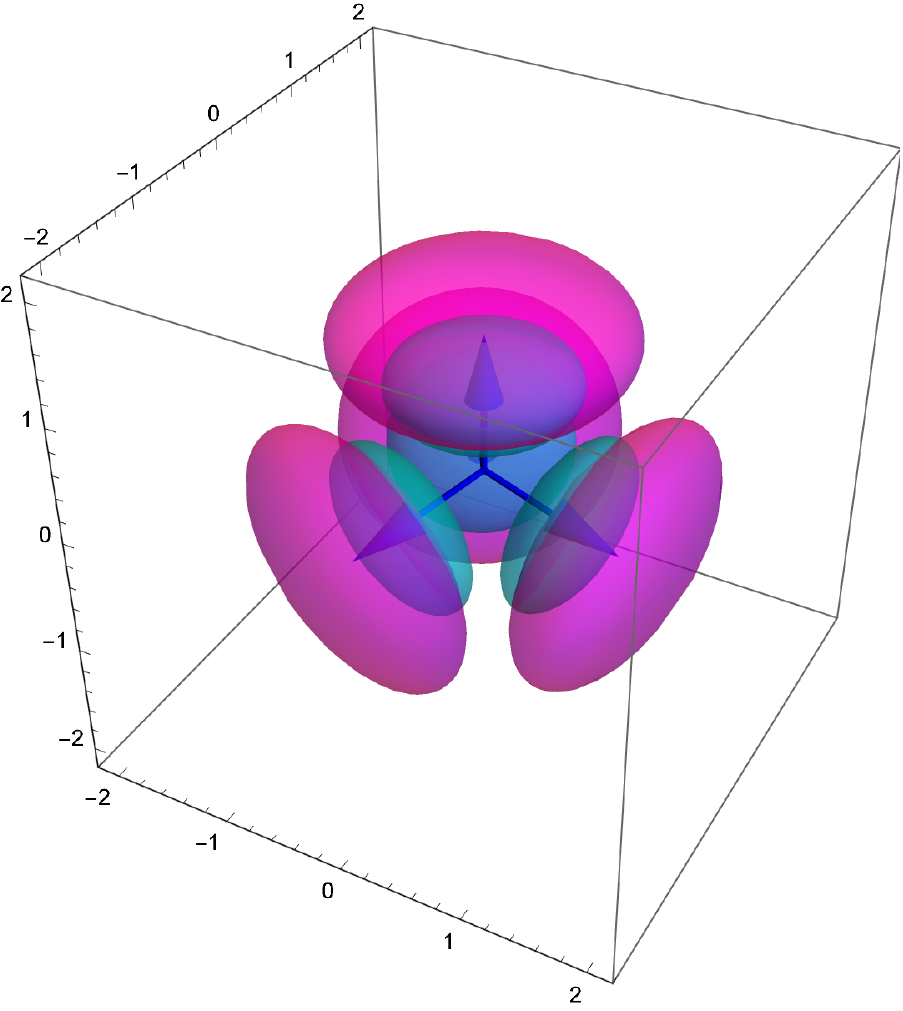}  \\
    \hspace*{1.2in}  (a) \hfill (b)   \hspace*{1.2in} \\
    } 
\caption[]{\ifnum\ShowFiles=1 {\bf S3-projection-1.pdf, S3-projection-4.pdf. }\fi
 \footnotesize  
 {\bf 3D subspace showing three axes of singularities.}
  In this 3D subspace of quaternion space, there are partial spheres instead
  of partial circles, but the singularity occurs in the same way, as the sphere
  closes in on the origin, normalization is impossible. 
  The four axes correspond to the $q_0$, $q_1$, $q_2$, and $q_3$ quaternion
  directions projected down symmetrically to the directions of the  four vertices of
  a 3D regular tetrahedron.   While there are 8 actual axes in 4 pairs, corresponding to the
  two pairs of axes $x=\pm 1$ and $y=\pm1$ in \Fig{2Dquaternionsingularities.fig},  here for readability
  we can only show the four positive axes directions whose manifold patches cover
  the solid ball that is the ``Northern hemisphere'' of $\Sphere{3}$;  the unshown ``Southern
  hemisphere'' is the second solid ball that, sewn onto the Northern hemisphere along
  the $\Sphere{2}$ equator,  completes the full $\Sphere{3}$ manifold.
   (a) A single pair of solid-ball patches, the larger corresponding to the $+q_{0}$ direction
   of the solution space $q_{0}\left(q_0,\,q_1,\,q_2,\,q_3\right)$; the smaller ball
   is the normalized version collapsing to a patch on the actual $\Sphere{3}$ patch that
   normalizes without singularity in the neighborhood of $q_{0} \approx +1$.
   (b) The four pairs that cover the nonsingular patches around 
   $q_0\approx +1$, $q_1\approx +1$, $q_2\approx +1$, and $q_3\approx +1$.
   The actual mathematical balls correspond to a volume rendered solid, which
   is difficult to portray in a figure, so multiple level sets are shown for each ball
   to depict the continuous volume with a finite sampling of spherical surfaces.
  }
 \label{4D8arcs.fig}
\end{figure}

The full quaternion map from the unnormalized representation to the normalized
true quaternion sector is divided into eight distinct regions, in opposite signed
pairs that represent equivalent rotations due to the identification $R(q)=R(-q)$.
Instead of portions of ordinary spheres as in the top of
 Figure \ref{ZeroAdjSingAB.fig}, we have
portions of hyperpheres centered at  $(\pm 1, 0,0,0)$, $(0,\pm 1,0,0)$, 
$(0,0,\pm 1,0 )$, and $(0, 0,0,\pm1)$.  Instead of being partial hemispherical
surfaces, these are now solid balls, each corresponding  to a portion of a set
of overlapping hemispheres of the quaternion manifold $\Sphere{3}$.  These
are difficult to draw, but an attempt can be made by projecting the axes of
the 4D space into 3D in the symmetric directions of the vertices of a tetrahedron.
In Figure \ref{4D8arcs.fig}(a), we show first a collection of slices of the
solid ball at various radii in 4D, aligned with one axis, for a single choice
of the eight  unnormalized and normalized maps.   Then in  Figure \ref{4D8arcs.fig}(b),
we reduce the number of samples of the solid balls to one, but show
a representative pair of unnormalized and normalized slices for \emph{the
 four  positive unit 4D axes}; there is another opposite sign counterpart for each
of these four that is omitted for clarity.

Perhaps another useful way to look
at the algebraic implications of quaternion adjugate normalization anomalies is simply to write down 
the fourteen  $[q_{\mu}q_{\nu}]$
matrices that result for any combination of single, double, and triple choices of indices
for the  $q_{k}$ from the set $k \in \{0,1,2,3\}$.  For the four point pairs, with any three zeroes
chosen, there are no degrees of freedom left, only identity quaternions; for the six with
two zeroes chosen, there is a circle lying on $x^2+y^2=1$, while for the four remaining
singularities, there is a sphere $\Sphere{2}$ described by the constraint $x^2+y^2+z^2=1$.
The matrices here are essentially the algebraic    
 version of the graphics in \Fig{ZeroAdjSingAB.fig}:  
\begin{equation}
\begin{array}{rc c rc c rc c rc}
     {\scriptstyle 123:} \!\!  \!\! &   \left[ \begin{array}{cccc}
      \pm 1 & 0 & 0& 0\\
      0 & 0 & 0 &0\\
      0 & 0 & 0& 0\\
      0 &0 & 0& 0\\
      \end{array} \right] &\hspace{-.2in} &
      {\scriptstyle  023:} \!\! \!\! &  \left[ \begin{array}{cccc}
      0 & 0 & 0& 0\\
      0 & \pm 1 & 0 &0\\
      0 & 0 & 0& 0\\
      0 &0 & 0& 0\\
      \end{array} \right] &  \hspace{-.2in}&
      {\scriptstyle   013: } \!\! \!\!  &  \left[ \begin{array}{cccc}
     0 & 0 & 0& 0\\
      0 & 0 & 0 &0\\
      0 & 0 & \pm 1& 0\\
      0 &0 & 0& 0\\
      \end{array} \right] & \hspace{-.2in}&
       {\scriptstyle  012:} \!\! \!\!  &  \left[ \begin{array}{cccc}
      0 & 0 & 0& 0\\
      0 & 0 & 0 &0\\
      0 & 0 & 0& 0\\
      0 &0 & 0& \pm  1\\
      \end{array} \right]    \\[0.35in]
       01:   \!\! \!\!&   \left[ \begin{array}{cccc}
      0 & 0 & 0& 0\\
      0 & 0 & 0 &0\\
      0 & 0 & x& y\\
      0 &0 & y& x\\
      \end{array} \right] &\hspace{-.2in} &
      02 :  \!\! \!\!&  \left[ \begin{array}{cccc}
      0 & 0 & 0& 0\\
      0 & x &0 &y\\
      0 & 0& 0& 0\\
      0 &y & 0& x\\
      \end{array} \right] & \hspace{-.2in} &
       03:   \!\! \!\! &  \left[ \begin{array}{cccc}
     0 & 0 & 0& 0\\
      0 & x & y &0\\
      0 & y & x& 0\\
      0 &0 & 0& 0\\
      \end{array} \right] &\hspace{-.2in} & \\[0.35in]
       23:  \!\! \!\! &  \left[ \begin{array}{cccc}
      x & y & 0& 0\\
      y & x & 0 &0\\
      0 & 0 & 0& 0\\
      0 &0 & 0& 0\\
      \end{array} \right]   &\hspace{-.2in}&
       31:   \!\! \!\! &   \left[ \begin{array}{cccc}
      x & 0 & y& 0\\
      0 & 0 & 0 &0\\
      y & 0 & x& 0\\
      0 &0 & 0& 0\\
      \end{array} \right] &\hspace{-.2in} &
      12:  \!\! \!\! &  \left[ \begin{array}{cccc}
      x & 0 & 0& y\\
      0 & 0 & 0 &0\\
      0 & 0 & 0& 0\\
      y &0 & 0& x\\
      \end{array} \right]  &\hspace{-.2in}&  \\[0.35in]
     0:  \!\! \!\!  &  \left[ \begin{array}{cccc}
      0 & 0 & 0& 0\\
      0 & x & y & z\\
      0 & z & x& y\\
      0 &y  & z& x\\
      \end{array} \right]    &\hspace{-.2in}&
       1:  \!\! \!\! &  \left[ \begin{array}{cccc}
      x & 0 & y & z\\
      0 & 0 & 0& 0\\
      z & 0 & x& y\\
      y &0 & z& x\\
      \end{array} \right]       &\hspace{-.2in}&
       2:  \!\!  \!\!&  \left[ \begin{array}{cccc}
      x  & y  & 0& z \\
      z  & x & 0 & y\\
      0 & 0 & 0& 0\\
      y &z & 0& x\\
      \end{array} \right]    &\hspace{-.2in}&
     3:  \!\! \!\!&     \left[ \begin{array}{cccc}
      x  & y  & z & 0\\
      z  & x & y & 0\\
      y & z & x& 0\\
      0 &0 &0&0\\
      \end{array} \right]       \\
      \end{array} \ .
      \label{tableofSing.eq}
\end{equation}



\clearpage

\section{ Proposed Solution Procedure For Perspective}
\label{perspective_soln.app}

 {\bf Proposed Solution Procedure For Perspective.}  Without making any claims to an optimal solution, we proceed
to describe a \emph{reasonable} three-step procedure following the concepts laid out
in Section \ref{3DPose.sec}:
\begin{itemize}
\item {\bf  Find the best orthographic approximation to the needed rotation.}  From
the assumed input point-cloud data $\{\Vec{x}_{k}\}$ and the matched image-plane data
$\{\Vec{u}_{k}  \}$, we can calculate from \Eqn{PoseRotSolnPmatUN.eq}  a good approximation
to the needed rotation $\tilde{R}$ that is very accurate at the center of the projected
image, and is disturbed by the growing perspective projection distortion for points
farther from the projection center.  In many cases, this approximation itself will
already be quite good.  
\item {\bf Optimize  $\tilde{R}$.}  However, as before, due to the errors being introduced
relative to a perfect orthographic data set (in this case compounded by perspective
distortion), $\tilde{R}$ will not in general 
be an actual rotation; we proceed again to apply
the Bar-Itzhack optimization as in section \ref{3DPose.sec} to produce a quaternion
$q_{\opt}$ that corresponds to 
 the \emph{optimal}  pure rotation matrix $R_{\opt} = R(q_{\opt})$
closest to  the imperfect $\tilde{R}$.
\item{\bf Define the focal-length determination as a separate optimization step.}
If we want to provide an estimate for the camera position via the focal length,
we can add a final step to the procedure by simply inserting the now-constant matrix
\[R_{\opt} = \left[ 
    \begin{array}{c} P(q_{\opt}) \\ D(q_{\opt}) \\ \end{array}
        \right] \]
into \Eqn{3D2DFPoseLSQ.eq}.  This now becomes a loss function
$\mathbf{S}(\bar{f})$  depending \emph{only}
on the inverse focal length $\bar{f}$, or $\mathbf{S}(f)$  depending \emph{only}
on the    focal length $f$.   Upon differentiation with respect to $\bar{f}$, the first
context results in an equation  whose numerator is a polynomial  of order $(3K-2)$,
and can easily be solved numerically to give $\bar{f}$.  The alternative using
 $\mathbf{S}(f)$ results in the explicit solution \Eqn{3D2DSolveALTforF.eq} for $f$.
\end{itemize}
There also appears to be no obstacle
to using this as an improved starting point for iterative refinement schemes such as
\citet{LuHagerMj-FastPose-2000}.

    \end{document}